\pdfoutput=1
\documentclass[11pt]{article}
\usepackage[left=1in,right=1in,top=1in,bottom=1in]{geometry}
\usepackage{setspace}
\usepackage{amsmath,amssymb,amsthm,bbm}
\usepackage{natbib}
\usepackage{color,hyperref,graphicx}
\usepackage[justification=centering]{caption}
\hypersetup{pdfborder = {0 0 0},colorlinks=true,linkcolor=blue,urlcolor=blue,citecolor=blue}
\usepackage{booktabs}
\usepackage{cleveref}
\usepackage{etoolbox}

\newtheoremstyle{exampstyle}
  {}
  {}
  {}
  {}
  {\bfseries\color{blue}}
  {.}
  {.5em}
  {}
\theoremstyle{exampstyle}
\newtheorem{thm}{Theorem}
\newtheorem{defn}{Definition}
\newtheorem{coro}{Corollary}
\newtheorem{prop}{Proposition}
\newtheorem{Assumption}{Assumption}
\newtheorem{axm}{Axiom}
\newtheorem{myexp}{Example}
\newcommand{\exampleendsymbol}{%
  \ifhmode\unskip\nobreak\hfill\else\leavevmode\nobreak\hfill\fi
  \textcolor{blue}{\ensuremath{\blacktriangle}}%
}
\AtEndEnvironment{myexp}{\exampleendsymbol}
\newenvironment{myexpcont}
  {\par\addvspace{\topsep}\noindent\textbf{\color{blue}Example \themyexp\ (continued).}\ }
  {\exampleendsymbol\par\addvspace{\topsep}}

\allowdisplaybreaks[1]

\newcommand{\Indicator}{\mathbbm{1}}
\newcommand{\Op}{O_{\mathbb{P}}}
\newcommand{\Universe}{X}
\newcommand{\AttFilter}{\mu}
\newcommand{\ChoiProb}{\pi}
\renewcommand{\epsilon}{\varepsilon}
\begin{document}
\raggedbottom

\onehalfspacing
\hypersetup{pageanchor=false}

\title{\vspace{0in}Attention Overload\thanks{We thank
Chris Chambers,
Emel Filiz-Ozbay,
Whitney Newey,
Erkut Ozbay, and seminar participants at Bristol,
Cornell,
LSE,
MIT,
Pittsburgh,
Princeton,
Rice,
Rutgers,
UC Davis,
UCLA, and conference participants at the 2020 Risk, Uncertainty \& Decision Conference, the 2020 Southern Economic Association Annual Meeting, the 2022 Cowles Foundation Econometrics Conference, and the 2023 Science of Decision Making Conference for their comments. We also thank Zexuan Wang and Dianzhuo Zhu for generously sharing their experimental data. The editor and four anonymous reviewers provided excellent comments and suggestions. Cattaneo gratefully acknowledges financial support from the National Science Foundation (SES-2241575), and the John Simon Guggenheim Memorial Foundation through a 2026 Guggenheim Fellowship.}}

\author{Matias D. Cattaneo\thanks{Department of Economics, Princeton University.}\and
        \hspace{-.05in}Paul H.Y. Cheung\thanks{Jindal School of Management, University of Texas at Dallas.}\and
        \hspace{-.05in}Xinwei Ma\thanks{Department of Economics, UC San Diego.}\and
        \hspace{-.05in}Yusufcan Masatlioglu\thanks{Department of Economics, University of Maryland.}}
\maketitle

\vspace{0in}

\begin{abstract}
    We introduce an Attention Overload Model (AOM) in which alternatives compete for attention, so each alternative's consideration probability weakly decreases as the choice problem expands. This nonparametric restriction has a search-capacity foundation. We identify exactly which preference rankings are compatible with observed choices and establish sharp bounds on latent attention. We then consider heterogeneous preferences in settings where alternatives are presented in a list, establishing nonparametric identification results for attention and the distribution of preferences. We also develop high-dimensional inference and finite-sample methods for these models. Using the travel-mode experiment of \citet*{Wang-Zhu_2025_WP}, we illustrate the methods and find that recovered pairwise preference shares closely match subjects' self-reported rankings.
\end{abstract}


\thispagestyle{empty}
\clearpage

\doublespacing

\setcounter{page}{1}
\hypersetup{pageanchor=true}
\section{Introduction}\label{section:Introduction}

This paper studies choice when alternatives compete for limited attention. We define an alternative's \textit{attention frequency} as the probability that it enters the consideration set and impose \textit{attention overload}: holding an alternative fixed, its attention frequency cannot increase when more rivals become available. If attention were deterministic, an item considered in a large supermarket would also be considered in a smaller store containing that item. The restriction is nonparametric and is motivated by evidence that larger choice sets can dilute item-level inspection when time and cognitive resources are limited \citep*{Reutskaja2009SatisfactionSatiate,Visschers-Hess-Siegrist_2010_PHN,Reutskaja_et_al_2011_AER,Geng_2016_EI,Thomas-Molter-Krajbich_2021_eLife}. For example, \citet*{Reutskaja_et_al_2011_AER} show that the mean share of items viewed falls as the number of alternatives increases. \citet*{Thomas-Molter-Krajbich_2021_eLife} provide complementary eye-tracking evidence that the probability that an individual item is inspected decreases as the choice set expands. We investigate the implications of this restriction and characterize what observed choices reveal about underlying preference orderings and attention frequencies.

Our identification strategy exploits menu variation in settings where preferences remain stable: changes in inventory or eligibility, randomized display sets, search results, product assortments, or repeated cross-sections exposed to different subsets of a common universe.\footnote{We do not treat changes in prices or product characteristics as menu variation. Comparability can be maintained by conditioning on these characteristics or redefining alternatives as product--state pairs. Section~\ref{section: empirical application} uses stable travel-mode categories under its stated assumptions on attribute profiles and preferences.} Attention overload is especially well suited to environments in which limited time and cognitive resources cause alternatives to compete for inspection. Its directional content makes the restriction both economically interpretable and directly testable, allowing the data to distinguish attention dilution from settings in which menu expansion raises item-level consideration. In the appendix, we show that the restriction arises naturally from a familiar mechanism featuring heterogeneous search orders and cognitive capacities.

Existing attention models restrict attention behavior by imposing assumptions on the attention rule. We differ from this literature by placing attention frequency at the center of our analysis, an empirically relevant object \citep{Reutskaja_et_al_2011_AER,caplin2011search,Thomas-Molter-Krajbich_2021_eLife}. Although existing models do not impose restrictions directly on this object, their assumptions nevertheless have implications for how attention frequencies vary across menus. We show that these implications need not align with Attention Overload. In the Independent Attention model of \citet{Manzini-Mariotti_2014_ECMA} and the Elimination-by-Aspects model of \citet{Aguiar_2017_EL}, attention frequencies are menu invariant, so these models correspond to a knife-edge case of Attention Overload. By contrast, the parametric model of \citet{Brady-Rehbeck_2016_ECMA} and the nonparametric Random Attention Model (RAM) of \citet*{Cattaneo-Ma-Masatlioglu-Suleymanov_2020_JPE} can generate attention frequencies that increase as the menu expands, the opposite of the comparative static underlying Attention Overload. These models therefore do not encode the empirically motivated dilution force that is central to our framework. Because the models restrict different features of attention, they can rationalize different choice data, reveal different preferences, and address different empirical questions. Section \ref{section: Comparison to Other Limited Attention Models} makes these distinctions explicit.

Motivated by these theoretical developments, a growing empirical literature establishes both the relevance and the testability of random attention.  Experiments that observe consideration directly document widespread stochastic limited consideration, with monotonicity among the more demanding implications \citep*{Ellis-FilizOzbay-Ozbay_2025_WP,Chadd_2023_JEBO,KhanKrajbichRaymondSundaresanVeiga2026}. \citet*{Wang-Zhu_2025_WP} find that RAM accepts the preference ranking implied by subjects' self-reports in a travel-mode experiment while rejecting rankings that place an objectively dominated option first; a health-preference experiment likewise finds largely consistent revealed preferences under RAM and RUM \citep*{Wang-Sicsic-Chauvin_2026_WP}. Choice overload experiments and field evidence also connect larger menus to limited or costly search and reduced purchase frequency \citep*{Dean-Ravindran-Stoye_2025_WP,deLara-Dean_2024_WP,Natan_2024_MS}. Frame- and covariate-based studies reject full-attention random utility in some settings, support latent consideration, and show that accounting for heterogeneous choice sets can materially affect recovered preferences and welfare conclusions \citep*{Aguiar2023QE,Abaluck_Adams_2021,Barseghyan-Coughlin-Molinari-Teitelbaum_2021_ECMA,Barseghyan-Molinari-Thirkettle_2021_AER}. Together, they show that limited consideration is empirically consequential.\footnote{Section SA-1 provides a fuller discussion of this evidence and broader comparisons with related models.}

Our first contribution is a revealed-preference and revealed-attention theory for AOM under homogeneous preferences. We show that a candidate preference represents the observed choice rule if and only if $\succ$-Regularity holds, and we give a direct construction of a rationalizing attention rule. $\succ$-Regularity relaxes classic regularity of stochastic choice in a way that captures the attention-dilution: inferior alternatives may become more likely to be chosen as the menu expands. The characterization sharply identifies all compatible preferences. It also yields computationally attractive implications based on binary comparisons and a mixed-integer formulation that represents the identified set implicitly without enumerating all $|X|!$ rankings. We derive sharp component-wise bounds on attention frequencies, which is the main interest of this paper. Thus, the model recovers an economically meaningful feature of latent attention without parametrizing the consideration-set distribution.

Our second contribution is a new treatment of heterogeneous preferences. We introduce a list-based attention-overload model with heterogeneous preferences (H-LAO) for environments in which the researcher observes a presentation order, such as ranked search results or advertisements. H-LAO permits an unrestricted marginal distribution over strict preferences and imposes Sequential Path Independence (SPI) on a \textit{list-based} attention rule. SPI is a cross-menu restriction requiring that the probability of reaching the next item on a list equal the probability of reaching the current item times a continuation probability. We provide a novel construction of the stochastic choice rule that would prevail under full attention. H-LAO is then sharply characterized by a mild choice overload condition and the standard Block--Marschak inequalities applied to this recovered full-attention rule. We provide sufficient conditions for point identification, partial-identification results when these conditions are not met, and illustrate both with an example.\footnote{The Supplemental Appendix discusses limited menu data, dependence between stopping and preferences, and violations of the maintained support condition.}

Our third contribution is econometric. For homogeneous AOM, we develop estimation and inference for compatible preferences and attention bounds when the number of menus and inequalities is large. The methods use high-dimensional Gaussian approximation, feasible correlated-Gaussian and universal critical values, and test inversion; binary screening reduces both statistical and computational burden. The approximation error depends logarithmically on the number of comparisons and on the smallest effective menu sample size, rather than on a fixed-dimensional asymptotic approximation. We also give a common finite-sample projection method for both models. For H-LAO, exact projection gives confidence sets for preference distributions under the benchmark and both sensitivity specifications. Explicit pairwise intervals remain valid without dividing by weak reach, and dependence-robust preference-event bounds are obtained by linear programming. A studentized inversion of the undivided pairwise moment improves precision when its standard error is estimable, while a fully linear projection procedure covers the model without Sequential Path Independence.

We also contribute to the empirical literature by providing evidence from the independently collected travel-mode experiment of \citet*{Wang-Zhu_2025_WP}. Its complete menu manipulation, dominated outside option, repeated choices, information-search records, and ex post rankings allow us to compare RAM and AOM, apply H-LAO, and provide an external comparison of recovered pairwise preference shares with separately elicited reports that are not used to estimate H-LAO. Subject-clustered specification diagnostics do not reject H-LAO, and simultaneous confidence intervals for all six recovered--reported pairwise differences contain zero. Due to space constraints, Sections SA-5 and SA-6 of the Supplemental Appendix report implementation benchmarks and simulation evidence on size, coverage, power, robustness, and computational performance across complete and incomplete menu designs.

The paper lies at the intersection of revealed preference, discrete choice, limited consideration, and econometrics. See \citet*{Matzkin_2007_Handbook}, \citet*{Matzkin_2013_ARE}, and \citet*{Molinari_2020_HandbookCh} for reviews of nonparametric and partial identification, and \citet*{deClippel-Rozen_2024_JEL} and \citet*{Strzalecki_2025_SCT} for broader treatments of boundedly rational and stochastic choice. Our distinctive behavioral primitive is the marginal attention received by each alternative. Restricting this object across menus yields revealed-preference content and sharp attention-frequency bounds under homogeneous preferences; with an observed list and explicit data and support conditions, it also permits recovery of reach and the full-attention stochastic choice rule under heterogeneous preferences. On the econometric side, we develop high-dimensional inference for the resulting choice-probability inequalities and finite-sample projection methods for partially identified preference and attention objects, including procedures valid under weak reach.

The rest of the paper proceeds as follows. Section \ref{section:Choice under Attention Overload} develops the homogeneous-preference theory and compares AOM with other random-attention models. Section \ref{section: Heterogeneous Preference over List} develops H-LAO. Section \ref{subsection:Econometric Methods single preference} presents estimation and inference. Section \ref{section: empirical application} reports the empirical application. The Supplemental Appendix contains additional literature, proofs, alternative inference procedures, simulation evidence, and computational details. We also provide an \texttt{R} package and replication code.

\section{Choice under Attention Overload}\label{section:Choice under Attention Overload}

This section studies homogeneous preferences and monotonicity of attention frequency; Section \ref{section: Heterogeneous Preference over List} allows preferences to be heterogeneous. Let $\Universe$ be a finite universe of alternatives and $\mathcal{D}$ a collection of nonempty menus. We permit incomplete menu data, so $\mathcal{D}$ may be a strict subset of $2^X\setminus\{\emptyset\}$. A \textit{choice rule} is a map $\ChoiProb:X\times\mathcal{D}\to[0,1]$ such that $\ChoiProb(a|S)=0$ for $a\notin S$ and $\sum_{a\in S}\ChoiProb(a|S)=1$. We interpret $\ChoiProb(a|S)$ as the population probability of choosing $a$ from $S$ and assume it is identified, or consistently estimable, for observed menus.

An important feature of our model is that consideration sets can be random. An \textit{attention rule} is a map $\AttFilter:2^X\times\mathcal{D}\to[0,1]$ such that $\AttFilter(T|S)=0$ for every $T\nsubseteq S$ and $\sum_{T\subseteq S}\AttFilter(T|S)=1$ for every $S\in\mathcal{D}$. The value $\AttFilter(T|S)$ is the probability of considering exactly $T\subseteq S$ when the menu is $S$. This formulation includes deterministic attention rules; for example, $\AttFilter(S|S)=1$ represents full attention. Our discussion on attention overload leads to the following novel theoretical construct.  Defined  with respect to an attention rule $\mu$,  we summarize the attention received by an individual alternative by adding the probabilities of all consideration sets containing it.

\begin{defn}[Attention Frequency]\label{definition:Attention Frequency} 
    Given $\mu$, the \textit{attention frequency} map $\phi_\mu:X\times\mathcal{D}\to[0,1]$ is $\phi_\mu(a|S):=\sum\limits_{T\subseteq S:\,a\in T}\mu(T|S)$.
\end{defn}

$\phi_\mu(a|S)$ is the probability that $a$ attracts attention in $S$. Whenever $\mu$ is clear from context, we omit the subscript. Attention frequency is binary under a deterministic attention rule but may take any value in $[0,1]$ under stochastic attention. When decision makers face an abundance of options, the alternatives compete for attention. We capture this competition by requiring an alternative's attention frequency to weakly decrease when the menu expands. We call this property \textit{Attention Overload}, the paper's central nonparametric restriction.

\begin{Assumption}[Attention Overload]\label{assumption:Attention Overload}
   For any $T,S\in\mathcal D$ with $a \in T \subseteq S$, $ \phi_\mu( a| S) \leq \phi_\mu(a|T)$.
\end{Assumption}

Attention overload is intuitively motivated by competition among alternatives and is supported by the empirical evidence discussed in the Introduction. As choice sets become larger, two empirical patterns are observed: (i) aggregate search effort increases, and (ii) decision avoidance becomes more prevalent. We now clarify how attention overload relates to these observations.

First, research on consumer search \citep{Reutskaja2009SatisfactionSatiate,caplin2011search,Thomas-Molter-Krajbich_2021_eLife} shows that individuals may exert greater search effort as menu size increases and, consequently, inspect more alternatives on average. At first glance, this pattern may appear to contradict attention overload. An increase in aggregate search effort, however, does not imply that any particular alternative becomes more likely to be considered.

To illustrate this distinction, suppose that from a menu of size $N$, a decision maker first draws a search quota  $q \leq N$ from an arbitrary distribution with the mean $k_N$, and conditional on $q$, inspects a uniformly selected subset of that size. Then, the attention frequency of every alternative  is
$\frac{k_N}{N}$.\footnote{Conditional on a search quota $q$, the probability that any given alternative is inspected is $
\binom{N-1}{q-1}/\binom{N}{q}=\frac{q}{N}$. Averaging over the distribution of $q$  gives $k_N/N$.} Hence,  Attention Overload is equivalent to, $ \frac{k_N}{N}\geq\frac{k_M}{M}$ for two nested menus with sizes $N <M$. Thus, total expected search $k_N$ may increase with menu size while Attention Overload continues to hold. Note that there is no restriction in the distributional assumption on search quota other than the mean. What matters is that the average number of inspected alternatives does not increase proportionally faster than the size of the menu.\footnote{Interestingly, this is precisely the pattern documented by \citet{Thomas-Molter-Krajbich_2021_eLife}, who show that participants inspect more alternatives as choice-set size increases ($k_N$ increasing in $N$), while each individual alternative becomes less likely to be inspected  and the fraction of the menu that is inspected declines ($\frac{k_N}{N}$ is decreasing).} %

Second, the literature also documents that individuals may become more likely to avoid making a choice when presented with a larger set of options \citep{Iyengar-Lepper_2000_JPSP}. This phenomenon is commonly referred to as \emph{choice overload}. At first glance, attention overload and choice overload may appear to go hand in hand. The two concepts, however, are logically distinct. 

We use the example above with two alternatives ($N=2$) to illustrate this point. In our framework, the empty consideration set can be interpreted as the DM not considering any alternatives (decision avoidance). We will keep the attention frequency the same, while varying the probability of empty consideration set depending on the distributional assumption on search quota. Consider two distribution assumptions. In the first one, the search quota is either $0$ or $2$ with equal probability, and in the second one, the search quota is exactly $1$. Note that, in both cases, the average number of products considered is $1$ and the attention frequency of each alternative will be the same, $\frac{1}{2}$. However, the probability of decision avoidance differs: It is $\frac{1}{2}$ or $0$, respectively. Attention Overload therefore does not determine the probability of decision avoidance.

Finally, before formally introducing the model, we provide a microfoundation for Attention Overload based on salience ordering and search quota. Suppose each decision maker draws a menu-independent salience ordering and a quota, and considers top alternatives in that more salient up to her quota capacity. If an item is considered in a larger menu, eliminating other items can only move it earlier in the search order, so it remains reached in every smaller menu containing it. This item-level monotonicity holds even under heterogeneous quota and salience rankings. The Supplemental Appendix formally states and proves this search foundation. Section \ref{section: Comparison to Other Limited Attention Models} compares Assumption \ref{assumption:Attention Overload} with other attention restrictions.

Given the discussion above, for simplicity, we assume $\mu(\emptyset|S)=0$. To model choice overload alongside attention overload, one can instead allow the consideration set to be empty and additionally require $\mu(\emptyset|S)$ to weakly increase as the menu expands.\footnote{The characterization below continues to hold if one allows $\mu(\emptyset|S)\neq 0$ and interprets it as the choice probability of the outside option, as in the literature (e.g., \citealp{manzini2024_approval,Brady-Rehbeck_2016_ECMA,Aguiar_2017_EL}), because its proof places no restriction on $\mu(\emptyset|S)$. Separately, Section SA-2.8 studies attention overload conditional on engagement.} Given a nonempty realized consideration set, the decision maker chooses its best alternative according to a strict preference ordering $\succ$. Combining this behavior with attention overload gives the following representation.

\begin{defn}[Attention Overload Representation]\label{definition:Attention Overload Representation}
    An \textit{Attention Overload Model} (AOM) is a pair $(\succ,\AttFilter)$ consisting of a preference ordering $\succ$ over $X$ and an attention rule $\AttFilter$ satisfying attention overload (Assumption \ref{assumption:Attention Overload}). The AOM $(\succ,\AttFilter)$ represents a choice rule $\ChoiProb$ if
    \begin{align*}
        \ChoiProb(a|S)=\sum_{T\subseteq S}\Indicator(\text{$a$ is $\succ$-best in $T$})\cdot\AttFilter(T|S)
    \end{align*}
    for all $a\in S$ and $S\in\mathcal D$. We call $\ChoiProb$ \textit{AOM-compatible} if some AOM represents it. Analogously, $\succ$ represents $\ChoiProb$ if there exists an attention rule $\AttFilter$ such that the AOM $(\succ,\AttFilter)$ represents $\ChoiProb$.
\end{defn}

AOM applies to structurally comparable menus: the same label $a$ represents the same relevant alternative, and the population preference ordering remains stable, as $S$ changes. Natural sources of such variation include inventory or eligibility changes, randomized displays, and repeated cross sections drawn from the same population. When prices or characteristics vary, comparability can be preserved by conditioning on them or redefining alternatives as product-state pairs. Applications with endogenous menu selection similarly require comparable preference and attention populations across menus. Under these conditions, cross-menu choice variation isolates the modeled attention response; the econometric analysis conditions on the realized menu assignments.

The unknown components of the model are the attention rule $\mu$ and preference ordering $\succ$; standard choice data identify only $\ChoiProb$. The analysis first characterizes all compatible preferences and then studies what can be learned about attention frequencies without identifying the probability assigned to every consideration set.

\subsection{Behavioral Implications and Revealed Preferences}\label{subsection:Characterization}

We begin with characterization and preference elicitation. For a candidate preference $\succ$, the central question is whether some attention rule satisfying attention overload can reproduce the observed choice pattern. Directly searching over attention rules is cumbersome when the universe is large. Our first result replaces that search with testable inequalities and thereby characterizes the complete set of compatible preferences. To motivate, we consider a few behavioral implications of the AOM.

Assume that $(\succ,\mu)$ represents $\ChoiProb$, and let $U_\succeq(a)=\{b\in X:b\succeq a\}$ denote the weak upper contour set of $a$. Classical regularity requires that removing alternatives can only help each alternative that remains: $\pi(a|T)\geq\pi(a|S)$ whenever $T,S\in\mathcal D$ and $a\in T\subseteq S$. This is too strong when attention is scarce. Shrinking the menu can free attention that flows to alternatives that $a$ would not have beaten, so $a$'s own choice probability need not rise. Attention overload instead requires the smaller menu to generate enough choice probability on $a$ or alternatives preferred to it. The following observable condition therefore bounds the choice share of $a$'s upper contour set rather than only $a$'s own share.

\begin{axm}[$\succ$-Regularity]\label{axiom:Attention Compensation}
    For all $T,S\in\mathcal D$ with $a\in T\subseteq S$, $\ChoiProb(U_\succeq (a)|T) \geq \ChoiProb(a|S)$.
\end{axm}

To establish necessity, observe that choice of $a$ requires attention to $a$, while attention overload gives $\phi(a|T)\geq\phi(a|S)$ for observed menus $T,S\in\mathcal D$ with $T\subseteq S$. Moreover, whenever $a$ is considered in $T$, the chosen alternative belongs to $U_\succeq(a)$. Hence \begin{align*}
    \ChoiProb(U_\succeq(a)|T)
    \geq \phi(a|T)
    \geq \phi(a|S)
    \geq \ChoiProb(a|S), 
\end{align*} which is $\succ$-Regularity.

$\succ$-Regularity coincides with classical regularity when $a$ is the best item in $T$, permits an individual lower-ranked item to violate classical regularity because choices of better items provide compensation, and is vacuous when $a$ is worst in $T$. Given a candidate $\succ$, the condition is directly testable from choice probabilities. The following result shows that it is also sufficient, so testing $\succ$-Regularity is equivalent to testing whether $\succ$ can rationalize $\ChoiProb$ under attention overload without constructing an attention rule.

\begin{thm}[Characterization]\label{theorem:Characterization}
    $\ChoiProb$ has an AOM representation with $\succ$ if and only if  $\ChoiProb$ satisfies $\succ$-Regularity.
\end{thm}

We proved necessity above; the supplemental appendix proves sufficiency by directly constructing a rationalizing attention rule. Thus, $
    \mathcal{P}_\pi:=\{\succ:\pi\text{ satisfies $\succ$-Regularity}\}$ is exactly the identified set of compatible preferences. The characterization also yields revealed preferences. Alternative $b$ is revealed preferred to $a$ if $b\succ a$ for every preference representing $\ChoiProb$. Theorem \ref{theorem:Characterization} motivates the following binary relation based directly on $\succ$-Regularity:
\begin{align*}
    \text{$b \text{P}_\ChoiProb a$\quad if $b \succ a$ for all preference orderings such that $\ChoiProb$ satisfies $\succ$-Regularity.}
\end{align*}
We remark that the relation $\text{P}_\ChoiProb$ does not require constructing an attention rule for each candidate preference. Checking $\succ$-Regularity uses only observed choice probabilities, a fact we exploit for inference in Section \ref{subsection:Econometric Methods single preference}.

\begin{coro}[Revealed Preference]\label{thm:revealed preference}
    Let $\ChoiProb$ be AOM-compatible. Then, $b$ is revealed preferred to $a$ if and only if $b\text{P}_\ChoiProb a$.
\end{coro}

Although Theorem \ref{theorem:Characterization} and Corollary \ref{thm:revealed preference} avoid constructing the underlying attention rule for revealed-preference analysis, they still require checking every possible preference ordering, which becomes computationally expensive as $|X|$ grows. Two approaches substantially reduce this computational burden.

The first approach admits a mixed-integer linear formulation that avoids explicitly enumerating preference orderings. For distinct $a,b\in X$, let $z_{ba}=1$ mean that $b\succ a$, and let $\mathcal Z_\pi$ contain all binary vectors $\mathbf{z}=(z_{ba})_{b\ne a}$ satisfying
\begin{align*}
    z_{ab}+z_{ba}&=1 &&\text{for all distinct $a,b$,}\\
    z_{ab}+z_{bc}-1&\leq z_{ac} &&\text{for all distinct $a,b,c$,}\\
    \pi(a|S)&\leq \pi(a|T)
    +\sum_{b\in T\setminus\{a\}}\pi(b|T)z_{ba}
    &&\text{for all $a\in T\subseteq S$ with $T,S\in\mathcal D$.}
\end{align*}
\begin{coro}[Mixed-Integer Characterization]\label{coro:mip-characterization}
The map taking a strict ordering $\succ$ to its binary incidence vector $\mathbf{z}$ is a bijection between $\mathcal P_\pi$ and $\mathcal Z_\pi$. Consequently, $\pi$ is AOM-compatible if and only if $\mathcal Z_\pi$ is nonempty. If $\pi$ is AOM-compatible, then $b\text{P}_\pi a$ if and only if $\min_{z\in\mathcal Z_\pi}z_{ba}=1$.
\end{coro}

The corollary replaces a search over $|X|!$ candidate rankings with a mixed-integer linear feasibility problem. It uses $|X|(|X|-1)$ binary variables, cubically many ordering constraints, and one linear inequality for each observed $a\in T\subseteq S$. Binary screens can first fix individual $z_{ba}$ coordinates, while nonbinary violations add restrictions that eliminate incompatible rankings. Pairwise revealed preference is then obtained by testing whether the opposite comparison remains feasible. For perspective, with ten alternatives, the formulation uses $90$ directed binary variables, only $45$ of which are independent after imposing pairwise completeness, instead of enumerating $10!=3{,}628{,}800$ rankings. The Supplemental Appendix proves the formulation and discusses its implementation; Example \ref{Exp_1} below illustrates the mechanics.

Our second approach provides an optimization-free screening device when the object of interest is a single pairwise comparison. As we will show below, regularity violations involving binary menus provide a simple and computationally attractive device for revealed preference analysis, although they do not exhaust the identifying power of attention overload. Alternatively, the result can also be used as an additional screening step if the goal is to obtain the entire set of compatible preferences $\mathcal{P}_{\pi}$.

Specifically, if $a,b\in S$ and $\ChoiProb(a|S)>\ChoiProb(a|\{a,b\})$, then every representation must rank $b$ above $a$. Otherwise $a\succ b$, so $a$ is best in the binary menu and $\phi(a|\{a,b\})=\ChoiProb(a|\{a,b\})$. But then
\begin{align*}
    \phi(a|\{a,b\})
    =\ChoiProb(a|\{a,b\})
    <\ChoiProb(a|S)
    \leq\phi(a|S),
\end{align*}
contradicting attention overload. The next proposition formalizes this observation.

\begin{prop}[Regularity Violation at Binary Comparisons]\label{lemma:Regularity Violation at Binary Comparisons}
    Let $(\succ,\mu)$ be an AOM representing $\ChoiProb$, and let $a,b\in S$. If $\ChoiProb(a|S)>\ChoiProb(a|\{a,b\})$, then $b\succ a$.
\end{prop}
                 
Proposition \ref{lemma:Regularity Violation at Binary Comparisons} gives an easy-to-implement revealed-preference test without requiring a complete representation. With a nonbinary smaller menu $T$, the conclusion is weaker: a violation $\pi(a|S)>\pi(a|T)$ implies only that some alternative in $T$ is preferred to $a$. Indeed, if $a$ were best in $T$, then
\begin{align*}
    \phi(a|T)=\pi(a|T)<\pi(a|S)\leq\phi(a|S),
\end{align*}
again contradicting attention overload.

\begin{prop}[Regularity Violation at Bigger Choice Problems]\label{lemma:Regularity Violation at Bigger Sets}
    Let $(\succ,\mu)$ be an AOM representing $\ChoiProb$, and let $a\in T\subset S$. If $\ChoiProb(a|S)>\ChoiProb(a|T)$, then there exists $b\in T$ such that $b\succ a$.
\end{prop}

Both propositions use regularity violations, and Proposition \ref{lemma:Regularity Violation at Bigger Sets} specializes to Proposition \ref{lemma:Regularity Violation at Binary Comparisons} when $T$ is binary. The following example uses these propositions to screen preference orderings before applying Theorem \ref{theorem:Characterization}; regularity violations alone do not exhaust the identifying power of attention overload.

\begin{myexp}\label{Exp_1} Consider the following choice data:

\begin{center}
    \begin{tabular}{r|cccc} 
    \toprule 
    $\ChoiProb(\cdot|S)$ & $a$ & $b$ & $c$ & $d$\\ 
    \midrule 
    $\{a, b, c, d\}$ & 0.05 & 0.1 & 0.1 & 0.75 \\ [-.5em]
    $\{a, b, c\}$ & 0.8  & 0.2 & 0 & -- \\[-.5em]
    $\{ b, c, d\}$ & --  & 0.7 & 0.3 & 0 \\[-.5em]
    $\{a, b\}$ & 0.9  & 0.1 & --  & -- \\
    \bottomrule 
    \end{tabular}
\end{center}
There are $4!=24$ candidate preference orderings. First, applying Proposition \ref{lemma:Regularity Violation at Binary Comparisons} from $\{a,b,c\}$ to $\{a,b\}$ rules out every ordering with $b\succ a$, leaving $12$ candidates. Two further violations involve nonbinary smaller menus. From $\{a,b,c,d\}$ to $\{a,b,c\}$, $c$ violates regularity, so Proposition \ref{lemma:Regularity Violation at Bigger Sets} rules out the four candidates in which $c$ is preferred to both $a$ and $b$. Analogously, the violation for $d$ from $\{a,b,c,d\}$ to $\{b,c,d\}$ rules out the four candidates in which $d$ is preferred to both $b$ and $c$. Four orderings remain: $a\succ b\succ c\succ d$, $a\succ b\succ d\succ c$, $a\succ c\succ b\succ d$, and $a\succ c\succ d\succ b$. Finally, checking $\succ$-Regularity requires both $b\succ d$ and $c\succ d$, leaving only $a\succ b\succ c\succ d$ and $a\succ c\succ b\succ d$.
\end{myexp}

The example therefore has two AOM-compatible preferences, $a\succ_1 b\succ_1 c\succ_1 d$ and $a\succ_2 c\succ_2 b\succ_2 d$. The revealed-preference relation $\text{P}_\ChoiProb$ is complete except for the comparison between $b$ and $c$. Starting from $24$ candidates, the binary screen eliminates $12$, the nonbinary violations eliminate another $8$, and the full characterization leaves exactly these two. Example \ref{Exp_1} thus shows how the results can be applied sequentially to reduce the computational burden.

Corollary \ref{coro:mip-characterization} reaches the same conclusion without enumerating the $24$ rankings. For these data, every $\mathbf{z}\in\mathcal Z_\pi$ satisfies (i) $z_{ab}=z_{ac}=z_{ad}=z_{bd}=z_{cd}=1,$ and (ii) $z_{bc}+z_{cb}=1$.
Thus, $\mathcal Z_\pi$ contains exactly two incidence vectors, corresponding to $\succ_1$ and $\succ_2$. Minimizing each coordinate over $\mathcal Z_\pi$ gives one for the five common comparisons above, whereas
\begin{align*}
    \min_{z\in\mathcal Z_\pi}z_{bc}
    =\min_{z\in\mathcal Z_\pi}z_{cb}=0.
\end{align*}
The mixed-integer formulation therefore reveals every comparison except that between $b$ and $c$, exactly as the direct characterization does.

\subsection{Attentive at Binaries}\label{subsection:Attentive at Binaries}

Revealed-preference analysis can be strengthened by limiting extreme inattention at binary menus. Because competition for attention is weaker in binary menus, one may expect decision makers to be more attentive when only two alternatives are available. We formalize this additional restriction by bounding the probability of considering only one of the two available alternatives.
\begin{align}\label{eq:attentive at binaries}
    \text{Fix $\eta\in[0.5,1]$,\quad and }\eta \geq \max\Big\{\mu(\{a\}|\{a,b\}),\mu(\{b\}|\{a,b\}) \Big\}\ \text{for every $\{a,b\}\in\mathcal D$}.
\end{align}
This condition limits extreme inattention without imposing full attention. A larger $\eta$ weakens the restriction, and $\eta=1$ is vacuous. Even at $\eta=0.5$, the full binary menu may receive zero attention: both singleton consideration sets can have probability $0.5$.

Condition \eqref{eq:attentive at binaries} generates additional revealed preferences. For every observed binary menu, define $a\mathrel{\text{P}_B}b$ whenever $\ChoiProb(a|\{a,b\})>\eta$. The choice probability of $a$ then cannot be generated entirely by singleton consideration because $\mu(\{a\}|\{a,b\})\leq\eta$. Hence $\mu(\{a,b\}|\{a,b\})>0$, and choosing $a$ when both alternatives are considered reveals $a\succ b$.

The parameter $\eta$ can be interpreted as the analyst's degree of caution in drawing welfare conclusions. At $\eta=1$, binary choice alone yields no strict comparison, so $\text{P}_B=\emptyset$. The familiar threshold $\eta=0.5$ \citep{marschak1959binary,fishburn1998stochastic} gives the strongest binary revealed-preference relation in this class, although it need not identify the complete ordering.

The necessity and sufficiency of this additional ordering restriction are formalized next.

\begin{coro}[Attentive-at-Binaries Characterization]\label{coro:attentive-binaries-characterization}
Fix $\eta\in[0.5,1]$. A strict ordering $\succ$ belongs to an AOM representation satisfying \eqref{eq:attentive at binaries} if and only if $\pi$ satisfies $\succ$-Regularity and $\succ$ extends $\mathrm P_B$.
\end{coro}

Thus, $\mathrm P_B$ can be used with Propositions \ref{lemma:Regularity Violation at Binary Comparisons} and \ref{lemma:Regularity Violation at Bigger Sets} to screen the candidate orderings further. The result applies to an incomplete domain: it restricts only the binary menus that belong to $\mathcal D$. We revisit Example \ref{Exp_1} to illustrate.

\begin{myexpcont}
Assume that we observe additional data at $\{b,c\}$: $\ChoiProb(b|\{b,c\})=0.7$. Then, by Condition \eqref{eq:attentive at binaries}, we conclude that the only possible preference consistent with the observed choice data is $a \succ b \succ c \succ d$ whenever $\eta \in [0.5,0.7)$, thereby achieving point identification of the preference ordering.
\end{myexpcont}

\subsection{Revealed Attention}\label{subsection:Revealed Attention}

Attention overload restricts marginal attention across menus but generally does not identify the probability assigned to each consideration set. The same choice rule can therefore admit multiple attention rules with different attention frequencies. This section derives observable lower and upper bounds on those frequencies and characterizes the corresponding joint identified set. These objects can be useful, for example, when a firm studies product visibility or a policymaker asks whether consumers attend to higher-quality options.

We first formalize the joint identified set. For a candidate preference $\succ$, let $\mathcal M_\pi(\succ)$ contain all attention rules $\mu$ such that (i) the choice equations in Definition \ref{definition:Attention Overload Representation} hold for every observed menu and (ii)
$\phi_\mu(a|R)\leq\phi_\mu(a|S)
    \text{ for every }S,R\in\mathcal D
    \text{ with }a\in S\subseteq R$. For fixed $\succ$, these are finitely many linear equalities and inequalities in $\mu$.

\begin{prop}[Joint Identified Set]\label{prop:joint-identified-set}
The sharp identified set for the preference and attention rule is $\mathcal I_{\mu,\succ}(\pi) = \bigcup_{\succ\in\mathcal P_\pi} \big(\mathcal M_\pi(\succ)\times\{\succ\}\big)$.
Moreover, $\mathcal M_\pi(\succ)$ is a nonempty polytope if and only if $\succ\in\mathcal P_\pi$. Sharp bounds on any linear functional of $\mu$ are obtained by minimizing and maximizing that functional over $\mathcal M_\pi(\succ)$ for each $\succ\in\mathcal P_\pi$, and then taking the extrema across preferences.
\end{prop}

This proposition gives an operational joint characterization. Theorem \ref{theorem:Revealed Attention} below provides closed-form solutions for the important marginal-attention coordinates, avoiding the need to solve the linear programs separately.

Consider bounding $\phi$ from below first. For any observed superset $R\in\mathcal D$ of $S$, attention overload implies that $\ChoiProb(a|R)\leq \phi(a|R)\leq \phi(a|S)$. Therefore, for any $S\in\mathcal D$, $\phi(a|S) \geq \max_{R\in\mathcal D:\,R\supseteq S} \ChoiProb(a|R)$. This lower bound on attention uses only the observed choice rule and does not require knowledge of the underlying attention rule. The preference-dependent upper bound is obtained analogously from the observed subsets $T\in\mathcal D$ of $S$ that contain $a$. These observations give the following theorem.

\begin{thm}[Revealed Attention]\label{theorem:Revealed Attention}
    Let $(\succ,\mu)$ be an AOM representing $\ChoiProb$. Then, for every $S\in\mathcal D$ and $a\in S$,  \begin{align*}
    \max_{R\in\mathcal D:\,R\supseteq S} \ChoiProb(a|R)
    \leq \phi_\mu(a|S)
    \leq \min_{T\in\mathcal D:\,a\in T\subseteq S} \ChoiProb(U_\succeq (a)|T).
    \end{align*}
\end{thm}

The theorem is conditional on a compatible preference $\succ$. Its lower bound is common to every $\succ\in\mathcal{P}_\pi$, while its upper bound can vary with $\succ$. This distinction matters when preferences are only partially identified and is made precise below.

Three extreme cases help interpret Theorem \ref{theorem:Revealed Attention}. If both endpoints are one, $a$ receives full attention in $S$; if both are zero, $a$ is revealed unattended; and if the interval is exactly $[0,1]$, the data reveal nothing about its attention frequency. These are the only possibilities under deterministic choice, the setting studied by \citet*{Lleras_et_al_2017_JET}. The theorem supplies a common characterization for deterministic and stochastic choice.

Stochastic choice also permits partial revealed attention: the lower endpoint can be strictly positive or the upper endpoint strictly below one without the two coinciding. We illustrate this possibility by revisiting Example~\ref{Exp_1}.

\begin{myexpcont}
For the menu $\{a,b,c,d\}$, Theorem \ref{theorem:Revealed Attention} gives $\phi(a|\{a,b,c,d\})=0.05$, places both $\phi(b|\{a,b,c,d\})$ and $\phi(c|\{a,b,c,d\})$ in $[0.1,0.25]$, and places $\phi(d|\{a,b,c,d\})$ in $[0.75,1]$, where the extrema range only over the menus observed in the example.
\end{myexpcont}

Attention can be point identified for selected alternatives. For example, suppose in addition that $R\in\mathcal D$, $R\supseteq\{a,b,c,d\}$, and $\ChoiProb(c|R)=0.25$. The preference-free lower bound on $\phi(c|\{a,b,c,d\})$ is then $0.25$, equal to its upper bound, so this attention frequency is point identified.

The lower and upper endpoints can be viewed as pessimistic and optimistic evaluations of attention. A remaining question is whether the coordinatewise pessimistic evaluations can be attained jointly by one attention rule satisfying attention overload. The next result answers affirmatively; the proof of Theorem \ref{theorem:Characterization} also constructs the analogous optimistic rule.

\begin{thm}[Pessimistic Evaluation for Attention] \label{theorem:Pessimistic Evaluation for Attention} 
    Let $(\succ,\mu)$ be an AOM representing $\ChoiProb$. Then there exists a pessimistic attention rule $\mu^*$ such that $(\succ,\mu^*)$ is also an AOM representing $\ChoiProb$. That is, for every $S\in\mathcal D$ and $a\in S$, $\phi_{\mu^*}(a|S)=\max\limits_{R\in\mathcal D:\,R\supseteq S}\ChoiProb(a|R)$.
\end{thm}

This theorem concerns the pessimistic evaluation; the proof of Theorem \ref{theorem:Characterization} also constructs the analogous optimistic attention rule. Consequently, for each fixed $\succ\in\mathcal{P}_\pi$, the two bounds in Theorem \ref{theorem:Revealed Attention} are sharp componentwise. When preferences are not point identified, the projection of the joint set in Proposition \ref{prop:joint-identified-set} onto attention frequencies is $
    \mathcal{I}_\phi(\pi)
    =\{\phi_\mu:(\mu,\succ)\in\mathcal I_{\mu,\succ}(\pi)\}.$ Projecting this set onto a single coordinate $\phi(a|S)$ gives the common lower endpoint $    \max_{R\in\mathcal D:\,R\supseteq S}\pi(a|R)$ and the upper endpoint $\max_{\succ\in\mathcal{P}_\pi}  \min_{T\in\mathcal D:\,a\in T\subseteq S}\pi(U_\succeq(a)|T)$.
These projected bounds do not imply that every vector formed by independently selecting values inside the coordinatewise intervals belongs to $\mathcal{I}_\phi(\pi)$, nor do they identify the complete attention rule $\mu$. Our sharpness claim concerns the identified preference set and the stated componentwise attention functionals.

\subsection{Comparison to Other Random Attention Models}\label{section: Comparison to Other Limited Attention Models}

Random-attention models differ in the object they discipline. Independent attention and elimination by aspects impose parametric structure on marginal or category-level attention \citep{tversky1972elimination,Manzini-Mariotti_2014_ECMA,Aguiar_2017_EL}. Logit attention specifies probabilities for consideration sets \citep{Brady-Rehbeck_2016_ECMA}. RAM instead imposes the nonparametric set-level restriction
\begin{align*}
    \mu(D|S)\leq\mu(D|S\setminus\{b\})
    \qquad\text{when }b\notin D\subseteq S.
\end{align*}
AOM places no monotonicity restriction on the probability of a particular consideration set. It requires only that the attention frequency $\phi(a|S)=\sum_{D:\,a\in D}\mu(D|S)$ not increase with the menu. Set-level RAM monotonicity and marginal attention overload therefore neither imply nor are implied by one another, and they have distinct implications for choice behavior. The following example illustrates their different revealed-preference implications. Consider
\begin{align*}
\begin{array}{c|ccc}
\toprule
S & \pi(a|S) & \pi(b|S) & \pi(c|S)\\ \midrule 
\{a,b,c\} & 0.4 & 0.3 & 0.3\\[-.5em]
\{a,b\}   & 0.8 & 0.2 & \text{--}\\[-.5em]
\{a,c\}   & 0.8 & \text{--} & 0.2\\[-.5em]
\{b,c\}   & \text{--} & 0.5 & 0.5  \\ \bottomrule 
\end{array}
\end{align*}

Under RAM, the fall in $b$'s choice probability when $c$ is removed reveals $b\succ c$, while the analogous fall for $c$ when $b$ is removed reveals $c\succ b$.\footnote{These conclusions follow from Lemma 1 of \citet*{Cattaneo-Ma-Masatlioglu-Suleymanov_2020_JPE}.} Under AOM, the same choice patterns reveal $a\succ b$ and $a\succ c$. Thus, RAM ranks the alternative whose choice probability falls above the removed alternative, whereas AOM reveals that the alternative whose choice probability falls is outranked by an alternative remaining in the smaller menu. The data therefore violate RAM through a revealed-preference cycle but admit either $a\succ b\succ c$ or $a\succ c\succ b$ under AOM. Section SA-1 also gives a four-alternative choice rule that passes RAM but fails AOM, establishing observable non-nesting in the other direction.

Two extreme cases further illustrate the distinction between attention models at the attention level. In a larger menu $S$, Ann considers every item, while Ben considers only $a$. For a smaller menu $a\in T\subset S$, attention overload requires Ann to retain full item-level attention and requires Ben to keep considering $a$, but it permits Ben to add items that were previously ignored. RAM does the reverse in these examples: it leaves Ann's behavior in $T$ unrestricted but forces Ben's consideration set to remain $\{a\}$. Table \ref{table:comparisons} also reports the corresponding predictions of common parametric models.

\begin{table}[ht!]\centering{\small
    \def\arraystretch{1.2}
\resizebox{\textwidth}{!}{
\begin{tabular}{l|c|c}
\toprule
                     & Ann                    & Ben                        \\
                          \cline{2-3}
                            &  \multicolumn{2}{c}{ Larger set $S$ } \\
                            \cline{2-3}
     &\ \ \ \ \ Full attention \ \ \   \ \            &  Limited attention \\
                    & $\phi(x|S)=1$  $\forall x \in S$             &$\phi(a|S)=1$ and $\phi(x|S)=0$ $\forall x\neq a$  \\
                    \cline{2-3}
                   &  \multicolumn{2}{c}{Predictions for  a smaller set $a \in T \subset S$ } \\
                   \midrule

\cite{Manzini-Mariotti_2014_ECMA} &      $\phi(x|T)=1$  $\forall x \in T$             &$\phi(a|T)=1$ and $\phi(x|T)=0$ $\forall x\neq a$ \\
\cite{tversky1972elimination,Aguiar_2017_EL}       &     $\phi(x|T)=1$  $\forall x \in T$      &    $\phi(a|T)=1$ and $\phi(x|T)=0$ $\forall x\neq a$         \\
\cite{Brady-Rehbeck_2016_ECMA}        &  No restriction   &   $\phi(a|T)=1$ and $\phi(x|T)=0$ $\forall x\neq a$       \\
\cite{Cattaneo-Ma-Masatlioglu-Suleymanov_2020_JPE}      &   No restriction   &    $\phi(a|T)=1$ and $\phi(x|T)=0$ $\forall x\neq a$       \\
\cite{Demirkan-Kimya_2020_JME}      &     No restriction  & No restriction     \\
\midrule
       Attention Overload Model (AOM)          &        $\phi(x|T)=1$  $\forall x \in T$       &   $\phi(a|T)=1$; otherwise unrestricted    \\
                       \bottomrule
\end{tabular}
}
\caption{Predictions for a smaller menu $T$, given Ann's full attention and Ben's singleton consideration in the larger menu $S$.}
\label{table:comparisons}}
\end{table}

The independent-attention and elimination-by-aspects models hold item or category attention parameters fixed across menus. They therefore preserve Ann's full attention but cannot let Ben begin considering previously ignored items merely because the menu shrinks. Menu-dependent independent attention \citep{Demirkan-Kimya_2020_JME} can allow either direction and does not by itself impose attention overload. Logit attention and RAM allow Ann to change arbitrarily in the smaller menu, while their set-level structure keeps Ben at $\{a\}$ in this example.

The different primitives also explain the different roles played by regularity. In \cite{Cattaneo-Ma-Masatlioglu-Suleymanov_2020_JPE}, violations of classical choice regularity are informative about preferences under RAM's set-level monotonicity. AOM does not simply require the opposite data pattern. Its characterization uses the preference-indexed $\succ$-Regularity condition: choice of $a$ in a larger menu is bounded by choice from the entire upper contour set of $a$ in a smaller menu. Classical regularity is recovered only when $a$ is best in that smaller menu; lower-ranked alternatives may violate it. Conversely, a classical regularity violation at a binary menu reveals that the other binary alternative must be preferred to $a$ under AOM. These implications arise from marginal competition for attention, not from relabeling the RAM restriction.

The appropriate model is application-dependent. RAM is natural when removing unavailable alternatives should not reduce the probability of drawing a fixed surviving consideration set; AOM is natural when adding rivals dilutes item-level visibility. When both restrictions are plausible they can be imposed jointly. When they conflict, auxiliary attention measures or their distinct choice implications can guide specification testing.

\section{Heterogeneous Preferences}\label{section: Heterogeneous Preference over List}

The baseline AOM attributes stochastic choice to heterogeneous attention under a common preference. This section allows both preferences and attention to vary across a population. The objective is to learn the distribution of preferences from aggregate choice probabilities while retaining the attention-overload mechanism.

Without further structure, suppose there exists a probability distribution $\tau$ over rational preferences and an attention rule $\mu$ satisfying attention overload such that
\begin{align*}
    \ChoiProb(a|S)= \sum_{T:\, a\in T\subseteq S}\tau (\{ \succ \in \mathcal{P} : a \text{ is $\succ$-best in } T \})\cdot\AttFilter(T|S)
\end{align*}
for all $a\in S$ and $S\in \mathcal{D}$. Here $\tau$ captures preference heterogeneity and $\mu$ attention heterogeneity, with the two independent conditional on the environment being studied.

An observed list supplies the additional structure needed to separate preference and attention heterogeneity. Many choice environments present alternatives sequentially, allowing the list to discipline the path of attention without restricting which preference orderings may occur. This structure distinguishes combinations of preference and attention heterogeneity that the unrestricted mixture alone would leave observationally equivalent.
In addition, as is standard in the literature, we introduce an outside option $o^*$, chosen if and only if the consideration set is empty. Therefore, we expand the choice rule so that $\pi(\cdot|S)$ is defined as a probability distribution over $S\cup\{o^*\}$. For convenience, we maintain the  support condition that the outside option is not chosen with probability one, so that $\pi(S|S)>0$ for every nonempty $S$.

Throughout this section, the structural H-LAO model is defined on the full menu domain $2^X\setminus\{\emptyset\}$, while $\mathcal D\subseteq 2^X\setminus\{\emptyset\}$ denotes the collection of menus for which choice probabilities are observed. Thus, the underlying choice and attention processes are defined for every nonempty menu, even when the researcher observes only their restriction to $\mathcal D$. This separation between the domain of the primitives and observability will make the partial-identification result more convenient to state.

\subsection{List-Based Attention Overload} 

Consumers often encounter an observed order: search results, ranked advertisements, product recommendations, ballots, or alternatives on a screen. 
In real-world applications, the presentation order can plausibly carry informational value. Here, instead of explicitly modeling how the list is generated, we take the observed list as the economically relevant state variable and impose structure on how attention is allocated conditional on that list. Ordered search has a long history in decision theory \citep{rubinstein2006model,horan2010sequential,guney2014theory,Yildiz2016TE,aguiar2016satisficing,kovach2020satisficing,Ishiietal_2021_JME,tserenjigmid2021order,YEGANE2022,KOSHEVOY2023,manzini2024_approval,caplin2011search,caplin2011search_TE}. We condition on a presentation order observed by the researcher and common within the population whose aggregate choices define $\pi$. When different groups face different lists, the analysis applies within list strata and therefore accommodates multiple observed presentation regimes.

As a running example, consumers browse search results from top to bottom and stop at heterogeneous positions. Once a consumer stops, her consideration set consists of the available items encountered up to that point. This prefix structure makes the mapping from attention frequencies to the distribution of stopping points one-to-one. Formally, let $\{a_1,a_2,\dots,a_{|X|}\}$ denote the observed list, where $j<k$ means that $a_j$ appears earlier than $a_k$. Each menu $S\subseteq X$ inherits this order and is written $\{a_{s_1},\dots,a_{s_{|S|}}\}$.

Our new attention rule imposes a sequential structure on consideration set formation. Any consideration set must respect the underlying order: if $a_k$ is included, then every feasible alternative preceding $a_k$ must also be included. In other words, once an alternative is considered, all earlier alternatives in the list are necessarily considered as well. Consequently, given $S$, the admissible consideration sets take the following form, and every other subset receives zero attention probability:
\begin{align*}
\emptyset, \ \{a_{s_1}\}, \ \{a_{s_1},a_{s_2}\}, \ \{a_{s_1},a_{s_2},a_{s_3}\}, \ \ldots
\end{align*}
For $j=1,\dots,|S|$, define the $j$th prefix and suffix of $S$ by $S_{:j}:=\{a_{s_1},\dots,a_{s_j}\}$ and $S_{j:}:=\{a_{s_j},\dots,a_{s_{|S|}}\}$, respectively.
Thus, $S_{:j}\subseteq S_{:k}$ if and only if $j\leq k$, whereas $S_{j:}\subseteq S_{k:}$ if and only if $j\geq k$.
We call an attention rule \textit{list-based} if, for every nonempty $S\subseteq X$, it assigns positive probability only to $\emptyset$ and the prefixes in $S$. The key consequence of the list structure is a one-to-one mapping between attention frequencies and attention rules.
Formally, a map $\phi:X\times(2^X\setminus\{\emptyset\})\to[0,1]$ is a \textit{list-based attention frequency} if, for every $S=\{a_{s_1},\dots,a_{s_{|S|}}\}$, $1\geq\phi(a_{s_1}|S)\geq\cdots\geq \phi(a_{s_{|S|}}|S)\geq0$.
Every list-based attention frequency induces the unique list-based attention rule
\begin{align*}
    \mu(S_{:j}|S) = \phi(a_{s_j}|S)-\phi(a_{s_{j+1}}|S),
    \qquad
    \mu(\emptyset|S) = 1-\phi(a_{s_1}|S),
\end{align*}
where the terminal convention is $\phi(a_{s_{|S|+1}}|S)=0$. Without the prefix structure in attention rule, the same attention frequencies may be generated by many different distributions over consideration sets.

The list-based attention frequency remains highly permissive. By itself, it does not imply attention overload: when the menu expands, the attention frequency of a given alternative may still increase. We therefore retain attention overload as a cross-menu restriction by imposing two assumptions on the attention frequencies.

\begin{Assumption}[Sequential Path Independence (SPI)]\label{PI}
For every nonempty $S=\{a_{s_1},\dots,a_{s_{|S|}}\}\subseteq X$ and every $j\neq1$, $\phi(a_{s_j}|S)=\phi(a_{s_{j-1}}|S)\phi(a_{s_j}|S_{j:})$.
\end{Assumption}

The first assumption, SPI, is a cross-menu condition on sequential search. The probability of reaching the next item equals the probability of reaching the current item times a continuation probability that depends only on the remaining alternatives. This is a recursive multiplicative restriction on the attention-frequency function.

On a suffix-closed domain,\footnote{A domain $\mathcal{D}$ is suffix-closed  (prefix-closed) if the suffixes (prefixes) of $S$ belong to $\mathcal{D}$ whenever $S \in \mathcal{D}$. } let $q(S)=\phi(a_{s_1}|S)$ denote first-position reach. Iterating SPI gives, for every menu $S$ and position $j$,
\begin{align}\label{eq:spi-product}
    \phi(a_{s_j}|S)=\prod_{k=1}^{j}q(S_{k:}).
\end{align}
Conversely, any assignment $q:\mathcal D\to[0,1]$ on a suffix-closed domain defines through \eqref{eq:spi-product} a list-based attention frequency satisfying SPI. On the full-menu domain, the model therefore retains $2^{|X|}-1$ unrestricted first-position probabilities. Arbitrary $q$ does not by itself impose attention overload, which remains the separate cross-menu restriction above. The following examples show that familiar stopping processes satisfy SPI.

\begin{myexp}[Constant Stopping Attention]\label{Exp Constant Attention Frequency}
    This example gives a parsimonious model with a single parameter describing the attention rule. At each stage, search continues with probability $\alpha\in[0,1]$, so the stopping hazard is $1-\alpha$ and $q(S)=\alpha$ for every menu $S$. Then $\phi(a_{s_j}|S)=\alpha^j$. For $j<|S|$, $\mu(S_{:j}|S)=(1-\alpha)\alpha^j$, while $\mu(S|S)=\alpha^{|S|}$. Notably, $\phi(a_{s_j}|S)$ depends only on the position $j$, not on the size of $S$.
\end{myexp}

\begin{myexp}[Rank-Dependent Attention]\label{Exp Rank-Dependent Attention Frequency}
    In this example, continuation depends on the number of alternatives remaining, capturing the idea that a lengthy remaining list may deter consumers from proceeding. Let $q(R)=h(|R|)$ for some weakly decreasing function $h:\{1,\dots,|X|\}\to[0,1]$. Then, $\phi(a_{s_j}|S)=\prod_{m=|S|-j+1}^{|S|}h(m)$.     Because $h$ is weakly decreasing, the attention frequency is weakly decreasing as the menu expands. There are at most $|X|$ parameters in this class.
\end{myexp}

\begin{myexp}[Alternative-Dependent Attention]\label{Exp Alternative-Dependent Attention Frequency}
    Unlike the previous two examples, continuation here depends on the specific alternative rather than only on its position. Let $\beta(a)\in[0,1]$ be the item-specific conditional factor for continuing far enough to inspect $a$ after reaching its predecessor, and set $q(S)=\beta(a_{s_1})$. Then $\phi(a_{s_j}|S)=\prod_{k\leq j}\beta(a_{s_k})$. For $j<|S|$, $\mu(S_{:j}|S)=\left(1-\beta(a_{s_{j+1}})\right)\prod_{k\leq j}\beta(a_{s_k})$,
    while $\mu(S|S)=\prod_{k\leq |S|}\beta(a_{s_k})$. This class uses only $|X|$ parameters.
\end{myexp}

Finally, we need one more condition to guarantee attention overload. 

\begin{Assumption}[Weak Attention Overload]\label{AO_1st}
For  $a=a_{s_1}=a_{t_1} \in S \subseteq T$,  $\phi(a|S) \geq \phi(a|T)$.
\end{Assumption}

Because the structural menu domain is full, Sequential Path Independence and Weak Attention Overload jointly imply Attention Overload.

\subsection{List-Based Attention Overload with Heterogeneous Preferences}

We are now ready to define the heterogeneous model. As in RUM, any strict preference ordering may occur. Let $\mathcal{P}$ denote the set of strict preference orderings over $X$.
 
\begin{defn}[List-Based Attention Overload with Heterogeneous Preferences]\label{def: List-based Preference Ordering}
A choice rule $\ChoiProb$ has a \textit{List-Based Attention Overload with Heterogeneous Preferences} representation (H-LAO) if there exists a list-based attention frequency $\phi$ satisfying Assumptions \ref{PI} and \ref{AO_1st}, and a distribution $\tau$ on $\mathcal P$, such that
\begin{align*}
\ChoiProb(a|S)
=
\sum_{\substack{1\leq j\leq |S|\\a\in S_{:j}}}
\tau(\{\succ\in\mathcal P:a\text{ is $\succ$-best in }S_{:j}\})
\underbrace{\big(\phi(a_{s_j}|S)-\phi(a_{s_{j+1}}|S)\big)}_{\AttFilter(S_{:j}|S)},
\end{align*}
and $\ChoiProb(o^*|S)=1-\phi(a_{s_1}|S)$, for every nonempty $S\subseteq X$ and every $a\in S$.
\end{defn}

Benchmark H-LAO permits an unrestricted marginal distribution $\tau$ over $\mathcal P$ and heterogeneous stopping positions, while maintaining conditional independence between preference and stopping type. RUM is the special case in which every consumer reaches the end of the list. Limited attention allows H-LAO to accommodate behavior inconsistent with RUM without tying admissible preferences to positions on the list. We later relax independence and derive a sharp identified set under arbitrary menu-specific dependence between preferences and stopping, followed by a second sharp set that also removes Sequential Path Independence.

We now discuss related models in the literature. The closest list-based benchmark is the Random Depth Model of \citet{Ishiietal_2021_JME}, which uses exogenous variation in list order for a fixed menu; H-LAO instead holds the observed order fixed and exploits variation across menus to study attention overload. Subsequent work by \citet{Masatlioglu-Vu_2024_JET} orders heterogeneous consideration sets by attentiveness and identifies their composition and frequency. Its primitive is expansion of consideration sets across attention types, whereas H-LAO disciplines cross-menu reach along an observed list and permits an unrestricted marginal preference distribution. The set-monotone and stable RAUM of \citet{KashaevAguiar2022JET} more generally combines random utility with RAM-style random attention and consequently delivers partial identification. H-LAO's stronger sequential and cross-menu structure recovers attention and the full-attention stochastic choice rule under positive reach. Other approaches use parametric consideration restrictions or enriched data such as mixture choice \citep{Dardanoni-Manzini-Mariotti-Tyson_2020_ECMA,Dardanoni2023JPE}. We retain a broader comparison in the Supplemental Appendix.
 
\subsection{Behavioral Characterization}

The characterization relies on recovering the attention frequencies and using them to reconstruct the choice probabilities under full attention. First, we define, for every $S$, $\varphi(a_{s_1}|S)=1-\pi(o^*|S)$. Intuitively, $\varphi(a_{s_1}|S)$ equals the probability that the decision-maker attends to at least one alternative in $S$, since by assumption they choose $o^*$ if and only if they do not consider any alternative. This allows us to recursively define, for all $S=\{a_{s_1},a_{s_2},\dots,a_{s_{|S|}}\}$ and $j \neq 1$,
\begin{align*}
\varphi(a_{s_j}|S)=\varphi(a_{s_{j-1}}|S)\varphi(a_{s_j}|S_{j:}).
\end{align*}
If the model is correctly specified, $\varphi$ coincides with the underlying $\phi$, and hence all attention frequencies are recovered from the choice data.\footnote{We discuss the implication of incomplete observed menus in the next section.} The recursion ensures that $\varphi$ satisfies \Cref{PI}, but attention overload remains an additional testable cross-menu restriction.

Iterating the recursion gives
\begin{align}\label{eq:recover-reach-incomplete}
    \varphi(a_{s_j}|S)=\prod_{k=1}^j\left[1-\pi(o^*|S_{k:})\right].
\end{align}
Every factor lies in $[0,1]$, so $1\geq\varphi(a_{s_1}|S)\geq\cdots\geq\varphi(a_{s_{|S|}}|S)\geq0$ holds by construction. Under the maintained support condition, every factor in \eqref{eq:recover-reach-incomplete} is strictly positive whenever the corresponding suffix is observed. In particular, under full menu data, terminal reach is strictly positive, so the recursion defining $f$ is well-defined. 

To guarantee that the recovered attention frequency satisfies the required cross-menu restriction, it is enough to impose the following observable condition on outside-option choice. This type of monotonicity is commonly used as a measure of choice overload (see \citealp{chernev2015choice} for a review).

\begin{axm}[Weak Choice Overload] \label{axm: Choice Overload} For  $a_{s_1}=a_{t_1} \in S \subseteq T$, $\pi(o^*|T) \geq \pi(o^*|S)$.\end{axm}

The remaining testability question concerns the full-attention preference component.
To calculate it, we define an auxiliary stochastic choice rule $f$ recursively.
The base case is $f(x|\{x\}):=1$.
For $S=\{a_{s_1},\dots,a_{s_{|S|}}\}$ with $|S|\geq2$ and $k=2,\dots,|S|$, let\footnote{The recursion defining $f$ requires the denominator $\varphi(a_{s_{|S|}}|S)$ to be nonzero. This is ensured by our maintained support condition $\pi(S|S)>0$ for every nonempty $S$. We discuss the corresponding result when this condition is relaxed in the Supplementary Appendix.}
\begin{align*}
f(a_{s_k}|S)
:=
\frac{1}{\varphi(a_{s_{|S|}}|S)}
\left(
\pi(a_{s_k}|S)
-
\sum_{j=k}^{|S|-1}
\big(
\varphi(a_{s_j}|S)-\varphi(a_{s_{j+1}}|S)
\big)
f(a_{s_k}|S_{:j})
\right).
\end{align*}
Finally, define $f(a_{s_1}|S)=1-\sum_{k=2}^{|S|}f(a_{s_k}|S)$.
Thus, $f$ adds up to one by construction. When $\pi$ is generated by the model, $f(a|S)$ is the probability that $a$ is chosen under full attention and therefore records the stochastic-choice implications of $\tau$.
If the model holds, $f$ must be rationalizable by at least one distribution $\tau$ over strict preferences. Under the finite-universe, full-menu, adding-up, and strict-ordering conditions used here, the Block--Marschak characterization of \citet{Falmagne_1978_JMP} says that this is equivalent to nonnegativity of the Block--Marschak polynomials derived from $f$.

\begin{axm}[Non-negativity of BM] \label{axm: non-negativity of BM}
    For all $T$ and all $x \in T$, $
        \sum_{S:\ S \supseteq T } (-1)^{|S\setminus T|}f(x|S) \geq 0.$
\end{axm}
Notably, this condition also ensures that $f$ is non-negative. We are then ready to state the characterization theorem.

\begin{thm}[Characterization]\label{Thm:H-LAOchar} Let $\mathcal{D}=2^X \setminus \{\emptyset\}$. Suppose that $\pi(R|R)>0$ for every nonempty $R\subseteq X$.
The choice rule $\pi$ has a  H-LAO representation if and only if $\pi$ satisfies \Cref{axm: Choice Overload} and \ref{axm: non-negativity of BM}.
\end{thm}

Theorem \ref{Thm:H-LAOchar} is an if-and-only-if, and hence sharp, behavioral characterization of H-LAO under full menu data.
It separates limited attention from preference heterogeneity: the construction of $\varphi$, together with Weak Choice Overload, guarantees Sequential Path Independence and Attention Overload, while BM non-negativity tests whether the implied full-attention choice rule is generated by a distribution of preferences.
The next result states what is point identified under those conditions. 

\begin{thm}[Identification]\label{Prop:Identification Full Data} Let $\mathcal{D}=2^X \setminus \{\emptyset\}$ and $(\phi,\tau)$ be a H-LAO representation of $\pi$. Then $\phi=\varphi$ and $\tau(\{\succ \in \mathcal{P}: a \text{ is } \succ\text{-best in } T \})=f(a|T)$.
\end{thm}

Thus, full-menu data point identify the attention frequency $\phi$ and the full-attention stochastic choice rule $f$. They need not uniquely identify the complete distribution $\tau$ over rankings. Once $f$ has been recovered, however, the remaining conclusions are standard consequences of finite random-utility theory \citep{Falmagne_1978_JMP,turansick2022identification}. In particular, the sharp compatible set is
\begin{align*}
    \mathcal T(f)=\Bigg\{\tau\in\Delta(\mathcal P):
    f(a|S)=\sum_{\succ\in\mathcal P}
    \Indicator\{a\text{ is $\succ$-best in }S\}\tau(\succ)
    \text{ for every }a\in S\subseteq X\Bigg\},
\end{align*}
and probabilities of preference events are bounded by minimizing and maximizing their $\tau$-mass over this polytope. Falmagne's Mobius inversion also recovers the probability of each lower contour set and hence every alternative's rank distribution. For example,
\begin{align*}
    \tau(\{\succ:a\succ b\})=f(a|\{a,b\}),
    \qquad
    \tau(\{\succ:a\text{ is top ranked}\})=f(a|X).
\end{align*}
The H-LAO contribution is the recovery of $f$ from limited-attention choice data; the Supplemental Appendix records the standard random-utility formulas needed to translate the recovered rule into preference-event information.

\begin{myexp}[Recovering attention and preference types]\label{exp:H-LAO-recovery}
Let the observed list be $a,b,c$. Suppose continuation at each stage is geometric with probability $0.8$, and let $\tau$ assign probabilities $0.5$, $0.3$, and $0.2$ to $a\succ b\succ c$, $b\succ c\succ a$, and $c\succ a\succ b$, respectively. The resulting choice probabilities are reported below, rounded to two decimal places; subsequent calculations use the unrounded probabilities.
\begin{center}
\begin{tabular}{lcccc}
\toprule
$\pi$ [$f$] & $a$ & $b$ & $c$ & $o^*$\\
\midrule
$\{a\}$       & $0.80 \ [1.00]$ & --       & --       & $0.20 \ [0.00]$\\
$\{b\}$       & --       & $0.80 \ [1.00]$ & --       & $0.20 \ [0.00]$\\
$\{c\}$       & --       & --       & $0.80 \ [1.00]$ & $0.20 \ [0.00]$\\
$\{a,b\}$     & $0.61 \ [0.70]$ & $0.19 \ [0.30]$ & --       & $0.20 \ [0.00]$\\
$\{a,c\}$     & $0.48 \ [0.50]$ & --       & $0.32 \ [0.50]$ & $0.20 \ [0.00]$\\
$\{b,c\}$     & --       & $0.67 \ [0.80]$ & $0.13 \ [0.20]$ & $0.20 \ [0.00]$\\
$\{a,b,c\}$   & $0.51 \ [0.50]$ & $0.19 \ [0.30]$ & $0.10 \ [0.20]$ & $0.20 \ [0.00]$\\
\bottomrule 
\end{tabular}   
\end{center}
Outside-option choices recover first-position attention of $0.8$ in every menu. Sequential Path Independence then gives reach probabilities $0.8$, $(0.8)^2$, and $(0.8)^3$ along the three-item list. Substituting these values into the recursion recovers the full attention choice probabilities illustrated above in square brackets.
Note that $f(a|X) < \pi(a|X)$ while $f(b|X) > \pi(b|X)$. In this three-alternative example, $\mathcal T(f)$ is a singleton and recovers the three stated preference types. With larger universes, $\mathcal T(f)$ need not be a singleton even though $f$ and the standard lower-contour and rank distributions remain identified.
\end{myexp}

\subsection{Identification under Limited Data}\label{subsection:hlao data requirements}

In the previous section, we consider identification under full domain. Here, we outline the sufficient conditions for identification under limited data, and provide partial identification results. First, recall that a domain $\mathcal{D}$ is suffix-closed (prefix-closed) if the suffixes (prefixes) of $S$ belong to $\mathcal{D}$ whenever $S\in\mathcal{D}$.

\begin{prop}[Sufficient Condition for Identification: Attention and Full-attention choice]
Let $\mathcal D$ be suffix-closed and $(\phi,\tau)$ be a H-LAO
representation of $\pi$. Then, $\phi(a|S)$ is point identified for every
$S\in\mathcal D$ and $a\in S$. If, in addition, $\mathcal D$ is also
prefix-closed,
$\tau(\{\succ\in\mathcal P:x\text{ is }\succ\text{-best in }S\})$
is point identified for every $S\in\mathcal D$ and $x\in S$.
\end{prop}

Thus suffix data identify attention frequencies, whereas recovery of the full-attention choice probabilities additionally requires the relevant smaller-prefix problems. The latter condition can be checked recursively and requires observation of the relevant contiguous list intervals.

For later use, define the attention-recoverable domain
\begin{align*}
\mathcal D_{\phi}
=
\left\{
S\in\mathcal D:
S_{j:}\in\mathcal D
\text{ for every }j=1,\dots,|S|
\right\}.
\end{align*}
Let $\mathcal D_f$ be the smallest collection of menus that contains every singleton and has the following recursive property: a menu $S\in\mathcal D_{\phi}$ with $|S|\geq2$ belongs to $\mathcal D_f$ whenever $\varphi(a_{s_{|S|}}|S)>0$ and every proper prefix $S_{:j}$ belongs to $\mathcal D_f$. Thus $\varphi(a|S)$ is recovered for $S\in\mathcal D_{\phi}$, while $f(a|S)$ is recovered for $S\in\mathcal D_f$.

We now provide an additional identification result for binary comparisons. Such comparisons may be particularly relevant to policymakers seeking to determine the proportion of the population that prefers one policy option to another. This information can help assess the relative support for competing policies and inform decisions between alternative interventions. Suppose $a$ precedes $b$ and the menus $\{a,b\}$ and $\{b\}$ are observed. Then  $\left[1-\pi(o^*|\{a,b\})\right]
     \left[1-\pi(o^*|\{b\})\right]$ 
is the probability of reaching $b$ in the binary menu. Then, one can show that 
    $$\tau(\{\succ:b\succ a\})
    =\frac{\pi(b|\{a,b\})}{\left[1-\pi(o^*|\{a,b\})\right]
     \left[1-\pi(o^*|\{b\})\right]}$$

\begin{prop}[Sufficient Condition for Identification: Pairwise Preferences]
Let $\mathcal{D}$ contain $\{a,b\}$,  $\{a\}$ and $\{b\}$, and $(\phi,\tau)$ be a H-LAO representation of $\pi$. Then,  $\tau(\{\succ \in \mathcal{P}: b \succ a \})$ is point identified.
\end{prop}

Hence singleton and binary menus identify all pairwise shares.
This conclusion uses only binary and singleton menus rather than suffix and prefix closure.
In general, when the aforementioned domain conditions are not satisfied, attention or preference need not be point identified.
Nevertheless, useful partial-identification bounds remain available.

\begin{prop}[Partial Identification of Attention]\label{thm:partial-identification-attention}
Let $(\phi,\tau)$ be an H-LAO representation of $\pi$, let $S\subseteq X$ be nonempty, and let $a\in S$. Then
\begin{align*} \max_{\substack{R\in\mathcal D_{\phi}:R\supseteq S}}
    \varphi(a|R)
    \leq \phi(a|S) \leq
    \min_{\substack{T\in\mathcal D_{\phi}:a\in T\subseteq S}}
    \varphi(a|T)
\end{align*}
A bound is omitted if its index set is empty.
\end{prop}

The proposition bounds the attention frequency of $a$ even when the target menu $S$ is unobserved.\footnote{Although $S$ need not belong to the observed domain $\mathcal D$, the structural H-LAO attention frequency is defined on the full menu domain.} If $S\in\mathcal D$ and the suffixes required to recover $\varphi(a|S)$ are observed, both bounds include $S$ and collapse to the identified value. For example, suppose that every menu in \Cref{exp:H-LAO-recovery} is observed except $\{a,c\}$. Then, the lower bound is given by $\varphi(c|\{a,b,c\})=(1-\pi(o^*|\{a,b,c\}))(1-\pi(o^*|\{b,c\}))(1-\pi(o^*|\{c\}))=0.8^3$, and the upper bound is given by $\varphi(c|\{c\})=(1-\pi(o^*|\{c\}))=0.8$, whereas the actual $\phi(c|\{a,c\})$ is $0.8^2$. This example shows that neither bound requires data on $\{a,c\}$.

We now discuss the identification of the distribution of types under limited data. We exploit two distinct sources of identifying power. First, because $f$ is random-utility rationalizable, it satisfies the classic regularity so that $f(a|T)\geq f(a|S)\geq f(a|R)$ for $a\in T\subseteq S\subseteq R$. Second, the list structure can strengthen the lower bound. When $S$ is a prefix of $R$, define
\begin{align*}
    \Psi(a,S,R)
    =
    \sum_{\substack{1\leq j\leq |R|:\\a\in R_{:j}\subsetneq S}}
    f(a|R_{:j})\mu(R_{:j}|R).
\end{align*}

Write $\ell(S):=a_{s_{|S|}}$ for the last alternative of $S$. For $a\in S$, define the eligible-superset collection
\begin{align*}
\mathcal R_{\Psi}(a,S)
=\Big\{R\in\mathcal D_{\phi}:&\ S\text{ is a prefix of }R,\
\varphi(\ell(S)|R)>0, R_{:j}\in\mathcal D_f
\text{ whenever }a\in R_{:j}\subsetneq S\Big\}.
\end{align*}

With this defined, we are ready to state the proposition.

\begin{prop}[Partial Identification of $\tau$]\label{thm:partial-identification-f}
Let $(\phi,\tau)$ be an H-LAO representation of $\pi$, let $S\subseteq X$ be nonempty, and let $a\in S$. Write $\theta(a,S):=\tau(\{\succ:a\text{ is $\succ$-best in }S\})$. Then
\begin{align*}
\max\left\{
  \max_{R\in\mathcal R_{\Psi}(a,S)}
  \frac{\pi(a|R)-\Psi(a,S,R)}
       {\varphi(\ell(S)|R)},
  \quad
  \max_{\substack{R\in\mathcal D_f:R\supseteq S}}f(a|R)
  \right\} 
  \leq \theta(a,S)
  \leq \min_{\substack{T\in\mathcal D_f:a\in T\subseteq S}}f(a|T).
\end{align*}
Each extremum is omitted if its index set is empty.
\end{prop}

To utilize this result, suppose in \Cref{exp:H-LAO-recovery} every menu is observed except $\{a,b\}$. Then, the first lower bound for $\tau(\{\succ:a\text{ is $\succ$-best in }\{a,b\}\})$ can be calculated as  \begin{align*}
    \frac{\pi(a|\{a,b,c\})-(\varphi(a|\{a,b,c\})-\varphi(b|\{a,b,c\}))}{\varphi(b|\{a,b,c\})}=\frac{0.5056-(0.8-0.8^2)}{0.8^2}=0.540
\end{align*}
whereas the second lower bound is not applicable since $f(a|\{a,b,c\})$ is not well-defined if $\{a,b\}$ is missing from $\mathcal{D}$. On the other hand, suppose in \Cref{exp:H-LAO-recovery} every menu is observed except $\{a,c\}$. Then, the first lower bound for $\tau(\{\succ:a\text{ is $\succ$-best in }\{a,c\}\})$ does not apply since there is not a set $R$ so that $\Psi(a,\{a,c\},R)$ is defined. Nonetheless, the second lower bound is well-defined since $f(a|\{a,b,c\})$ is calculated as $0.5$ from $\pi$ given the dataset.

The Supplemental Appendix reports two nested sensitivity envelopes around benchmark H-LAO. The set $\mathcal T_{\mathcal D}^{\mathrm{dep}}(\pi)$ retains recovered prefix masses but permits arbitrary menu-specific dependence between preferences and stopping. The set $\mathcal T_{\mathcal D}^{\mathrm{noPI}}(\pi)$ additionally treats the prefix masses as latent and drops Sequential Path Independence. Both are sharp for their stated repeated-cross-section specifications, and
$\mathcal T_{\mathcal D}(\pi)
 \subseteq\mathcal T_{\mathcal D}^{\mathrm{dep}}(\pi)
 \subseteq\mathcal T_{\mathcal D}^{\mathrm{noPI}}(\pi)$
whenever the three sets are defined. This nesting makes the identifying content of preference--stopping independence and Sequential Path Independence transparent. All definitions, proofs, multiple-list results, and projection procedures appear in the Supplemental Appendix.

\section{Econometric Methods}\label{subsection:Econometric Methods single preference}

This section develops estimation and inference for both models. We begin with homogeneous AOM, using Theorem \ref{theorem:Characterization}, Propositions \ref{lemma:Regularity Violation at Binary Comparisons} and \ref{lemma:Regularity Violation at Bigger Sets}, and Theorem \ref{theorem:Revealed Attention} to test candidate preferences and form confidence bounds for attention. We then turn to H-LAO and construct confidence sets for preference distributions and attention that remain valid under weak reach.

Related econometric work identifies preferences in the presence of costly search using rich covariate variation and derivative restrictions \citep{Abaluck-Compiani-Zhang_2026_JPE}. Our procedures instead take menu and list variation as the source of identification and conduct simultaneous inference directly on the resulting choice-probability restrictions. Section SA-1 compares additional approaches to latent consideration and preference heterogeneity.

\subsection{Inference for Homogeneous AOM}

For homogeneous AOM, we observe choices for $n$ units indexed by $i=1,\dots,n$. Unit $i$ faces menu $Y_i\in\mathcal D$ and chooses $y_i\in Y_i$, where $\mathcal D\subseteq2^X\setminus\{\emptyset\}$ is the collection of observed menus.

\begin{Assumption}[Choice Data]\label{assumption:Choice Data}
    The data consist of a random sample of menus and observed choices $\{(Y_i,y_i):1\leq i\leq n\}$ with $Y_i\in\mathcal{D}$ and $\mathbb{P}[y_i=a|Y_i=S]=\ChoiProb(a|S)$.
\end{Assumption}

We conduct inference conditionally on the realized menu assignments $(Y_1,\dots,Y_n)$, or equivalently on the menu-specific sample sizes and observation indices. Conditional on these assignments, choices from different menus are independent and $N_S=\sum_{i=1}^n \Indicator(Y_i=S)$, the sample size for the choice problem $S$, is fixed. All probability statements below use this conditional interpretation; integrating over the menu assignments gives the corresponding unconditional conclusions. We restrict $\mathcal{D}$ to menus with $N_S\geq2$ and consider sequences for which the relevant menu-specific sample sizes diverge. Throughout this section, every logarithmic factor is understood to be truncated below at one, that is, $\max\{\log(\cdot),1\}$.

A given preference ordering $\succ$ is compatible with our AOM if and only if $\succ$-Regularity holds. For inference, define the nonredundant comparison index
\begin{align*}
    \mathcal I_\succ(\mathcal D)
    =\{(a,S,T):T,S\in\mathcal D,\ T\subset S,\ a\in T,\ U_\succeq(a)\cap T\neq T\}.
\end{align*}
The last condition is equivalent to requiring that $a$ not be the $\succ$-worst alternative in $T$, as otherwise $\pi(U_\succeq(a)|T)=1$ and the corresponding restriction $\pi(a|S)\leq1$ is automatically satisfied. We therefore collect the informative inequality constraints in the following hypothesis:
\begin{align}\label{eq:hypotheses}
    \mathsf{H}_0: \max_{(a,S,T)\in\mathcal I_\succ(\mathcal D)} {D}(a|S,T) \leq 0,\quad
    \text{where } {D}(a|S,T) = \pi(a|S) - \pi(U_{\succeq}(a)|T).
\end{align}
To construct a test statistic for the hypotheses in \eqref{eq:hypotheses}, we can replace the unknown choice probabilities by their estimates. Specifically, for any event $A\subseteq S$, let $\widehat{\ChoiProb}(A|S)$ be an estimated choice probability, with $\sigma^2(A|S)$ and $\widehat{\sigma}^2(A|S)$ denoting its true and estimated variances:
\begin{align*}
    \widehat{\ChoiProb}(A|S)
    &=\frac{1}{N_S}\sum_{i=1}^n \Indicator(Y_i=S,y_i\in A),\\
    \sigma^2(A|S)&=\frac{\pi(A|S)(1-\pi(A|S))}{N_S},\ \ 
    \widehat{\sigma}^2(A|S)=\frac{\widehat{\pi}(A|S)(1-\widehat{\pi}(A|S))}{N_S}.
\end{align*}
We write $\widehat{\ChoiProb}(a|S)$, $\sigma^2(a|S)$, and $\widehat{\sigma}^2(a|S)$ as shorthand for the singleton event $A=\{a\}$.

Now define the test statistic
\begin{align*}
\mathsf{T}(\succ) = \max\left\{\max_{(a,S,T)\in\mathcal I_\succ(\mathcal D)}{\widehat{D}(a|S,T)}/{\widehat{\sigma}(a|S,T)}\ ,\ 0\right\},
\end{align*}
where $\widehat\sigma(a|S,T)$ is the standard error of $\widehat D(a|S,T)$, so $\widehat{\sigma}^2(a|S,T)=\widehat{\sigma}^2(a|S)+\widehat{\sigma}^2(U_{\succeq}(a)|T)$. Truncating the maximum at zero prevents rejection when every estimated difference is nonpositive. The notation $\mathsf T(\succ)$ emphasizes that the inequalities and the statistic depend on the candidate ordering.

The standardized procedure is used for comparisons with positive oracle variance. We interpret $0/0$ as zero and a nonzero numerator divided by zero as signed infinity, leaving the other comparisons active. Gaussian simulation sets coordinates with zero estimated standard errors to zero. Under the conditions below, the probability of a zero estimated standard error vanishes. The plug-in covariance matrix used for Gaussian simulation is computed from the menu-level multinomial covariance matrices and is therefore positive semidefinite.

The validity argument has two steps: a uniform Gaussian coupling and a distributional comparison for the feasible critical value. For one comparison, $\widehat D(a|S,T)/\sigma(a|S,T)$ is approximately Gaussian with unit variance and mean $D(a|S,T)/\sigma(a|S,T)\leq0$ under the null. The challenge is to approximate all comparisons simultaneously when their number may be large and their standard errors are estimated. The following theorem supplies that coupling.

\begin{thm}[Uniform Gaussian Coupling]\label{theorem:Uniform Gaussian Approximation}
Assume Assumption \ref{assumption:Choice Data} holds and all oracle variances entering the comparisons are positive. Let $\mathfrak{c}_\succ=|\mathcal I_\succ(\mathcal D)|$ be the number of nonredundant comparisons in \eqref{eq:hypotheses}, and
\begin{align*}
   \mathfrak{c}_\sigma = \min_{(a,S,T)\in\mathcal I_\succ(\mathcal D)}\Bigg\{\min\Big\{ \frac{N_S}{\log(N_S)}, \frac{N_T}{\log(N_T)} \Big\} \cdot \max\Big\{ \sigma(a|S),\ \sigma(U_{\succeq}(a)|T)\Big\}\Bigg\},
\end{align*}
Suppose $\mathfrak{c}_\sigma\to\infty$ and $\log(\mathfrak{c}_\succ)/\mathfrak{c}_\sigma\to0$. Then, on a possibly enlarged probability space, there exist mean-zero Gaussian random variables $\{\mathsf{Z}_\succ(a|S,T):(a,S,T)\in\mathcal I_\succ(\mathcal D)\}$ such that (i) their covariance structure matches that of the centered and oracle-standardized comparisons indexed by $\mathcal I_\succ(\mathcal D)$; and (ii)
\begin{align*}
\max_{(a,S,T)\in\mathcal I_\succ(\mathcal D)} \Big| \frac{\widehat{D}(a|S,T) -D(a|S,T)}{\widehat{\sigma}(a|S,T)} - \mathsf{Z}_\succ(a|S,T) \Big| = \Op\left( \frac{\log(\mathfrak{c}_{\succ})}{\mathfrak{c}_{\sigma}} \right).
\end{align*}
\end{thm}

The coupling error depends on the number of comparisons only logarithmically, so the approximation permits many inequalities relative to the effective menu sample sizes and provides an explicit sample-size/dimension benchmark. Because the universal constants are not calibrated, the Supplemental Appendix complements the theorem with simulations documenting finite-sample performance.

The definition of $\mathfrak{c}_\sigma$ keeps each comparison's menu sample sizes paired with its own standard errors before taking the minimum over comparisons. It therefore avoids multiplying the smallest sample size from one comparison by the least favorable standard error from an unrelated comparison. Intuitively, $\mathfrak{c}_\sigma$ measures the smallest effective sample size on a square-root scale, up to logarithmic factors. When menu sample sizes are comparable and the relevant choice indicators have variances of order $\pi(1-\pi)$, it simplifies to
\begin{align*}
\mathfrak{c}_\sigma \propto  \min_{T,S\in\mathcal{D}} \min\Big\{ \frac{\sqrt{\pi (1-\pi) N_S}}{\log(N_S)},\ \frac{\sqrt{\pi (1-\pi) N_T}}{\log(N_T)} \Big\} = \min_{S\in\mathcal{D}}\frac{\sqrt{\pi (1-\pi) N_S}}{\log(N_S)}.
\end{align*}
Since $\pi (1-\pi) N_S$ is related to the expected number of times an alternative is selected, a Gaussian approximation to a multinomial distribution typically requires this effective sample size to diverge (even in a low and fixed dimensional setting). 

There is consequently no explicit requirement that $|X|$ be $o(n)$ or $o(\sqrt n)$. The formal restriction operates through $\log(\mathfrak c_\succ)$ and the smallest effective menu sample size. With complete menu data, the number of menus and inequalities grows combinatorially in $|X|$, making adequate per-menu sampling increasingly demanding. For example, if $n$ observations are allocated approximately evenly across the $2^{|X|}-1$ nonempty menus, then $N_S\approx n/(2^{|X|}-1)\approx n/2^{|X|}$. Thus, even the basic requirement $N_S\to\infty$ entails $2^{|X|}=o(n)$, so $|X|$ can grow only roughly logarithmically with $n$; the theorem's additional logarithmic terms tighten this benchmark. This feature motivates our methods for incomplete domains and binary screening: the six-alternative simulation has 664 nonredundant inequalities under the complete domain but as few as 15 under the reported sparse design. Binary screening avoids forming all subset relations, after which Corollary \ref{coro:mip-characterization} represents the surviving rankings through a mixed-integer feasibility problem instead of enumerating them. Section SA-5 of the Supplemental Appendix gives the implementation hierarchy, optimization dimensions, solver certificates, and reproducible scalability benchmarks for both AOM and H-LAO.

Theorem \ref{theorem:Uniform Gaussian Approximation} attains the nearly optimal inverse-root-effective-sample-size scale. The proof represents each menu-level choice indicator as a function of a scalar uniform variable, making the centered comparisons an empirical process indexed by functions of total variation at most two. A bounded-variation strong coupling \citep[Lemma SA26]{Cattaneo-Feng-Underwood_2024_JASA}, followed by control of the estimated standard errors, gives the result; see also \citet[Remark 1]{cattaneo2025strong}. Direct distributional approximations provide another route \citep[see Section 5 of][]{chernozhukov2023nearly}. Our theorem does not require a lower bound on the minimum eigenvalue of the covariance matrix and is therefore closer in scope to \citet*{Chernozhukov-Chetverikov-Kato-Koike_2022_AOS}; applying their general result directly would yield an error of order $\mathfrak c_\sigma^{-1/2}$ up to logarithmic factors.

The next step is to construct critical values with a feasible Gaussian approximation. Conditional on the data, we simulate an auxiliary Gaussian vector independently of the sample, with covariance equal to the estimated correlation matrix of the standardized comparisons. Let $\widehat{\mathsf{Z}}_\succ(a|S,T)$ denote its coordinates (see the Supplemental Appendix for the precise formula), and define
\begin{align*}
\widehat{\mathsf{T}}^{\mathtt{G}}(\succ) = \max\left\{\max_{(a,S,T)\in\mathcal I_\succ(\mathcal D)} \widehat{\mathsf{Z}}_\succ(a|S,T)\ ,\ 0\right\}.
\end{align*}
Define $\widehat{\mathrm{cv}}(\alpha,\succ)=\inf\{t:\mathbb{P}[\widehat{\mathsf T}^{\mathtt G}(\succ)\leq t|\text{Data}]\geq1-\alpha\}$. The next theorem bounds the test's size distortion.

\begin{thm}[Preference Elicitation with Theorem \ref{theorem:Characterization}]\label{theorem:Valid Critical Values}
Assume the conditions of Theorem \ref{theorem:Uniform Gaussian Approximation} hold, fix $\alpha\in(0,1/2)$, and suppose the remainder displayed below converges to zero. Then, under the null hypothesis $\mathsf{H}_0$ in \eqref{eq:hypotheses},
\begin{align*}
\mathbb{P}\big[ \mathsf{T}(\succ) > \widehat{\mathrm{cv}}(\alpha,\succ) \big] \leq \alpha + c \sqrt{\frac{\log^{\frac{5}{2}}(\mathfrak{c}_{\succ})\log^{\frac{1}{2}}(\mathfrak{c}_{\sigma})}{\mathfrak{c}_\sigma}}
\end{align*}
for some constant $c$.
\end{thm}

Standard test inversion gives the asymptotically valid confidence set $\mathsf{CS}(1-\alpha)=\{\succ:\mathsf{T}(\succ)\leq\widehat{\mathrm{cv}}(\alpha,\succ)\}$. Condition \eqref{eq:attentive at binaries} can be incorporated by adding the probability comparisons implied by the $\eta$-constrained binary relation.

The size bound above uses the fully general Gaussian comparison inequality in Proposition 2.1 of \citet*{Chernozhukov-Chetverikov-Kato-Koike_2022_AOS} and therefore delivers a conservative fourth-root error remainder without imposing nonsingularity. When the oracle correlation matrix is bounded away from singularity, Theorem 2.3 of \citet{lopes2022central} gives nearly linear covariance-error dependence; the Supplemental Appendix verifies this condition for the important binary-screening design under balanced standard errors. It also gives a universal critical value and a finite-sample projection procedure based on simultaneous menu-probability bands, neither of which requires covariance comparison, as well as an alpha-split hybrid of the universal and correlated-Gaussian critical values. Generalized moment selection can improve power by reducing the influence of inequalities that are far from binding. The all-inequalities zero-centered critical value (labeled \texttt{LF} in the software) is the procedure covered directly by Theorem \ref{theorem:Valid Critical Values}. The simulations report this baseline and the GMS refinement of \citet{Andrews-Soares_2010_ECMA} from the same Gaussian draws. We retain the correlated-Gaussian result as the default because it uses the dependence structure and is generally more powerful.

Pairwise revealed preference can be assessed by checking whether, for example, $a\succ b$ for every $\succ\in\mathsf{CS}(1-\alpha)$. This approach uses the full identifying power of Theorem \ref{theorem:Characterization} but can be computationally costly. Proposition \ref{lemma:Regularity Violation at Binary Comparisons} supplies a simpler screen when binary menus are observed or pairwise comparisons are the primary target. As Example \ref{Exp_1} illustrates, applying this screen first can sharply reduce the number of orderings that must be tested against $\succ$-Regularity.

We fix two alternatives, say $a$ and $b$, and let $\mathcal{D}_{ab}$ be the collection of choice problems containing both $a$ and $b$, excluding the binary comparison; that is, $\mathcal{D}_{ab} = \{ S\in \mathcal{D}:\ S \supsetneq \{a,b\}\}$. We also recall the simplified notation $\pi(a|b) = \pi(a|\{a,b\})$ and $N_{ab}=N_{\{a,b\}}$ for binary comparisons. Then, we may deduce the preference ordering $b\succ a$ if we are able to reject $\mathsf{H}_0: \max_{S\in\mathcal{D}_{ab}} D(a|S,b) \leq 0$, where we define $D(a|S,b) = \ChoiProb(a|S) - \ChoiProb(a|b)$.
The corresponding test statistic is
\begin{align*}
\mathsf{T}(ab) &= \max\left\{\max_{S\in\mathcal{D}_{ab}}{\widehat{D}(a|S,b)}/{\widehat{\sigma}(a|S,b)}\ ,\ 0\right\}, 
\end{align*}
where $\widehat{D}(a|S,b) = \widehat{\ChoiProb}(a|S) - \widehat{\ChoiProb}(a|b)$, and $\widehat{\sigma}(a|S,b)$ is its standard error. We employ the same technique to construct a critical value. Specifically, let $\widehat{\mathsf{Z}}_{ab}(S)$ for $S\in\mathcal{D}_{ab}$ be a collection of Gaussian random variables with zero mean, unit variance, and covariance matching the estimated correlation matrix of $\widehat{D}(a|S,b)$. The critical value is $\widehat{\mathrm{cv}}(\alpha,ab) = \inf\{ t : \mathbb{P}[\widehat{\mathsf{T}}^{\mathtt{G}}(ab) \leq t|\text{Data}]\geq 1-\alpha \}$ with $\widehat{\mathsf{T}}^{\mathtt{G}}(ab) = \max\{\max_{S\in \mathcal{D}_{ab}} \widehat{\mathsf{Z}}_{ab}(S),0\}$. The next result follows from Theorem \ref{theorem:Valid Critical Values}.

\begin{coro}[Preference Elicitation with Proposition \ref{lemma:Regularity Violation at Binary Comparisons}]\label{theorem:Valid Critical Values 1}
Assume the conditions of Theorem \ref{theorem:Valid Critical Values} hold for the binary-screening comparisons. Define $\mathfrak{c}_\sigma$ as
\begin{align*}
\mathfrak{c}_\sigma = \min_{S\in\mathcal{D}_{ab}}\Bigg\{\min\Big\{ \frac{N_S}{\log(N_S)}, \frac{N_{ab}}{\log(N_{ab})} \Big\} \cdot \max\Big\{ \sigma(a|S), \sigma(a|b)\Big\}\Bigg\}.
\end{align*}
Then, under the null hypothesis that 
$\mathsf{H}_0: \max_{S\in\mathcal{D}_{ab}} D(a|S,b) \leq 0$, the size distortion satisfies 
\begin{align*}
\mathbb{P}\big[ \mathsf{T}(ab) > \widehat{\mathrm{cv}}(\alpha,ab) \big] \leq \alpha + c \sqrt{\frac{\log^{\frac{5}{2}}(|\mathcal{D}_{ab}|)\log^{\frac{1}{2}}(\mathfrak{c}_{\sigma})}{\mathfrak{c}_\sigma}}
\end{align*}
for some constant $c$. If, in addition, there is a constant $C<\infty$ such that
\begin{align*}
    C^{-1}\leq\frac{\sigma(a|S)}{\sigma(a|b)}\leq C
    \qquad\text{for every }S\in\mathcal D_{ab},
\end{align*}
and $\log^{5/2}(|\mathcal D_{ab}|\mathfrak c_\sigma)/\mathfrak c_\sigma\to0$, then the fixed-pair size distortion satisfies the sharper bound
\begin{align*}
\mathbb{P}\big[ \mathsf{T}(ab) > \widehat{\mathrm{cv}}(\alpha,ab) \big]
\leq \alpha+c(1+C^2)
\frac{\log^{5/2}(|\mathcal D_{ab}|\mathfrak c_\sigma)}{\mathfrak c_\sigma}.
\end{align*}
\end{coro}

The comparisons $\widehat D(a|S,b)$ share only the binary-menu estimate $\widehat\pi(a|b)$; the estimates $\widehat\pi(a|S)$ are independent across $S\in\mathcal D_{ab}$. If $\sigma(a|S)\asymp\sigma(a|b)$ uniformly over $\mathcal D_{ab}$, this diagonal-plus-rank-one structure keeps the minimum eigenvalue of the correlation matrix bounded away from zero. The nondegenerate Gaussian comparison therefore removes the outer square root in the fixed-pair error rate, as formalized in Corollary \ref{theorem:Valid Critical Values 1}. The Supplemental Appendix provides the covariance calculation and rate derivation. A screen that stacks multiple ordered pairs remains covered by the singularity-robust bound unless nondegeneracy of its joint correlation matrix is established separately.

We use the lower attention bound $\phi(a|S)\geq\max_{R\in\mathcal D:\,R\supseteq S}\pi(a|R)$ from Theorem \ref{theorem:Revealed Attention} to construct a one-sided confidence bound; the upper bound is handled analogously. Maximizing noisy estimates creates upward selection bias and can overstate the lower bound. We therefore construct $\underline{\widehat\phi}(a|S)$ with coverage approaching $1-\alpha$.

Our construction is based on computing the maximum over a collection of adjusted empirical choice probabilities. This is a finite intersection-bound problem. As emphasized by \citet{Chernozhukov-Lee-Rosen_2013_ECMA}, directly taking a supremum or infimum of noisy estimates creates selection bias and can make the estimated identified set too small. Our correction follows the same general principle, but specializes it to a finite collection of menu-specific multinomial probabilities. This yields a pivotal Gaussian maximum because the menu subsamples are conditionally independent; their framework also accommodates continua of inequalities, nonparametric first stages, and adaptive inequality selection. To be precise, we define
\begin{align*}
\underline{\widehat{\phi}}(a|S) &= \max_{R\supseteq S,R\in\mathcal{D}}\Big\{ \widehat{\ChoiProb}(a|R) - \mathrm{cv}(\alpha,\underline{{\phi}}(a|S))\cdot\widehat{\sigma}(a|R) \Big\}.
\end{align*}
Here, $\mathrm{cv}(\alpha,\underline{{\phi}}(a|S))$ is the $(1-\alpha)$-quantile of $\max_{R\supseteq S,R\in\mathcal{D}}\mathsf{Z}_{a,S}(R)$, where $\mathsf{Z}_{a,S}(R)$ are independent standard Gaussian random variables. As such, the critical value is pivotal. If any required estimated standard error is zero, we conservatively set $\underline{\widehat\phi}(a|S)=0$. Otherwise, the construction uses the normal approximation $\widehat{\ChoiProb}(a|R) \overset{\mathrm{a}}{\thicksim} \text{Normal}(\ChoiProb(a|R),\ \sigma^2(a|R))$. Conditional on the menu assignments, the estimated choice probabilities are mutually independent because they are constructed from different subsamples, and
\begin{align*}
&\ \mathbb{P}\Big[ \widehat{\ChoiProb}(a|R) - \mathrm{cv}(\alpha,\underline{{\phi}}(a|S))\cdot\widehat{\sigma}(a|R) \leq \ChoiProb(a|R),\ \forall R\supseteq S,\ R\in\mathcal{D} \Big]\\
&\quad = \mathbb{P}\Big[ \max_{R\supseteq S,R\in\mathcal{D}}\frac{ \widehat{\ChoiProb}(a|R) - \ChoiProb(a|R)}{\widehat{\sigma}(a|R)} \leq \mathrm{cv}(\alpha,\underline{{\phi}}(a|S)) \Big]\\
&\quad  \approx \mathbb{P}\Big[ \max_{R\supseteq S,R\in\mathcal{D}}\mathsf{Z}_{a,S}(R) \leq \mathrm{cv}(\alpha,\underline{{\phi}}(a|S)) \Big] = 1-\alpha.
\end{align*}
where $\approx$ denotes a large-sample approximation. Thus, with probability approaching $1-\alpha$, every adjusted estimate lies below its population choice probability, and their maximum lies below the population attention bound. The next theorem makes this argument precise.
 
\begin{thm}[Attention Frequency Elicitation with Theorem \ref{theorem:Revealed Attention}]\label{theorem:Valid Lower Bound for Attention}
Let $\ChoiProb$ be AOM-compatible, assume Assumption \ref{assumption:Choice Data} holds, and suppose $0<\pi(a|R)<1$ for every $R\in\mathcal D$ with $R\supseteq S$ entering the construction. Define
\begin{align*}
\mathfrak{c}_{\supseteq S}=\big|\{R\in\mathcal{D}:R\supseteq S\}\big|
\end{align*}
as the number of supersets of $S$, and 
\begin{align*}
    \mathfrak{c}_\sigma = \min_{R\supseteq S,R\in\mathcal{D}} \frac{\sqrt{N_R \pi(a|R)(1-\pi(a|R))}}{\log(N_R)}.
\end{align*}
Suppose $\mathfrak{c}_\sigma\to\infty$ and $\log(\mathfrak{c}_{\supseteq S})/\mathfrak{c}_\sigma\to0$. Then,
\begin{align*}
    \mathbb{P}\big[ \phi(a|S) \geq \underline{\widehat{\phi}}(a|S) \big] \geq 1 - \alpha - c \frac{\log^{\frac{3}{2}}(\mathfrak{c}_{\supseteq S})\log(\mathfrak{c}_{\sigma})}{\mathfrak{c}_\sigma}
\end{align*}
for some constant $c$. 
\end{thm}

\subsection{Estimation and Inference for Heterogeneous H-LAO}\label{subsection: hlao inference main}

H-LAO (partially) identifies the preferences distribution rather than selecting one candidate ordering. Estimation and statistical inference again start from the empirical choice rule. Replacing $\pi$ by $\widehat\pi$ in \eqref{eq:recover-reach-incomplete} gives plug-in estimates of all reach probabilities and prefix masses on $\mathcal D_\phi$. On menus in $\mathcal D_f$, the recursion also gives a plug-in estimate of the full-attention rule $f$, after which preference-event bounds are obtained by linear programming over $\mathcal T(f)$. On suffix-closed domains, the direct polytope $\mathcal T_{\mathcal D}(\widehat\pi)$ instead uses undivided observed-choice equations. It is the appropriate formulation near zero reach or when the proper-prefix support needed for $\mathcal D_f$ is unavailable.

We extend Assumption \ref{assumption:Choice Data} by allowing the observed outcome set to be $\mathcal Y(S)=S\cup\{o^*\}$. Let
\begin{align*}
    d_{\mathcal D}=\sum_{S\in\mathcal D}|\mathcal Y(S)|,
    \qquad
    \epsilon_S(\alpha)
    =\sqrt{\frac{\log(2d_{\mathcal D}/\alpha)}{2N_S}},
\end{align*}
and form simultaneous cell bands
\begin{align*}
    \underline p(a|S)&=\max\{0,\widehat\pi(a|S)-\epsilon_S(\alpha)\},\qquad
    \overline p(a|S)=\min\{1,\widehat\pi(a|S)+\epsilon_S(\alpha)\}.
\end{align*}
Hoeffding's inequality and a union bound imply that all true menu-outcome probabilities belong to these bands with probability at least $1-\alpha$, conditionally on the menu assignments. Let $\mathcal C^{\mathrm H}_{1-\alpha}$ be the set of choice rules satisfying the bands and the menu-specific adding-up restrictions. To distinguish the data-generating object from the choice-rule arguments over which the projection is computed, write $p_0=\pi$ for the true choice rule under the maintained model and use $p$ for a generic candidate choice rule. More generally, let $\mathcal C_{1-\alpha}$ be any confidence region for $p_0$ satisfying $\mathbb P[p_0\in\mathcal C_{1-\alpha}]\geq1-\alpha$; the Hoeffding region is a finite-sample example.

The population identified sets below are defined in Propositions SA-4--SA-6 of the Supplemental Appendix. For a suffix-closed $\mathcal D$, define
\begin{align*}
    \widehat{\mathcal T}^{\mathrm{ind}}_{1-\alpha}
    &=\bigcup_{p\in\mathcal C_{1-\alpha}}
      \mathcal T_{\mathcal D}(p),\qquad
    \widehat{\mathcal T}^{\mathrm{dep}}_{1-\alpha}
    =\bigcup_{p\in\mathcal C_{1-\alpha}}
      \mathcal T^{\mathrm{dep}}_{\mathcal D}(p).
\end{align*}
On any observed-menu domain, also define
\begin{align*}
    \widehat{\mathcal T}^{\mathrm{noPI}}_{1-\alpha}
    =\bigcup_{p\in\mathcal C_{1-\alpha}}
      \mathcal T^{\mathrm{noPI}}_{\mathcal D}(p).
\end{align*}
For the first two sets, the identified set is taken to be empty when recovered attention is invalid.

\begin{thm}[Projection Inference for H-LAO]\label{theorem: projection inference hlao}
Suppose $\mathbb P[p_0\in\mathcal C_{1-\alpha}]\geq1-\alpha$. If $p_0$ has an independent H-LAO representation with preference distribution $\tau_0$, then
\begin{align*}
    \mathbb P\left[\tau_0\in
    \widehat{\mathcal T}^{\mathrm{ind}}_{1-\alpha}\right]
    \geq1-\alpha.
\end{align*}
The analogous conclusion holds for $\widehat{\mathcal T}^{\mathrm{dep}}_{1-\alpha}$ when preferences and stopping may be dependent, and for $\widehat{\mathcal T}^{\mathrm{noPI}}_{1-\alpha}$ on any observed-menu domain when Sequential Path Independence is also removed. Projecting the corresponding confidence set gives simultaneous coverage for any collection of preference-event probabilities. All conclusions hold in finite samples when $\mathcal C_{1-\alpha}=\mathcal C^{\mathrm H}_{1-\alpha}$.
\end{thm}

The reason is direct: on the primitive-region coverage event, $p=p_0$ is included in the union and sharp identification places $\tau_0$ in the corresponding population polytope. Applying the same union construction to the reach maps recovered from each candidate $p$ gives simultaneous confidence sets for attention frequencies under Sequential Path Independence. Without Sequential Path Independence, one instead projects the feasible latent prefix masses. Exact independent projection preserves preference--stopping independence but leads to polynomial feasibility constraints when $p$ varies.

Pairwise inference has a particularly transparent weak-reach implementation. If $a$ precedes $b$, H-LAO implies the undivided moment
\begin{align*}
    \pi(b|\{a,b\})
    =r_b(a,b)\tau(\{\succ:b\succ a\}),
\end{align*}
where $r_b(a,b)=[1-\pi(o^*|\{a,b\})][1-\pi(o^*|\{b\})]$. Let $[\underline r_b(a,b),\overline r_b(a,b)]$ be the product bounds induced by the outside-option bands. A simultaneous confidence set for the pairwise share is
\begin{align*}
    \widehat C^{\mathrm H}_{ba}
    =\left\{\theta\in[0,1]:
    [\underline r_b(a,b)\theta,\overline r_b(a,b)\theta]
    \cap[\underline p(b|\{a,b\}),\overline p(b|\{a,b\})]\neq\emptyset\right\}.
\end{align*}
When $\overline r_b(a,b)>0$, its endpoints are
\begin{align*}
    \max\left\{0,\frac{\underline p(b|\{a,b\})}{\overline r_b(a,b)}\right\}
    \qquad\text{and}\qquad
    \min\left\{1,\frac{\overline p(b|\{a,b\})}{\underline r_b(a,b)}\right\},
\end{align*}
where the second ratio is $+\infty$ if $\underline r_b(a,b)=0$. If the upper reach bound is zero and the inside-choice band contains zero, the confidence set is $[0,1]$. Thus, the procedure remains valid at zero reach instead of dividing by a weak denominator.

\begin{coro}[Finite-Sample Pairwise Inference under Weak Reach]\label{coro: weak reach pairwise inference main}
Suppose the independent H-LAO model holds. Then, simultaneously for every pair supported by the observed singleton and binary menus,
\begin{align*}
    \mathbb P\left[
    \tau_0(\{\succ:b\succ a\})\in\widehat C^{\mathrm H}_{ba}
    \text{ for every such pair }(a,b)
    \right]
    \geq1-\alpha.
\end{align*}
Empty intervals provide a finite-sample specification check.
\end{coro}

The preceding interval protects uniformly against a weak or zero denominator but can be conservative. A sharper alternative directly studentizes the same undivided moment. Write
\begin{align*}
    \widehat g_{ba}(\theta)
    =\widehat\pi(b|\{a,b\})
    -\theta\big[1-\widehat\pi(o^*|\{a,b\})\big]
    \big[1-\widehat\pi(o^*|\{b\})\big]
\end{align*}
and let $\widehat V_{ba}(\theta)$ be its plug-in delta-method variance, computed from the multinomial covariance matrix for $\{a,b\}$ and the binomial variance for $\{b\}$. For a fixed collection $\mathcal B$ of ordered pairs, define
\begin{align*}
    \widehat C^{\mathrm S}_{ba}
    =\left\{\theta\in[0,1]:
    \widehat g_{ba}(\theta)^2
    \leq \left[\Phi^{-1}\left(1-\frac{\alpha}{2|\mathcal B|}\right)\right]^2
    \widehat V_{ba}(\theta)
    \right\}.
\end{align*}
If $\widehat V_{ba}(\theta)=0$ for every $\theta\in[0,1]$, set $\widehat C^{\mathrm S}_{ba}=[0,1]$ by convention.

\begin{coro}[Studentized Pairwise Inference]\label{coro: studentized pairwise inference main}
Suppose the independent H-LAO model holds and $\mathcal B$ is fixed. For every $(b,a)\in\mathcal B$, suppose the binary- and singleton-menu sample sizes diverge, the variance at the true share is positive and consistently estimated, and the Lyapunov and product-remainder conditions in the Supplemental Appendix hold. Then
\begin{align*}
    \liminf_{\min_{(b,a)\in\mathcal B}(N_{ab}\wedge N_b)\to\infty}
    \mathbb P\left[
    \tau_0(\{\succ:b\succ a\})\in\widehat C^{\mathrm S}_{ba}
    \text{ for every }(b,a)\in\mathcal B
    \right]
    \geq1-\alpha.
\end{align*}
\end{coro}

Because the null is imposed before studentization, this Anderson--Rubin/Fieller construction never divides the estimator by reach and remains informative under weak positive reach whenever the moment variance is estimable. The finite-sample projection set $\widehat C^{\mathrm H}_{ba}$ remains the appropriate safeguard at exact or nearly degenerate zero-reach boundaries. The Supplemental Appendix gives the closed-form variance and proof. The formal result and simulations use Bonferroni calibration; the appendix separately describes a possible covariance-aware extension.

For general preference events under dependence, simultaneous product bounds on reach can be combined with the linear constraints defining $\mathcal T^{\mathrm{dep}}_{\mathcal D}(p)$. Minimizing and maximizing the event probability over this enlarged feasible set gives finite-sample-valid outer confidence bounds by linear programming. Under the model without Sequential Path Independence, prefix masses are latent variables rather than products of outside-choice probabilities. Combining their adding-up and attention-overload restrictions directly with the primitive cell bands and the constraints defining $\mathcal T^{\mathrm{noPI}}_{\mathcal D}(p)$ gives a fully linear, finite-sample projection procedure on any observed-menu domain. The Supplemental Appendix gives the exact projection results, the pairwise proofs, and both LP formulations. Correlated-Gaussian or multiplier-bootstrap regions for the primitive probabilities can replace the Hoeffding bands to improve power while preserving the same projection logic.

\section{Empirical Application}\label{section: empirical application}

We apply our theory and methods to the laboratory travel-mode experiment of \citet*{Wang-Zhu_2025_WP}. The application is useful for three reasons. First, menus are manipulated over a common four-alternative universe and a dominated ``Default'' is always available. Second, the experiment records information acquisition before choice, providing an observable measure of search and engagement with individual modes. Third, subjects report their preference rankings after completing the choice tasks, allowing an external comparison with the preference shares recovered from choice data alone.

\subsection{Data and Implementation}\label{subsection: empirical data and implementation}

The experiment contains 39 subjects, each completing 48 tasks, for 1,872 subject--task observations. In every task the inside menu is a subset of $\{A,B,C,D\}$, while a strictly dominated Default remains available. All 16 inside-menu subsets occur in the data. Twenty-one subjects face the complete 16-menu design. The other 18 face the 11 menus containing at least two inside alternatives. The workbook records menu availability, price and travel-time attributes, whether each attribute was displayed, the order of information acquisition, final choice, and an ex post ranking. Each of the 39 subjects reports one ex post ranking; 37 place all four inside alternatives above Default, while two place Default above an inside alternative.

Table \ref{table: empirical descriptive statistics} summarizes menu variation and information acquisition. Search intensity rises with menu size: the mean number of displayed attributes increases from $0.36$ on singleton menus to $1.44$ on the four-alternative menu, while the fraction of tasks with no displayed attribute falls from $0.75$ to $0.54$. Subjects can choose without displaying an attribute because they are told the ordinal attribute pattern and may rely on prior knowledge. A display therefore provides direct evidence of costly information acquisition about an alternative. Because a few Default choices follow an inside-option display, we interpret displays as recording pre-choice search rather than the final H-LAO consideration set. Conversely, failure to display is not evidence of inattention because subjects can rely on prior knowledge. Default choice outside the empty menu is uncommon, ranging from $0.016$ on singleton menus to $0.043$ on the grand menu.

\begin{table}[t]
\centering
\caption{Menu Size, Information Acquisition, and Default Choice}
\label{table: empirical descriptive statistics}
\begin{tabular}{ccccc}
\toprule
Inside-menu size & Observations & Mean displays & No-display share & Default-choice share \\
\midrule
$0$ & 63 & 0.000 & 1.000 & 1.000 \\
$1$ & 252 & 0.357 & 0.754 & 0.016 \\
$2$ & 648 & 0.832 & 0.628 & 0.031 \\
$3$ & 630 & 1.106 & 0.579 & 0.037 \\
$4$ & 279 & 1.441 & 0.538 & 0.043 \\
\bottomrule
\end{tabular}

\begin{flushleft}
\footnotesize Notes: The inside menu contains travel modes A--D. Default is always available and is the only alternative when the inside menu is empty. A display is one opened price or travel-time attribute. Statistics are computed from the 1,872 subject--task observations supplied by \citet*{Wang-Zhu_2025_WP}.
\end{flushleft}
\end{table}

Prices and travel times vary across tasks. Alternative A is generally fastest and most expensive, B is slowest and least expensive, and C and D are intermediate. Across observations in which each mode is available, mean prices for A--D are $80.6$, $24.2$, $40.5$, and $53.6$, while mean travel times are $115.1$, $392.8$, $310.6$, and $156.9$. The workbook contains 93 exact four-mode attribute profiles, and no exact profile occurs under more than one menu. Thus exact profile matching cannot remove attribute variation. Our analysis maintains that the distribution of price and travel-time profiles is independent of menu composition. Under this assumption, integrating over the profiles yields the menu-invariant marginal distribution of rankings over stable travel-mode categories required by H-LAO and incorporates attribute variation into the preference heterogeneity allowed by the model.

We use two analysis samples. The primary full-domain sample contains the 21 subjects who face every menu; it contributes 945 observations on the 15 nonempty inside menus. Following the original RAM exercise of \citet*{Wang-Zhu_2025_WP}, the dominance-consistent sample contains the 37 subjects whose ex post reports rank all four inside alternatives above Default. We retain their sample definition for the RAM replication, the matched RAM--AOM comparison, and the broader H-LAO analyses. For the pooled analyses, we maintain that assignment to the complete and reduced-menu designs is as good as random and that design version does not affect behavior conditional on the displayed menu; menu-specific choice and search frequencies therefore identify probabilities for a common population. The criteria serve different purposes: the samples overlap in 20 subjects because one complete-design subject is excluded from the dominance-consistent sample.

We code A--D as the inside alternatives, Default as the outside option, and A--B--C--D as the observed list order. Default is always available, objectively dominated, and ranked below every inside mode by 37 of the 39 subjects. These features make it a natural empirical counterpart to H-LAO's outside option. Accordingly, its choice represents no effective inside consideration at the final-choice stage, while pre-choice displays record earlier information acquisition. The empty inside menu is used to verify coding but omitted from H-LAO estimation because it selects Default mechanically. For the complete-design cohort, the 15 nonempty menus recover list reach and prefix-stopping probabilities under Sequential Path Independence. We then compute pairwise preference shares, the full-attention choice rule when positive reach permits it, Block--Marschak diagnostics, and preference-event bounds under independence and preference--stopping dependence. Within the dominance-consistent sample, the common-menu analysis restricts attention to the 11 menus shared across the two experimental designs and uses the direct observed-domain polytope and its no-Sequential-Path-Independence extension rather than extrapolating to unobserved singleton menus.

The homogeneous analysis applies the AOM preference and attention procedures to the same menu-choice frequencies and compares the compatible AOM rankings with the RAM rankings reported by \citet*{Wang-Zhu_2025_WP}. Following their original RAM analysis, this exercise adopts a common ranking over the travel-mode categories. H-LAO then allows the marginal distribution of these rankings to be heterogeneous. We report a benchmark that treats task-level observations as independent, but subjects are the independent sampling units because each contributes 48 choices. Our preferred analysis therefore permits arbitrary dependence among a subject's choices across tasks and menus. It conducts inference at the subject level and uses conservative finite-sample bounds for sparse outcomes. For the H-LAO pairwise shares, we construct simultaneous intervals by allowing the menu-specific choice probabilities to vary jointly within confidence bands and calculating the preference shares consistent with those bands and the model. The primitive region uses covariance-aware Gaussian bands for regular row-i.i.d. cells and cluster-multiplier bands for regular clustered cells, with exact-binomial and cluster-Hoeffding fallbacks, respectively, for nonregular cells. When both components are present, they split the common error budget. All simulated critical values in the application use 4,999 draws. Section SA-4.10 of the Supplemental Appendix gives the construction. The complete cohort has $G=21$ subjects and the dominance-consistent sample has $G=37$; at these cluster counts, the simulations in Section SA-6 show substantially better coverage for the projection intervals than for the shorter studentized alternative.

Observed information acquisition provides valuable additional evidence. For mode $a$ and menu $S$, define the observable search-and-choice proxy as the fraction of tasks in which the subject either displays at least one attribute of $a$ or chooses $a$. The proxy combines direct information search with final choice and therefore captures engagement that either measure alone may miss. We use its variation across menus as evidence complementary to the model-based attention estimates. Because displays record pre-choice search, the proxy is a measure of the decision process rather than a literal observation of the final structural consideration set.

\subsection{Results}\label{subsection: empirical results and qualifications}

For both RAM and AOM, we report the least-favorable (LF) Gaussian-maximum procedure and its generalized moment selection (GMS) refinement described in Section \ref{subsection:Econometric Methods single preference}.

We first reproduce the laboratory RAM exercise of \citet*{Wang-Zhu_2025_WP}. Following their design, we remove the two subjects with irrational reports, set attentive-at-binaries to $2/3$, and test the nine distinct rational reported rankings together with the two irrational rankings. Table \ref{table: empirical ram} reproduces their result under row-i.i.d. sampling: all nine rational rankings are retained and both irrational rankings are rejected. The conclusion is unchanged after clustering by subject at the $5\%$ level (the smallest GMS $p$-value among the rational rankings falls from $0.424$ to $0.066$).

\begin{table}[t]
\centering
\caption{Replication of the RAM Ranking Exercise}
\label{table: empirical ram}
\resizebox{\textwidth}{!}{%
\begin{tabular}{llcccc}
\toprule
& & \multicolumn{2}{c}{LF} & \multicolumn{2}{c}{GMS} \\
\cmidrule(lr){3-4}\cmidrule(lr){5-6}
Candidate rankings & Sampling & Nonrejected & Min. $p$-value & Nonrejected & Min. $p$-value \\
\midrule
Nine rational & Row-i.i.d. & 9 & 0.792 & 9 & 0.424 \\
Nine rational & Clustered & 9 & 0.186 & 9 & 0.066 \\
Two irrational & Row-i.i.d. & 0 & 0.000 & 0 & 0.000 \\
Two irrational & Clustered & 0 & 0.000 & 0 & 0.000 \\
\bottomrule
\end{tabular}
}
\begin{flushleft}
\footnotesize Notes: The sample contains the 37 subjects with rational ex post reports. The nine rational candidate rankings are the distinct reported rankings of A--D with Default placed last. The two irrational candidate rankings place Default first or above one inside mode. Attentive-at-binaries is set to $2/3$, as in \citet*{Wang-Zhu_2025_WP}. Nonrejection counts and minimum $p$-values are computed within each candidate-ranking group at level $0.05$.
\end{flushleft}
\end{table}

Table \ref{table: empirical aom} reports the AOM ranking tests. We treat Default as a fifth, always-available alternative and test all 24 rankings that place this dominated option last. In the complete cohort every ranking survives both GMS and least-favorable calibration under both sampling assumptions. On the 37-subject dominance-consistent sample used in Table \ref{table: empirical ram}, row-i.i.d. inference again retains all 24 rankings, while subject-clustered GMS eliminates eight and the least-favorable procedure retains all 24. The nine rational reported rankings in Table \ref{table: empirical ram} are included among the 24 AOM candidates. In the same-subject comparison, clustered RAM retains all nine reported rankings, while clustered AOM GMS retains eight of those nine and narrows the full set from 24 to 16 candidates. Thus, the data are consistent with AOM, and the subject-clustered GMS procedure is informative about homogeneous preferences. The broader complete-cohort set reflects its smaller sample, while the least-favorable procedure provides the deliberately conservative benchmark.

\begin{table}[t]
\centering
\caption{AOM Ranking Inference}
\label{table: empirical aom}
\resizebox{\textwidth}{!}{%
\begin{tabular}{llcccc}
\toprule
& & \multicolumn{2}{c}{LF} & \multicolumn{2}{c}{GMS} \\
\cmidrule(lr){3-4}\cmidrule(lr){5-6}
Sample & Sampling & Nonrejected & Min. $p$-value & Nonrejected & Min. $p$-value \\
\midrule
Complete & Row-i.i.d. & 24 & 0.987 & 24 & 0.562 \\
Complete & Clustered & 24 & 0.829 & 24 & 0.260 \\
Dominance-consistent & Row-i.i.d. & 24 & 0.794 & 24 & 0.225 \\
Dominance-consistent & Clustered & 24 & 0.159 & 16 & 0.024 \\
\bottomrule
\end{tabular}

}
\begin{flushleft}
\footnotesize Notes: ``Complete'' uses the 21 subjects who face every menu; ``Dominance-consistent'' uses the same 37 subjects as Table \ref{table: empirical ram}. Each sample tests all 24 rankings of A--D that place the dominated Default last. ``Nonrejected'' is the number of candidate rankings not rejected at the 5\% level. GMS uses generalized moment selection; LF uses the least-favorable Gaussian maximum. Clustered inference treats subjects as independent and uses subject multiplier draws. The minimum $p$-value is taken across the 24 candidate rankings.
\end{flushleft}
\end{table}

Table \ref{table: empirical hlao specification} reports the direct H-LAO specification diagnostics using least-favorable calibration. The largest recovered-attention discrepancy is $0.078$ in the complete cohort and $0.057$ in the dominance-consistent sample; the corresponding largest Block--Marschak discrepancies are $0.175$ and $0.072$. None is statistically significant under either the row-i.i.d. benchmark or subject-clustered inference. The data are therefore jointly consistent with H-LAO's attention-overload and recovered random-utility restrictions, including under our preferred subject-clustered inference.

\begin{table}[t]
\centering
\caption{H-LAO Specification Diagnostics}
\label{table: empirical hlao specification}
\resizebox{\textwidth}{!}{%
\begin{tabular}{llccccc}
\toprule
& & & \multicolumn{2}{c}{Row-i.i.d.} & \multicolumn{2}{c}{Clustered} \\
\cmidrule(lr){4-5}\cmidrule(lr){6-7}
Sample & Restriction & Max. discrepancy & Lower & $p$-value & Lower & $p$-value \\
\midrule
Complete & Attention overload & 0.078 & 0.000 & 0.451 & 0.000 & 0.510 \\
Complete & Block--Marschak & 0.175 & -0.102 & 0.940 & -0.034 & 0.890 \\
Complete & Omnibus & 0.175 & 0.000 & 0.451 & 0.000 & 0.510 \\
Dominance-consistent & Attention overload & 0.057 & 0.000 & 0.798 & 0.000 & 0.745 \\
Dominance-consistent & Block--Marschak & 0.072 & -0.053 & 0.915 & -0.016 & 0.274 \\
Dominance-consistent & Omnibus & 0.072 & 0.000 & 0.798 & 0.000 & 0.274 \\
\bottomrule
\end{tabular}
}
\begin{flushleft}
\footnotesize Notes: ``Complete'' uses the 21 subjects who face every menu; ``Dominance-consistent'' uses the 37 subjects retained in the original RAM analysis and combines the two experimental designs. A discrepancy is written so that a positive population value violates the indicated restriction. ``Lower'' is the simultaneous 95\% one-sided lower confidence endpoint. The table uses least-favorable (LF) calibration: all nondegenerate inequalities enter the zero-centered maximum, without GMS recentering. The row-i.i.d. columns treat task-level observations as independent; the clustered columns use subject influence vectors and multiplier calibration. The $p$-values are calibrated over the joint family of diagnostic inequalities using the same Gaussian or subject-cluster multiplier maxima as the confidence endpoints. Full-attention probability restrictions are included in the omnibus test but omitted as a separate row because their largest estimated violation is zero.
\end{flushleft}
\end{table}

Table \ref{table: empirical pairwise validation} compares the H-LAO estimates with the ex post reports in both samples. In the matched 20-subject complete-design validation sample, the mean absolute difference across the six pairwise shares is $0.038$ and the largest difference is $0.074$. In the dominance-consistent sample, the corresponding differences are $0.039$ and $0.099$. Every reported share lies within its clustered H-LAO interval, and all simultaneous difference intervals contain zero under both sampling schemes. Thus, the point estimates are close, and equality is not rejected by the simultaneous clustered intervals. The comparison uses information not employed in estimating H-LAO: H-LAO uses only menu availability, outside-option choice, inside choice, and the maintained list structure. Inclusion in the validation samples requires a usable dominance-consistent report, but the reported rankings are not used to estimate H-LAO.

\begin{table}[t]
\centering
\caption{Revealed and Self-Reported Pairwise Preference Shares}
\label{table: empirical pairwise validation}
\resizebox{\textwidth}{!}{%
\begin{tabular}{clcccccc}
\toprule
& & \multicolumn{2}{c}{Point estimates} & \multicolumn{2}{c}{Row-i.i.d.} & \multicolumn{2}{c}{Clustered} \\
\cmidrule(lr){3-4}\cmidrule(lr){5-6}\cmidrule(lr){7-8}
& & H-LAO & Reported & Projection CI & Difference CI & Projection CI & Difference CI \\
\midrule
\multicolumn{8}{l}{\textit{Complete-design validation sample}} \\
& $B\succ A$ & 0.874 & 0.800 & [0.657, 1.000] & [-0.206, 0.354] & [0.587, 1.000] & [-0.099, 0.247] \\
& $D\succ B$ & 0.400 & 0.400 & [0.196, 0.795] & [-0.333, 0.333] & [0.106, 1.000] & [-0.220, 0.220] \\
& $D\succ A$ & 0.814 & 0.800 & [0.633, 1.000] & [-0.257, 0.284] & [0.644, 1.000] & [-0.193, 0.220] \\
& $C\succ B$ & 0.576 & 0.600 & [0.360, 1.000] & [-0.360, 0.313] & [0.287, 1.000] & [-0.227, 0.179] \\
& $C\succ A$ & 0.847 & 0.900 & [0.678, 1.000] & [-0.274, 0.169] & [0.629, 1.000] & [-0.183, 0.078] \\
& $D\succ C$ & 0.267 & 0.200 & [0.082, 0.594] & [-0.213, 0.346] & [0.013, 1.000] & [-0.084, 0.217] \\
\addlinespace
\multicolumn{8}{l}{\textit{Dominance-consistent sample}} \\
& $B\succ A$ & 0.760 & 0.784 & [0.598, 1.000] & [-0.238, 0.190] & [0.519, 1.000] & [-0.161, 0.113] \\
& $D\succ B$ & 0.419 & 0.432 & [0.251, 0.745] & [-0.266, 0.240] & [0.196, 1.000] & [-0.162, 0.136] \\
& $D\succ A$ & 0.803 & 0.811 & [0.643, 1.000] & [-0.216, 0.199] & [0.652, 1.000] & [-0.144, 0.128] \\
& $C\succ B$ & 0.536 & 0.541 & [0.355, 0.923] & [-0.262, 0.252] & [0.308, 1.000] & [-0.160, 0.150] \\
& $C\succ A$ & 0.779 & 0.865 & [0.625, 1.000] & [-0.277, 0.105] & [0.576, 1.000] & [-0.216, 0.044] \\
& $D\succ C$ & 0.369 & 0.270 & [0.221, 0.639] & [-0.127, 0.326] & [0.155, 1.000] & [-0.052, 0.250] \\
\bottomrule
\end{tabular}
}
\begin{flushleft}
\footnotesize Notes: For the complete-design validation panel, both H-LAO estimates and reported shares use the 20 complete-design subjects with usable dominance-consistent reports. The 21-subject complete cohort remains the sample for the main H-LAO analysis. For the dominance-consistent sample, both estimates use the 37 subjects retained in the original RAM analysis. Difference is H-LAO minus Reported. The H-LAO intervals allow the menu-specific choice probabilities to vary jointly within simultaneous confidence bands and report all pairwise shares consistent with those bands and the model. The row-i.i.d. benchmark uses correlated-Gaussian bands for cells with at least five successes, at least five failures, and positive estimated variance, and exact binomial bands otherwise. Clustered inference treats subjects as the independent units, accounts for dependence among observations from the same subject, and uses multiplier bands when at least five clusters contribute successes, at least five contribute failures, and the estimated variance is positive; other cells use cluster-Hoeffding bands. When both components occur, each receives half of the five-percent familywise error budget. The difference intervals use a delta-method approximation, appropriate when reach is bounded away from zero, with Gaussian multiplier calibration jointly over the six comparisons. The clustered intervals account for covariance between choices and reports from the same subjects; the row-i.i.d. benchmark treats these components as independent. Because these intervals concern equality rather than one-sided inequalities, the GMS--LF distinction does not apply.
\end{flushleft}
\end{table}

Table \ref{table: empirical common menu} uses the 11 menus observed for every subject in the dominance-consistent sample and drops Sequential Path Independence. The simultaneous outer confidence bounds remain informative despite this weaker specification: every reversal of the baseline ordering $A\succ B\succ C\succ D$ has a strictly positive clustered lower endpoint. The upper endpoints equal one, so the common-menu design establishes the presence of each preference reversal without tightly measuring its population share.

\begin{table}[t]
\centering
\caption{Preference-Share Confidence Bounds without Sequential Path Independence}
\label{table: empirical common menu}
\begin{tabular}{lcc}
\toprule
 & \multicolumn{2}{c}{95\% outer confidence bounds} \\
\cmidrule(lr){2-3}
Preference event & Row-i.i.d. & Subject-clustered \\
\midrule
$B \succ A$ & [0.595, 1.000] & [0.521, 1.000] \\
$C \succ A$ & [0.623, 1.000] & [0.578, 1.000] \\
$D \succ A$ & [0.640, 1.000] & [0.653, 1.000] \\
$C \succ B$ & [0.352, 1.000] & [0.310, 1.000] \\
$D \succ B$ & [0.248, 1.000] & [0.198, 1.000] \\
$D \succ C$ & [0.220, 1.000] & [0.157, 1.000] \\
\bottomrule
\end{tabular}

\begin{flushleft}
\footnotesize Notes: Both columns use the same dominance-consistent sample of 37 subjects observed on the 11 menus with at least two inside alternatives. The entries are simultaneous 95\% outer confidence bounds. They allow the menu-specific choice probabilities to vary jointly within their simultaneous confidence region and report the most conservative preference bounds consistent with the model without Sequential Path Independence. A strictly positive lower endpoint establishes the presence, but not the precise population prevalence, of the corresponding preference reversal. The row-i.i.d. benchmark treats task-level observations as independent. The subject-clustered bounds account for dependence among observations from the same subject and use conservative finite-sample bounds for sparse outcomes.
\end{flushleft}
\end{table}

The search records provide separate descriptive evidence for the dominance-consistent sample. Table \ref{table: empirical search} reports larger-menu minus smaller-menu differences for 76 nested-menu comparisons. Attribute display alone is mixed: 43 differences are positive, 33 are negative, and none is significant after simultaneous calibration. Strikingly, the broader search-and-choice proxy---displayed or chosen---declines in 70 of the 76 comparisons, with a mean difference of $-0.255$; 36 declines are significant under subject-clustered inference, whereas no increase is significant. This widespread pattern provides independent descriptive evidence consistent with attention dilution. Because the proxy measures search and choice engagement rather than final structural consideration, we do not use it as a formal model test.

\begin{table}[t]
\centering
\caption{Observed Information Search across Nested Menus}
\label{table: empirical search}
\resizebox{\textwidth}{!}{%
\begin{tabular}{lrrrcccc}
\toprule
 & \multicolumn{3}{c}{Differences} & \multicolumn{2}{c}{Row-i.i.d.} & \multicolumn{2}{c}{Clustered} \\
\cmidrule(lr){2-4} \cmidrule(lr){5-6} \cmidrule(lr){7-8}
Attention proxy & Mean & Negative & Positive & Sig. negative & Sig. positive & Sig. negative & Sig. positive \\
\midrule
Displayed & 0.009 & 33 & 43 & 0 & 0 & 0 & 0 \\
Displayed or chosen & -0.255 & 70 & 6 & 38 & 0 & 36 & 0 \\
\bottomrule
\end{tabular}

}
\begin{flushleft}
\footnotesize Notes: The calculations use the 37-subject dominance-consistent sample. Each difference is the observed attention-proxy rate in the larger menu minus its rate in the smaller menu; negative differences therefore agree with attention dilution. Counts of significant negative and positive differences use simultaneous 95\% two-sided multiplier intervals across all 76 nested-menu comparisons. ``Displayed'' records whether at least one attribute of the mode was opened; ``Displayed or chosen'' records whether the mode was displayed or selected as the final choice.
\end{flushleft}
\end{table}

Four features of the design guide interpretation. First, price and travel-time profiles vary across tasks and cannot be held fixed by exact matching. The maintained menu-independent profile distribution incorporates this variation into the stable marginal distribution of travel-mode-category rankings used by H-LAO; homogeneous AOM provides the corresponding common-ranking benchmark. Variation in the underlying composition of observed categories can nevertheless weaken category-level random-utility implications \citep{Liao-Saito-Sandroni_2026_WP}. Second, the complete menu domain is available for 21 subjects. The common-menu no-Sequential-Path-Independence analysis uses the broader dominance-consistent sample and therefore changes both menu support and subject composition; we interpret it as a sensitivity analysis rather than a direct precision comparison. Third, subjects rather than task-level rows are the independent sampling units. We therefore base our preferred inferential conclusions on subject-clustered procedures and report row-i.i.d. results as benchmarks. Fourth, attribute displays measure costly pre-choice information search. We report displays separately and use displayed-or-chosen as a complementary descriptive measure of mode-level engagement.

The experiment's rare combination of full menu manipulation, a dominated outside option, repeated choices, observed information acquisition, and separately elicited rankings makes it unusually informative for comparing RAM, AOM, and H-LAO and for examining the preference and attention objects recovered by our methods. We use the application as an empirical illustration for stable travel-mode categories; because price and travel-time profiles vary across tasks, it is not a design-based test conditional on fixed profiles.

\section{Conclusion}

This paper introduces attention overload, a nonparametric restriction requiring an alternative's attention frequency to weakly decrease as its menu expands. Mixtures of menu-independent search priorities and inspection capacities provide a behavioral foundation for the restriction. Despite leaving the consideration-set distribution unrestricted, attention overload yields a sharp behavioral characterization, revealed-preference implications, and informative bounds on latent attention frequencies. The constructive characterization also supports econometric procedures that remain valid when the number of menus and inequalities is large, while its mixed-integer formulation represents the compatible-preference set implicitly and answers feasibility and pairwise-revelation queries without enumerating all rankings.

We also develop H-LAO, a list-based extension that accommodates heterogeneous preferences and attention. Outside-option choices reveal how far consumers proceed down a list. When the relevant menus are observed, Sequential Path Independence allows us to recover attention frequencies; with a positive probability of reaching the end of the list, additional menu variation recovers the choices that would prevail under full attention. Our identification analysis also characterizes what remains learnable when some menus are missing or consumers do not reach the end of a list. Under the stated support conditions, singleton and binary data can be enough to identify pairwise preference shares. The Supplemental Appendix develops confidence procedures for complete and incomplete menu data, including cases in which reach is small or zero, and linear-programming methods that allow preferences and stopping behavior to be dependent or relax Sequential Path Independence.

Independent empirical studies establish the relevance of random limited attention and show that its restrictions are empirically testable. Attention overload turns the item-level dilution mechanism into an application-specific comparative static whose validity can be assessed directly from choice data. The Supplemental Appendix reports simulations evaluating the finite-sample behavior and computational demands of the proposed procedures. In the travel-mode application, subject-clustered diagnostics do not reject H-LAO, its pairwise preference shares closely match subjects' separately elicited rankings, and all simultaneous confidence intervals for the differences contain zero. The displayed-or-chosen search-and-choice proxy declines in 70 of 76 nested-menu comparisons, with 36 significant declines and no significant increases under subject-clustered simultaneous inference. Together, the homogeneous and heterogeneous models show how economically interpretable cross-menu attention restrictions can recover preference and attention information from standard choice data without committing to a parametric consideration-set model.

\singlespacing
\addcontentsline{toc}{section}{References}
\bibliographystyle{econometrica}
\bibliography{CCMM_2026_AttOverload--bib}

\end{document}


\raggedbottom

\onehalfspacing
\hypersetup{pageanchor=false}

\title{\vspace{0in}Supplemental Appendix to ``Attention Overload''}
\author{Matias D. Cattaneo\thanks{Department of Economics, Princeton University.}\and
        \hspace{-.05in}Paul H.Y. Cheung\thanks{Jindal School of Management, University of Texas at Dallas.}\and
        \hspace{-.05in}Xinwei Ma\thanks{Department of Economics, UC San Diego.}\and
        \hspace{-.05in}Yusufcan Masatlioglu\thanks{Department of Economics, University of Maryland.}}
\maketitle

\begin{abstract}
    This Supplemental Appendix follows the organization of the main paper. It provides additional literature and model comparisons, proofs and extensions for the homogeneous and heterogeneous attention-overload models, econometric proofs and alternative inference procedures, and simulation and computational evidence.
\end{abstract}

\clearpage
\setcounter{tocdepth}{2}
\tableofcontents
\thispagestyle{empty}

\clearpage
\setcounter{page}{1}
\hypersetup{pageanchor=true}
\pagestyle{plain}

\doublespacing

\section{Additional Literature and Model Comparisons}\label{Appendix A.1}

Evidence relevant to AOM plays three distinct roles. First, limited and stochastic attention has been documented across investment, school choice, job search, grocery consumption, computer purchases, and airport choice \citep{Huberman2001ContagiousSoar,Laroche1984BrandChoice,Rosen1998CollegePerspective,Sheridan1975ComparativeBehavior,Demuynck2018RevealedConsideration,Goeree_2008_ECMA,Basar2004AArea}. The consideration-set literature likewise treats attention as a scarce and variable input into choice \citep*{Hauser_Wernerfelt_1990_JCR,Shocker1991ConsiderationSuggestions}.

Second, some evidence speaks directly to menu expansion and random-attention restrictions. Consumers may inspect fewer items as choice sets expand \citep*{Reutskaja2009SatisfactionSatiate,Reutskaja_et_al_2011_AER}, and eye-tracking evidence shows that the probability of inspecting an individual item decreases as the choice set expands \citep*{Thomas-Molter-Krajbich_2021_eLife}. Larger choice sets can also increase outside-option choice \citep*{Iyengar-Lepper_2000_JPSP}. Choice-overload experiments document limited and costly search in large menus \citep*{Dean-Ravindran-Stoye_2025_WP,deLara-Dean_2024_WP}, while field evidence links larger platform assortments to lower purchase frequency among incumbent consumers through costly attention \citep*{Natan_2024_MS}. Experiments that observe consideration directly document widespread stochastic limited consideration and assess restrictions imposed by leading attention models \citep*{Ellis-FilizOzbay-Ozbay_2025_WP,Chadd_2023_JEBO,KhanKrajbichRaymondSundaresanVeiga2026}. Applying RAM to a travel-mode experiment, \citet*{Wang-Zhu_2025_WP} find that it accepts the ranking implied by subjects' self-reports while rejecting rankings that place an objectively dominated option first. A health-preference menu experiment similarly finds that the revealed rankings under RAM and RUM are largely consistent \citep*{Wang-Sicsic-Chauvin_2026_WP}. Together, these studies establish that random attention has empirical content and motivate testing attention overload's item-level comparative static in each application.

Third, related evidence helps separate attention overload from other mechanisms. Laboratory evidence on rational inattention shows that subjects adjust information acquisition to incentives while rejecting commonly used linear-Shannon cost specifications \citep*{Dean-Neligh_2023_JPE}. \citet{Abaluck-Gruber_2023_REStud} find that reducing the number of offered health-insurance plans can improve choice quality but reject individual choice overload as the explanation. In an experiment with a default option, \citet{Chadd-FilizOzbay-Ozbay_2025_ExpEcon} find endogenous selection into costly search and that subjects who reject the default tend to discover higher-ranked alternatives. Recent theory obtains choice overload from still other mechanisms, including menu-dependent discrimination in threshold--Luce choice and minimax-regret search \citep*{Caliari-Petri_2025_WP,Auster-Che_2025_WP}. AOM is distinct: it imposes competition directly on each alternative's marginal attention frequency. These comparisons clarify both the empirical motivation for AOM and why its restriction should be tested in the application at hand.

The evidence above motivates AOM and H-LAO. We compare them with related models of stochastic choice. RUM assumes full attention but allows for preference heterogeneity. Early random limited-attention models typically maintained homogeneous preferences while allowing for attention heterogeneity; more recent work has begun to incorporate both sources of heterogeneity. Several related approaches have appeared since the first circulation of this paper in 2021. We organize the comparison by attention primitives, data requirements, and sources of identifying variation.

After this empirical overview, Section \ref{Related Lit: single preference} focuses on random-attention models with homogeneous preferences and compares them with AOM. Section \ref{Related Lit: multiple preference} considers models with heterogeneous preferences and compares them with H-LAO. Section \ref{Related Lit: other choice models} discusses other stochastic-choice models related to AOM or H-LAO.

\subsection{Random Attention Models with Homogeneous Preferences}\label{Related Lit: single preference}

The main paper compares the attention restrictions of AOM with other random-attention models using the examples of Ann and Ben. Here we compare their observable implications and show that the models can rationalize different choice rules.

We first compare two nonparametric models: the \textit{Random Attention Model} \citep[RAM,][]{Cattaneo-Ma-Masatlioglu-Suleymanov_2020_JPE} and AOM. Given the flexibility of their attention rules, the central challenge is revealed preference. Table \ref{example-SA1: Not RAM, Yes AOM} has an AOM representation but not a RAM representation, while Table \ref{example-SA1: Yes RAM, Not AOM} has a RAM representation but not an AOM representation.

\begin{table}[ht!]
    \begin{subtable}[h]{0.45\textwidth}
        \centering
        \begin{tabular}{r|ccc} \hline\hline
        $\ChoiProb(\cdot|S)$ & $a$ & $b$ & $c$ \\
        \hline
        $\{a,b,c\}$ & 0.4 & 0.3 & 0.3 \\
        $\{a,b\}$ & 0.8 & 0.2 & -- \\
        $\{a,c\}$ & 0.8 & -- & 0.2 \\
        $\{b,c\}$ & -- & 0.5 & 0.5\\  \hline\hline
        \end{tabular}
        \caption{}\label{example-SA1: Not RAM, Yes AOM}
    \end{subtable}
    \begin{subtable}[h]{0.45\textwidth}
        \centering
        \begin{tabular}{r|cccc} \hline\hline
        $\ChoiProb(\cdot|S)$ & $a$ & $b$ & $c$ & $d$ \\
        \hline
        $\{a,b,c,d\}$  & 1/2 & 1/2 & 0 & 0 \\
        $\{a,b,c\}$ & 0 & 2/3 & 1/3 & -- \\
        $\{a,b\}$ & 1/2 & 1/2 & -- & -- \\
        \hline\hline
        \end{tabular}
        \caption{}\label{example-SA1: Yes RAM, Not AOM}
    \end{subtable}
   \caption{Differences in explanatory power: RAM versus AOM}\label{example-SA1: AOM vs RAM}
\end{table}

Recall that RAM reveals $a\succ b$ if there exist $a,b\in S$ such that $\ChoiProb(a|S\setminus\{b\})<\ChoiProb(a|S)$ \citep{Cattaneo-Ma-Masatlioglu-Suleymanov_2020_JPE}. The choice rule in Table \ref{example-SA1: Not RAM, Yes AOM} therefore has no RAM representation because RAM reveals both $b\succ c$ and $c\succ b$. AOM instead reveals $a\succ c$ and $a\succ b$ and admits either $a\succ b\succ c$ or $a\succ c\succ b$.

There is no $3$-alternative example that RAM can explain but AOM cannot: with three alternatives, AOM explains a superset of the choice data explained by RAM. Table \ref{example-SA1: Yes RAM, Not AOM} therefore uses four alternatives. In that example, AOM would require both $a\succ b$ and $b\succ a$, so it cannot rationalize the data. RAM instead reveals $a\succ d$ and $b\succ c$, which is acyclic and rationalizes the data.

More generally, while AOM and RAM provide testable restrictions (inequalities) in terms of the choice rule, they can take very different forms. To compare, first recall from our main characterization result that a preference ordering has an AOM representation if and only if $\succ$-Regularity holds:
\begin{align*}
    \ChoiProb(a|S)\leq \ChoiProb(U_{\succeq}(a)|T)
    \qquad\text{for }T\subset S.
\end{align*}
In contrast, RAM imposes a regularity condition when removing ``better'' alternatives:
\begin{align*}
    \ChoiProb(a|S)\leq \ChoiProb(a|S\setminus\{b\})
    \qquad\text{whenever }b\succ a.
\end{align*}
Hence, the two underlying regularity violations used for identification are conceptually different.

In the realm of nonparametric attention models, we can also compare with the convex hull of deterministic consideration-set mappings. For example, the random competition filter (RCF) is an important special case of AOM. Let $\Gamma_j(\cdot)$ be deterministic consideration-set mappings that satisfy \textit{Competition Filter} \citep{Lleras_et_al_2017_JET}. An RCF model is specified by
\begin{align*}
    \mu_{\texttt{RCF}}(T|S)
    &=\sum_{j=1}^J \alpha_j \Indicator(\Gamma_j(S)=T),
    &
    \phi_{\texttt{RCF}}(a|S)
    &=\sum_{j:a\in \Gamma_j(S)} \alpha_j,
\end{align*}
where $\alpha_j\geq 0$ for $j=1,2,\dots,J$ and $\sum_{j=1}^J\alpha_j=1$. Random competition filters satisfy attention overload because, for every $a\in T\subseteq S$, $a\in \Gamma_j(S)$ implies $a\in \Gamma_j(T)$.

RCF also nests stochastic versions of bounded rationalization and imprecise narrowing down. In the former, the decision maker randomizes over rationales; in the latter, she randomizes over procedures for forming criteria. Because each realized rationale or procedure induces a deterministic competition filter, a menu-independent mixture over them is an RCF.

Several parametric models can also be compared with AOM. In the \textit{Independent Attention} model of \citet{Manzini-Mariotti_2014_ECMA}, each alternative $a$ has a menu-invariant consideration probability $\gamma(a)$, so $\phi_{\texttt{MM}}(a|S)=\gamma(a)$. Equality across menus satisfies attention overload, making independent attention a special case of AOM.

The \textit{Elimination by Aspects} attention model of \citet{Aguiar_2017_EL} is also nested. Each category $D$ has a fixed probability $m(D)$; when that category is available, the decision maker selects its best alternative and otherwise chooses the default. If $\mathcal C$ denotes the collection of categories, then $\phi_{\texttt{Aguiar}}(a|S)=\sum_{D\in\mathcal C:\,a\in D}m(D)$, which is menu invariant and therefore satisfies attention overload.

The \textit{Bounded Processing Capacity Rule} (BPCR) of \citet{MARCHANT2023102771} is another special case. The decision maker has capacity $k$: she considers the entire menu when $|S|\leq k$ and draws uniformly from the subsets of size $k$ when $|S|>k$. The attention frequency is
\begin{align*}
    \phi_{\texttt{MS}}(a|S)
    =\begin{cases}
        1, & |S|\leq k,\\
        \dfrac{\binom{|S|-1}{k-1}}{\binom{|S|}{k}}, & |S|>k.
    \end{cases}
\end{align*}
The second expression equals $k/|S|$, so attention frequency weakly decreases as the menu expands and BPCR satisfies attention overload.

The \textit{Logit Attention} model \citep[e.g.,][]{Brady-Rehbeck_2016_ECMA} need not satisfy attention overload. To construct a separating example, let the Luce weight $\ell(\cdot)$ on menus satisfy
\begin{align*}
    &\ell(abcd)=10,\quad
    \ell(bcd)=10,\quad
    \ell(bc)=4,\quad
    \ell(c)=2,\quad
    \ell(ab)=1,\quad
    \ell(b)=1,\\
    &\ell(S)=0
    \qquad\text{for all other nonempty }S\subseteq\{a,b,c,d\}.
\end{align*}
With $a\succ b\succ c\succ d$, these weights yield the following choice probabilities, listed in menu order:
\begin{align*}
    \pi(\cdot|abcd)&=(11/28,15/28,1/14,0),\\
    \pi(\cdot|abc)&=(1/8,5/8,1/4),\\
    \pi(\cdot|ab)&=(1/2,1/2).
\end{align*}
Under AOM, $\pi(b|abc)=5/8>1/2=\pi(b|ab)$ requires $a\succ b$, while $\pi(a|abcd)=11/28>1/8+1/4=\pi(a|abc)+\pi(c|abc)$ requires $b\succ a$. Hence AOM cannot rationalize these choices. Giving the other nonempty subsets sufficiently small positive weights makes the logit rule well defined on every menu and preserves both strict inequalities.
Conversely, the choice rule in Table \ref{example-SA1: Not RAM, Yes AOM} is inconsistent with RAM and hence with its logit-attention subclass \citep{Cattaneo-Ma-Masatlioglu-Suleymanov_2020_JPE}, but it is consistent with AOM.

\citet{Demirkan-Kimya_2020_JME} allow the independent-consideration parameter $\gamma(a,S)$ to depend on the menu. Because $\gamma(a,S)$ may exceed $\gamma(a,T)$ for $T\subset S$, their model can violate attention overload.

These comparisons locate AOM's contribution precisely. AOM is neither a relabeling of RAM nor another parametric consideration-set model: it places a nonparametric cross-menu restriction on item-level marginal attention. That restriction produces different revealed-preference implications from RAM and accommodates several familiar attention mechanisms while remaining empirically refutable.

\subsection{Random Attention Models with Heterogeneous Preferences} \label{Related Lit: multiple preference}

A growing literature combines random utility with random limited attention. Within the nonparametric identification paradigm, \citet{KashaevAguiar2022JET} generalize RAM by incorporating random utility in the set-monotone and stable RAUM. There are two key differences from H-LAO. First, H-LAO targets list-based attention overload, whereas set-monotone and stable RAUM builds on RAM. Second, benchmark H-LAO uses an observed list to discipline sequential attention while allowing an unrestricted marginal preference distribution; under positive reach and full menu data this additional structure point identifies attention and the full-attention stochastic choice rule, whereas their more general model yields partial identification. Two robustness envelopes below first relax benchmark independence between preferences and stopping and then remove Sequential Path Independence itself.

Two subsequent models are especially useful comparisons for the list structure. \citet{Masatlioglu-Vu_2024_JET} order heterogeneous consideration sets by an attentiveness index and identify their composition and frequency. Their primitive is an expansion of consideration sets across attention types, while H-LAO imposes cross-menu overload on list reach. \citet{Wei_2024_WP} instead identifies preferences and attention from stopping-time variation without requiring menu variation and permits dependence between the two. H-LAO uses menu and observed-list variation; its benchmark independence restriction is relaxed in the dependence-robust identified set.

Several papers instead accommodate heterogeneous preferences and random limited attention through parametric restrictions. For example, \citet{Aguiar2023QE} develops attention-index models, and \citet{Dardanoni-Manzini-Mariotti-Tyson_2020_ECMA} parameterizes cognitive capacity through consideration-set size. In contrast, H-LAO relies on a nonparametric attention-overload restriction.

A different strand uses enriched choice data to accommodate heterogeneous preferences and random limited attention. For example, \citet{Gibbard2021JMathE} studies frame-dependent choice, and \citet{Dardanoni2023JPE} uses mixture choice data. In contrast, AOM uses menu-indexed choice probabilities to identify compatible preferences and bound attention frequencies, while H-LAO combines aggregate choice probabilities with an observed presentation order to identify attention and the full-attention random-utility rule.

The discrete-choice literature has long emphasized preference heterogeneity, and some recent work also allows attention to vary. \citet*{Barseghyan-Coughlin-Molinari-Teitelbaum_2021_ECMA} study partial identification of preferences and consideration-set formation by restricting consideration-set size or the frequency of singleton consideration. \citet{Abaluck_Adams_2021} exploit asymmetries in cross-partial derivatives and identify consideration and preferences using rich exogenous covariate variation, while \citet{Abaluck-Compiani-Zhang_2026_JPE} use derivative restrictions to identify preferences in the presence of costly search. \citet*{Barseghyan-Molinari-Thirkettle_2021_AER} and \citet{Barseghyan-Molinari_2023_JBES} provide identification results for risk preferences with more restricted covariate variation. In contrast, AOM and H-LAO operate in an abstract choice domain and obtain identification from cross-menu attention restrictions and list structure.

Repeated-choice data provide another source of identification. \citet{Aguiar-Kashaev_2025_JBES} recover the joint distribution of preferences and latent choice sets when the panel is sufficiently long or the support of choice sets is sparse. \citet{Chib-Shimizu_2026_JASA} use time-invariant subject-specific consideration sets and a mixture representation to obtain scalable Bayesian estimation with many alternatives. These approaches recover persistent latent choice sets from within-subject choice histories; our methods instead use cross-menu choice restrictions and do not require a panel for the theoretical identification results.

The interpretation of latent consideration also depends on the maintained utility model. \citet{Allen_2026_JOE} shows that, with bounded utility shifters, an additive random-utility model with exogenous consideration can be observationally equivalent to full and extended additive random utility. That result concerns identification from utility-index variation without the cross-menu restrictions imposed here. It underscores why our attention conclusions derive from attention overload, list structure, and menu variation rather than from an unrestricted random-utility representation alone.

H-LAO's contribution is the recovery of item-specific reach and the latent full-attention stochastic choice rule from aggregate choices under explicit list, domain, and support conditions, together with sharp sensitivity envelopes when preference--stopping independence or Sequential Path Independence is relaxed. Once the full-attention rule has been recovered, the full toolkit of standard random-utility methods becomes available.

\subsection{Other Related Stochastic Choice Models} \label{Related Lit: other choice models}

AOM does not impose classical regularity on every alternative, but it contains several regular stochastic-choice models. Standard RUM is a leading example because classical regularity implies $\succ$-Regularity for a compatible ordering.

Once H-LAO recovers the full-attention rule, the remaining identification question is the standard one for finite random utility. \citet{Chambers-Turansick_2025_GEB} derive tight limits on the support restrictions capable of point identifying a random-utility model, while \citet{Caradonna-Turansick_2026_WP} characterize identified sets for a broader class of stochastic-choice models. These results reinforce our distinction between point recovery of the full-attention choice rule and generally set-valued recovery of the complete ranking distribution.

For example, \citet{Gul_Natenzon_Pesendorfer_2014_ECMA} study attribute rules in which the decision maker draws an attribute and selects an alternative possessing it. Every attribute rule is a RUM and therefore has an AOM representation. \citet{Fudenberg-Iijima-Strzalecki_2015_ECMA} introduce additive perturbed utility, under which deliberate randomization can make deterministic choice costly. Their choice rules satisfy regularity and hence also admit AOM representations.

Moreover, there are several stochastic choice models that allow for regularity violations. Intriguingly, we can show that some of them are AOM by directly checking $\succ$-Regularity. Important examples include \citet{Echenique-Saito-Tserenjigmid_2018_MSS} and \citet{Echenique-Saito_2019_ET}. In addition, \citet{Filiz-Ozbay2020ProgressiveChoice} introduces the \textit{Less-is-more Progressive Random Choice} model relying on the less-is-more choice function from \citet*{Lleras_et_al_2017_JET}. Since each less-is-more choice function can be mapped back to a competition filter, the model is essentially a special case of RCF and thus it can be rationalized by AOM.

H-LAO is also related to the literature on choice over lists. \citet{rubinstein2006model} develops deterministic choice from lists and random choice induced by randomly appearing lists. \citet{guney2014theory} and \citet{Yildiz2016TE} study successive-elimination procedures, including settings with an unobserved list and recursive random choice. Those papers capture elimination under a binary relation that need not be a preference, whereas H-LAO combines heterogeneous rational preferences with random limited attention. H-LAO is especially close to the \textit{Random Depth Model} of \citet{Ishiietal_2021_JME}, where a random depth determines how far down a list the decision maker explores. Their model studies a fixed menu with exogenous list variation, while H-LAO focuses on attention overload across menus.

Recent work also clarifies the consequences of observing coarsened alternatives. \citet{Kono-Saito-Sandroni_2026_AER} study random utility when choices of some alternatives are unobservable and aggregated into an outside option. \citet{Liao-Saito-Sandroni_2026_WP} allow the underlying composition of observed categories to vary across decision makers. H-LAO instead treats the listed alternatives and outside option as observed structural categories and requires the relevant preference comparisons to remain stable across menus. The empirical application states this requirement explicitly because price and travel-time profiles vary within its travel-mode categories.

\citet{Honda2021RandomCraving} studies a \textit{Random Craving Model}, a special case of RUM in which preferences are restricted to temptation orderings. It can therefore be viewed as a full-attention special case of H-LAO with a restricted preference distribution.

\section{Homogeneous Attention Overload}\label{section: homogeneous supplemental}

This section follows the homogeneous-preference analysis in Section 2 of the main paper. It presents the deferred proofs in the paper's order: Theorem 1, Corollaries 2 and 3, Proposition 3, and Theorems 2 and 3. Corollary 1 follows immediately from Theorem 1, while Propositions 1 and 2 are proved in the main text. The section then develops three additional results concerning a search-capacity foundation, decision avoidance, and observable separation from RAM.

\subsection{Proof of Theorem 1}

Necessity is established in the main paper, so it remains to prove sufficiency. Fix $S\in\mathcal D$ and enumerate it as $x_1\succ x_2\succ\cdots\succ x_m$. Because $S$ is fixed, the two objects defined below depend on $S$; for notational convenience, we suppress this dependence. First, define
\begin{align*}
q_i:=\max_{R\in\mathcal D:\,R\supseteq S}\pi(x_i|R).
\end{align*}
It is immediate that $q_1=\pi(x_1|S)$ by  $\succ$-Regularity. Because $S$ is included among its own observed supersets, and from $\succ$-Regularity, we know that the following inequality holds:
\begin{align*}
   \pi(x_i|S)  \leq q_i&\leq\sum_{j\leq i} \pi(x_j|S) \\
  \Rightarrow q_i - \pi(x_i|S)  & \leq \sum_{j < i} \pi(x_j|S)
\end{align*}
For $i\geq2$, define
\begin{align*}
    \lambda_i:=
    \begin{cases}
    \displaystyle\frac{q_i-\pi(x_i|S)}{\sum_{j<i}\pi(x_j|S)},& \text{if } \sum_{j<i}\pi(x_j|S)>0,\\[6pt]
    0,& \text{if }\sum_{j<i}\pi(x_j|S)=0.
    \end{cases}
\end{align*}
The preceding inequalities imply $\lambda_i\in[0,1]$; when the denominator is zero, they also imply $q_i=\pi(x_i|S)$.

Construct the attention rule as follows. For each consideration set $T\subseteq S$, distinguish the best element (according to $\succ$), say $x_i$, from the rest of the elements in the set, where each inferior alternative,  $x_j$ for $j>i$, appears independently with probability $\lambda_j$. In particular, suppose the best element of $T$ is $x_i$. Then, let
\begin{align*}
    \tilde\mu(T|S):=\pi(x_i|S)\prod_{   j>i: \  x_j \in T}\lambda_{j}  \prod_{j>i: \ x_j \notin T} (1-\lambda_{j})
\end{align*}

Therefore, the resulting attention rule rationalizes the observed choices:
\begin{align*}
    \sum_{T \subseteq S: \  x_i \text{ is } \succ \text{-best} } \tilde\mu(T|S) =\pi(x_i|S)\sum_{T \subseteq S: \  x_i \text{ is } \succ \text{-best} }  \left(\prod_{   j>i: \  x_j \in T}\lambda_{j}  \prod_{j>i: \ x_j \notin T} (1-\lambda_{j}) \right) =\pi(x_i|S)
\end{align*}
Its attention frequency for $x_i$ is
\begin{align*}
    \phi_{\tilde\mu}(x_i|S)
    =\pi(x_i|S)+\lambda_i\sum_{j<i}\pi(x_j|S)
    =q_i.
\end{align*}
The final equality follows from the definition of $\lambda_i$ when the sum is positive; when it is zero, the preceding inequalities give $q_i=\pi(x_i|S)$.
Finally, if $S,S'\in\mathcal D$, $S\subseteq S'$, and $x_i\in S$, then the collection of observed supersets of $S'$ is contained in the collection of observed supersets of $S$. Hence,
\begin{align*}
    \phi_{\tilde\mu}(x_i|S)=\max_{R\in\mathcal D:\,R\supseteq S}\pi(x_i|R)
    \geq\max_{R\in\mathcal D:\,R\supseteq S'}\pi(x_i|R)
    =\phi_{\tilde\mu}(x_i|S'),
\end{align*}
so the constructed attention rule satisfies attention overload. Repeating the construction menu by menu proves sufficiency and simultaneously establishes the pessimistic evaluation in Theorem 3.

For completeness, the same construction works for the optimistic evaluation after setting
\begin{align*}
    q_i=\min_{T\in\mathcal D:\,x_i\in T\subseteq S}\pi(U_\succeq(x_i)|T).
\end{align*}
The inclusion of $S$ and $\succ$-Regularity again imply
\begin{align*}
    \pi(x_i|S)\leq q_i\leq\sum_{j\leq i}\pi(x_j|S),
\end{align*}
and the collection of eligible subsets expands with the menu, so the resulting $q_i$ satisfies attention overload. More generally, the construction rationalizes any attention-frequency selection between the two bounds that itself satisfies attention overload.\qed

\subsection{Proof of Corollary 2}

For any strict ordering $\succ$, set $z_{ba}=1$ if and only if $b\succ a$. Completeness and asymmetry give $z_{ab}+z_{ba}=1$. Transitivity gives
\begin{align*}
    z_{ab}+z_{bc}-1\leq z_{ac}
\end{align*}
for every three distinct alternatives. Moreover,
\begin{align*}
    \pi(U_\succeq(a)|T)
    =\pi(a|T)+\sum_{b\in T\setminus\{a\}}\pi(b|T)z_{ba},
\end{align*}
so the remaining mixed-integer inequalities are exactly $\succ$-Regularity. Theorem 1 therefore maps every compatible ordering into $\mathcal Z_\pi$.

Conversely, take any $z\in\mathcal Z_\pi$ and define $b\succ_z a$ when $z_{ba}=1$. The binary restrictions make $\succ_z$ complete and asymmetric. The three-alternative inequalities make it transitive: if $a\succ_z b$ and $b\succ_z c$, then $z_{ab}=z_{bc}=1$, so $z_{ac}=1$ and $a\succ_z c$. Thus $\succ_z$ is a strict ordering. The final set of inequalities says precisely that $\pi$ satisfies $\succ_z$-Regularity, and Theorem 1 implies $\succ_z\in\mathcal P_\pi$. The two maps are inverse, proving the bijection and the model-feasibility claim. Finally, $b$ is revealed preferred to $a$ exactly when every feasible incidence vector has $z_{ba}=1$, which is equivalent to the stated minimization result.\qed

\subsection{Proof of Corollary 3}

Necessity follows from the argument in the main paper: if $\pi(a|\{a,b\})>\eta$, the bound on singleton consideration implies positive probability on $\{a,b\}$, so every representation satisfying attentive-at-binaries must have $a\succ b$.

For sufficiency, fix a strict ordering $\succ$ that satisfies $\succ$-Regularity and extends $\mathrm P_B$, and use the optimistic construction at the end of the preceding proof. Consider any observed binary menu $\{a,b\}\in\mathcal D$ and suppose without loss that $a\succ b$. For the lower-ranked item, every eligible subset in the optimistic minimum gives upper-contour probability one, so $q_b=1$. If $\pi(a|\{a,b\})>0$, the construction therefore sets $\lambda_b=1$ and $\tilde\mu(\{a\}|\{a,b\})=0$; if that probability is zero, the same equality holds directly. Moreover,
\begin{align*}
    \tilde\mu(\{b\}|\{a,b\})=\pi(b|\{a,b\})\leq\eta,
\end{align*}
because a strict reverse inequality would imply $b\mathrel{\mathrm P_B}a$, contradicting that $\succ$ extends $\mathrm P_B$. Thus both singleton-consideration probabilities satisfy the attentive-at-binaries restriction. The argument applies separately to every observed binary menu and does not require unobserved singletons or binaries.\qed

\subsection{Proof of Proposition 3}

Fix $\succ$. The defining constraints of $\mathcal M_\pi(\succ)$ are exactly nonnegativity and adding up, rationalization of the observed choice probabilities, and attention overload. Hence $\mu\in\mathcal M_\pi(\succ)$ if and only if $(\mu,\succ)$ is an AOM representation of $\pi$. Theorem 1 implies that such a $\mu$ exists if and only if $\succ\in\mathcal P_\pi$, proving the sharp joint-set formula and nonemptiness claim.

For fixed $\succ$, all constraints are linear and the variables are probabilities in a finite-dimensional product of simplices, so $\mathcal M_\pi(\succ)$ is a compact polytope. A linear functional therefore attains its minimum and maximum on each nonempty $\mathcal M_\pi(\succ)$. Since $X$ is finite, there are finitely many $\succ\in\mathcal P_\pi$; taking the smallest and largest conditional optima gives the sharp extrema over the union.\qed

\subsection{Proof of Theorem 2}
Let $(\succ,\mu)$ be an AOM representing $\ChoiProb$, and fix $S\in\mathcal D$ and $a\in S$. For every $R\in\mathcal D$ with $R\supseteq S$, attention overload and the fact that an alternative must be considered before it can be chosen imply
\begin{align*}
    \phi_\mu(a|S)\geq \phi_\mu(a|R)\geq \pi(a|R).
\end{align*}
Taking the maximum over the observed supersets $R$ gives the lower bound. Similarly, for every $T\in\mathcal D$ with $a\in T\subseteq S$,
\begin{align*}
    \phi_\mu(a|S)\leq \phi_\mu(a|T)\leq \pi(U_{\succeq}(a)|T).
\end{align*}
Taking the minimum over $T$ gives the upper bound.\qed

\subsection{Proof of Theorem 3}
The constructive proof of Theorem 1 sets $\phi(a|S)=\max_{R\in\mathcal D:\,R\supseteq S}\pi(a|R)$ and therefore establishes the pessimistic evaluation directly.\qed

The remaining subsections develop additional behavioral foundations and observable implications of homogeneous AOM.

\subsection{A Search-Capacity Foundation for Attention Overload}

Attention overload follows from a broad class of deletion-consistent search processes. Let a search type draw a menu-independent priority ordering $\rho$ of $X$ and a capacity $K\in\{0,1,\dots,|X|\}$. From menu $S$, the type considers the first $\min\{K,|S|\}$ available alternatives according to $\rho$. Both $\rho$ and $K$ may be random and arbitrarily dependent.

\begin{prop}[Random priority and capacity imply attention overload]\label{prop: search foundation}
Every mixture of menu-independent priority-and-capacity search types satisfies attention overload. If $K\geq1$ almost surely, it also satisfies the no-empty-consideration normalization of the homogeneous AOM.
\end{prop}

\begin{proof}
Fix a search type and observed menus $T,S\in\mathcal D$ with $a\in T\subseteq S$. If $a$ is among the first $K$ available alternatives in $S$, at most $K-1$ elements of $S$ precede it according to $\rho$. Removing the alternatives in $S\setminus T$ cannot increase the number of available alternatives preceding $a$. Hence $a$ is also among the first $K$ alternatives in $T$. The deterministic consideration indicator therefore satisfies
\begin{align*}
    \Indicator\{a\text{ considered in }S\}
    \leq
    \Indicator\{a\text{ considered in }T\}.
\end{align*}
Taking expectations over the joint distribution of $(\rho,K)$ gives $\phi(a|S)\leq\phi(a|T)$. If $K\geq1$, every nonempty menu generates a nonempty consideration set.
\end{proof}

The result includes heterogeneous inspection capacities and heterogeneous salience rankings. It also extends immediately to a menu-independent search sequence with nonnegative inspection costs and a fixed budget: deleting unavailable items weakly reduces the cost incurred before reaching any surviving item.

\subsection{Decision Avoidance as Monotone Engagement}

To incorporate the outside option in the homogeneous-preference model, consider a two-stage mechanism. The attention-overload restriction here applies conditional on engagement; allowing empty consideration under Assumption 1 of the main paper instead restricts unconditional attention frequencies. For every $S\in\mathcal D$, define the observed engagement probability
\begin{align*}
    q(S)=1-\pi(o^*|S),
\end{align*}
and suppose $q(S)>0$. Conditional inside-choice probabilities are
\begin{align*}
    p(a|S)=\frac{\pi(a|S)}{q(S)},
    \qquad a\in S.
\end{align*}
A \textit{decision-avoidance AOM} first engages with probability $q(S)$, where engagement is monotone in the sense that $q(T)\geq q(S)$ whenever $T,S\in\mathcal D$ and $T\subseteq S$, and chooses $o^*$ otherwise. Conditional on engagement, the decision maker follows a no-empty-consideration AOM with preference $\succ$.

\begin{prop}[Decision-avoidance characterization]\label{prop: decision avoidance}
The observed choice rule has a decision-avoidance AOM representation with preference $\succ$ if and only if
\begin{align*}
    q(T)\geq q(S)
    \qquad\text{for every }T,S\in\mathcal D\text{ with }T\subseteq S,
\end{align*}
and the conditional choice rule $p$ satisfies $\succ$-Regularity. Therefore,
\begin{align*}
    \mathcal P_{\pi}^{\mathrm{DA}}
    =\{\succ:p\text{ satisfies $\succ$-Regularity}\}
\end{align*}
is the sharp identified set of preferences whenever engagement is monotone.
\end{prop}

\begin{proof}
Necessity of monotone engagement follows from the first stage. Conditional on engagement, the model is exactly the homogeneous AOM, so Theorem 1 of the main paper implies $\succ$-Regularity of $p$.

Conversely, suppose engagement is monotone and $p$ satisfies $\succ$-Regularity. Theorem 1 of the main paper gives a no-empty attention rule $\mu^c$ satisfying attention overload and representing $p$ with $\succ$. Engage with probability $q(S)$ and, conditional on engagement, draw the consideration set from $\mu^c(\cdot|S)$. This reproduces
\begin{align*}
    \pi(a|S)=q(S)p(a|S),
    \qquad
    \pi(o^*|S)=1-q(S).
\end{align*}
Moreover, the unconditional inside-item attention frequency is
\begin{align*}
    \phi(a|S)=q(S)\phi^c(a|S).
\end{align*}
For $T,S\in\mathcal D$ with $a\in T\subseteq S$, both $q(S)\leq q(T)$ and $\phi^c(a|S)\leq\phi^c(a|T)$, so $\phi(a|S)\leq\phi(a|T)$. Thus the overall process also satisfies attention overload. Sharpness of the preference set follows from the conditional AOM characterization.
\end{proof}

This extension delivers an exact characterization when outside choice measures nonengagement and makes the choice-overload channel directly testable through $q(S)$. A genuine outside alternative that may be selected after inside items are considered represents a distinct mechanism, which can be separated from inattention with additional information or restrictions.

\subsection{Observable Separation from RAM}

The examples in Section \ref{Related Lit: single preference} can be summarized as a formal comparison of observable choice implications.

\begin{prop}[Observable non-nesting of AOM and RAM]\label{prop: observable nonnesting}
Neither the choice implications of AOM nor those of RAM contain the other across finite choice environments. The choice rule in Table \ref{example-SA1: Not RAM, Yes AOM} has an AOM representation but no RAM representation, while the choice rule in Table \ref{example-SA1: Yes RAM, Not AOM} has a RAM representation but no AOM representation. The two models can therefore produce different specification and revealed-preference conclusions from observed choice probabilities, not only from different latent attention rules.
\end{prop}

\begin{proof}
For Table \ref{example-SA1: Not RAM, Yes AOM}, RAM reveals both $b\succ c$ and $c\succ b$, so no strict preference can rationalize the data under RAM. The AOM characterization instead permits $a\succ b\succ c$ and $a\succ c\succ b$. For Table \ref{example-SA1: Yes RAM, Not AOM}, AOM would require both $a\succ b$ and $b\succ a$, whereas the RAM revealed-preference restrictions $a\succ d$ and $b\succ c$ are acyclic. The constructions and calculations are given in Section \ref{Related Lit: single preference}.
\end{proof}

\section{Heterogeneous-Preference List-Based Attention Overload}\label{section: hlao supplemental}

This section proves the H-LAO results stated in Section 3 of the main paper and then develops the direct observed-domain and robustness formulations. It also records the finite random-utility implications of the recovered full-attention rule and a descriptive measure of agreement with full-attention choice. Throughout this section, if $S=\{a_{s_1},\dots,a_{s_{|S|}}\}$ is written in the inherited list order, let $S_{:j}:=\{a_{s_1},\dots,a_{s_j}\}$ and $S_{j:}:=\{a_{s_j},\dots,a_{s_{|S|}}\}$ denote its $j$th prefix and suffix, respectively.

\subsection{Proof of Theorem 4}

We first prove necessity. Suppose that $\pi$ has a H-LAO representation $(\phi,\tau)$. The outside-option probability gives $\varphi(a_{s_1}|S)=1-\pi(o^*|S)=\phi(a_{s_1}|S)$. Sequential Path Independence and induction on $j$ then yield $\varphi(a_{s_j}|S)=\phi(a_{s_j}|S)$ for every $j$.

Because the domain is full, every suffix $S_{k:}$ is observed. By the maintained assumption that $\pi(S_{k:}|S_{k:})>0$, we have $1-\pi(o^*|S_{k:})>0$ for every $k$. Sequential Path Independence therefore implies
\begin{align*}
\varphi(a_{s_{|S|}}|S)
=
\phi(a_{s_{|S|}}|S)
=
\prod_{k=1}^{|S|}
\left[1-\pi(o^*|S_{k:})\right]
>0.
\end{align*}
Thus, the recursive definition of $f$ is well-defined.

Next, let $T\subseteq S$ have the same first alternative $a$. Weak Attention Overload gives $\phi(a|T)\geq\phi(a|S)$. Since $\pi(o^*|R)=1-\phi(a|R)$ whenever $a$ is the first alternative in $R$, it follows that $\pi(o^*|S)\geq\pi(o^*|T)$. Hence Weak Choice Overload holds. It remains to identify the full-attention choice probabilities.

\begin{claim}\label{Claim: f as RUM weight}
For all $a\in T$, we have
\begin{align*}
f(a|T)=\tau(\{\succ\in\mathcal P:a\text{ is $\succ$-best in }T\}).
\end{align*}
\end{claim}

\begin{proof}
We proceed by strong induction on $|T|$. The result is immediate for a singleton. Suppose it holds for every proper subset of $S=\{a_{s_1},\dots,a_{s_{|S|}}\}$. For $k=2,\dots,|S|$, the recursive definition of $f$, the induction hypothesis, and $\varphi=\phi$ give
\begin{align*}
f(a_{s_k}|S)
&=
\frac{1}{\phi(a_{s_{|S|}}|S)}
\Bigg(
\pi(a_{s_k}|S)
-\sum_{j=k}^{|S|-1}
\big(\phi(a_{s_j}|S)-\phi(a_{s_{j+1}}|S)\big)
\tau(\{\succ:a_{s_k}\text{ is $\succ$-best in }S_{:j}\})
\Bigg)\\
&=
\tau(\{\succ:a_{s_k}\text{ is $\succ$-best in }S\}).
\end{align*}
The last equality follows by substituting the H-LAO choice equation for $\pi(a_{s_k}|S)$: the terms corresponding to the proper prefixes cancel, leaving $\phi(a_{s_{|S|}}|S)$ times the probability that $a_{s_k}$ is $\succ$-best in $S$. The denominator is strictly positive by the argument above. Because both $f(\cdot|S)$ and the full-attention choice probabilities add up to one, the result for $a_{s_1}$ follows from the result for $k\geq2$.
\end{proof}

The claim implies the BM inequalities, proving necessity.

For sufficiency, suppose that $\pi$ satisfies Weak Choice Overload and non-negativity of the BM polynomials. Because the domain is full and $\pi(R|R)>0$ for every nonempty menu $R$, every suffix $S_{k:}$ satisfies $1-\pi(o^*|S_{k:})>0$. Hence the recursive construction of $\varphi$ gives $\varphi(a_{s_{|S|}}|S)>0$ for every $S$, so $f$ is well-defined.

Since $\varphi$ is recursively constructed according to the Sequential Path Independence formula, it satisfies Sequential Path Independence. Moreover, if $T\subseteq S$ have the same first alternative $a$, Weak Choice Overload gives $\pi(o^*|S)\geq\pi(o^*|T)$, and therefore $\varphi(a|T)=1-\pi(o^*|T)\geq1-\pi(o^*|S)=\varphi(a|S)$. Thus, $\varphi$ satisfies Weak Attention Overload.

Set $\tilde\phi=\varphi$ and let $\tilde\mu$ be the induced list-based attention rule, defined by $\tilde\mu(\emptyset|S)=1-\varphi(a_{s_1}|S)$ and $\tilde\mu(S_{:j}|S)=\varphi(a_{s_j}|S)-\varphi(a_{s_{j+1}}|S)$, with the terminal convention $\varphi(a_{s_{|S|+1}}|S)=0$. The recursive construction of $\varphi$ makes it weakly decreasing along the list, so these probabilities are nonnegative and add up to one. Hence $\tilde\mu$ is a well-defined list-based attention rule induced by an attention frequency satisfying the required restrictions.

The function $f$ adds up to one by construction, and the BM condition implies that it is random-utility rationalizable. Because $X$ is finite, all nonempty menus are observed, and preferences are strict, the Block--Marschak characterization in \citet{Falmagne_1978_JMP} implies that there exists a probability measure $\tilde\tau$ over $\mathcal P$ such that
\begin{align*}
f(a|T)
=
\tilde\tau(\{\succ\in\mathcal P:a\text{ is $\succ$-best in }T\}).
\end{align*}

It remains to show that $(\tilde\mu,\tilde\tau)$ explains the data. The outside-option probability is correct by construction. For $k=2,\dots,|S|$, the inside choice probability generated by this representation is
\begin{align*}
\tilde\pi(a_{s_k}|S)
&=
\varphi(a_{s_{|S|}}|S)
\tilde\tau(\{\succ\in\mathcal P:a_{s_k}\text{ is $\succ$-best in }S\})\\
&\quad+
\sum_{j=k}^{|S|-1}
\tilde\tau(\{\succ\in\mathcal P:a_{s_k}\text{ is $\succ$-best in }S_{:j}\})
\big(\varphi(a_{s_j}|S)-\varphi(a_{s_{j+1}}|S)\big)\\
&=
\varphi(a_{s_{|S|}}|S)f(a_{s_k}|S)
+
\sum_{j=k}^{|S|-1}
f(a_{s_k}|S_{:j})
\big(\varphi(a_{s_j}|S)-\varphi(a_{s_{j+1}}|S)\big)\\
&=
\pi(a_{s_k}|S),
\end{align*}
where the final equality follows from the definition of $f$. Finally, both the observed and constructed choice probabilities add up to one over $S\cup\{o^*\}$, so equality for the outside option and for $a_{s_2},\dots,a_{s_{|S|}}$ implies equality for $a_{s_1}$. This proves sufficiency.\qed

\subsection{Proof of Theorem 5}

The equality $\varphi=\phi$ follows directly from the recovery argument. The equality $f(a|T)=\tau(\{\succ\in\mathcal P:a\text{ is $\succ$-best in }T\})$ was established by the induction argument in the proof of Theorem 4.\qed

\subsection{Standard Random-Utility Implications}

Theorem 5 point identifies $f$. The remaining identification statements follow from standard finite random-utility theory \citep{Falmagne_1978_JMP,turansick2022identification}. In particular, the sharp set of distributions over rankings that generate the recovered rule is
\begin{align*}
\mathcal T(f)
=
\Bigg\{
\tau\in\Delta(\mathcal P):
f(a|S)
=
\sum_{\succ\in\mathcal P}
\Indicator\{a\text{ is $\succ$-best in }S\}\tau(\succ)
\text{ for every }a\in S\subseteq X
\Bigg\}.
\end{align*}
Indeed, every H-LAO representation must generate the recovered $f$, and combining any $\tau\in\mathcal T(f)$ with the identified attention rule reproduces $\pi$. The set is a nonempty compact polytope. Consequently, sharp bounds on any preference event are obtained by minimizing and maximizing its $\tau$-mass over $\mathcal T(f)$.

Full menu data also identify lower-contour and rank distributions. Let
\begin{align*}
L_\succ(a)=\{x\in X:a\succeq x\}
\end{align*}
be the weak lower contour set of $a$, and for $a\in T$ define the Block--Marschak polynomial
\begin{align*}
K_f(a,T)
=
\sum_{S:\,T\subseteq S\subseteq X}
(-1)^{|S\setminus T|}f(a|S).
\end{align*}

Moebius inversion gives
\begin{align*}
K_f(a,T)
=
\tau(\{\succ:L_\succ(a)=T\})
\qquad\text{for every }a\in T\subseteq X.
\end{align*}
Thus, if $|X|=p$ and rank one is best,
\begin{align*}
\Prob_\tau(\operatorname{rank}_\succ(a)=r)
=
\sum_{\substack{T:\,a\in T\subseteq X\\|T|=p-r+1}}
K_f(a,T),
\qquad r=1,\dots,p.
\end{align*}
Pairwise shares and top-rank shares satisfy
\begin{align*}
\Prob_\tau(a\succ b)=f(a|\{a,b\}),
\qquad
\Prob_\tau(a\text{ is top ranked})=f(a|X).
\end{align*}
The complete joint distribution over rankings may nevertheless remain set identified.

For completeness, for every $S$ containing $a$,
\begin{align*}
f(a|S)
=
\Prob_\tau(a\text{ is best in }S)
=
\Prob_\tau(S\subseteq L_\succ(a)).
\end{align*}
Applying Moebius inversion to these containment probabilities yields
\begin{align*}
\Prob_\tau(L_\succ(a)=T)
=
\sum_{S:\,T\subseteq S\subseteq X}
(-1)^{|S\setminus T|}
\Prob_\tau(S\subseteq L_\succ(a))
=
K_f(a,T).
\end{align*}
The rank formula follows by grouping lower contour sets according to cardinality. Pairwise and top-rank identities follow directly from the random-utility interpretation of $f$.

For completeness, the probability that observed choice agrees with full-attention choice provides a common descriptive summary for the two models. Under homogeneous AOM and a compatible preference $\succ$, the best item $a^\succ(S)$ is chosen if and only if it is considered, so the agreement probability is $\pi(a^\succ(S)|S)=\phi(a^\succ(S)|S)$. It is therefore point identified when the preference is identified and otherwise takes its sharp values over $\succ\in\mathcal P_\pi$. Under benchmark H-LAO, independence of preferences and stopping implies that the agreement probability equals
\begin{align*}
\sum_{\succ\in\mathcal P}
\tau(\succ)\phi(a^\succ(S)|S)
=
\sum_{a\in S}\phi(a|S)f(a|S).
\end{align*}
This probability is point identified wherever $\phi$ and $f$ are recovered and can otherwise be bounded using the corresponding sharp preference set. It provides a descriptive measure of agreement with full-attention choice and is kept distinct from cardinal welfare; the characterization and econometric results rely directly on the model's choice restrictions.
\subsection{Proof of Proposition 4}

Let $S=\{a_{s_1},\dots,a_{s_{|S|}}\}\in\mathcal D$. Since $\mathcal D$ is suffix-closed, every suffix $S_{k:}$ is observed. The outside-option probability identifies first-position reach in each suffix as $\phi(a_{s_k}|S_{k:})=1-\pi(o^*|S_{k:})$. Sequential Path Independence therefore gives $\phi(a_{s_j}|S)=\prod_{k=1}^j[1-\pi(o^*|S_{k:})]=\varphi(a_{s_j}|S)$ for every $j$. Since $S$ was arbitrary, $\phi(a|S)$ is point identified for every observed $S\in\mathcal D$ and $a\in S$.

Suppose, in addition, that $\mathcal D$ is prefix-closed. By the maintained assumption that $\pi(R|R)>0$ for every observed menu $R$, each factor $1-\pi(o^*|S_{k:})$ is strictly positive, so $\varphi(a_{s_{|S|}}|S)>0$ for every $S\in\mathcal D$. We can therefore recover $f$ recursively. The result is immediate for a singleton. Suppose $f$ has been identified for every observed menu of size strictly smaller than $|S|$. Prefix closure implies that every proper prefix $S_{:j}$ is observed, so the recursive definition of $f$ identifies $f(a_{s_k}|S)$ for $k=2,\dots,|S|$. The remaining value $f(a_{s_1}|S)$ is identified by adding up. By the same induction argument used in the proof of Theorem 4, under H-LAO, $f(x|S)=\tau(\{\succ\in\mathcal P:x\text{ is $\succ$-best in }S\})$. Thus these preference probabilities are point identified for every $x\in S$.\qed

\subsection{Proof of Proposition 5}

Suppose first that $a$ precedes $b$ on the observed list. Sequential Path Independence gives the probability of reaching $b$ in the binary menu as $r_b(a,b)=[1-\pi(o^*|\{a,b\})][1-\pi(o^*|\{b\})]$. By the maintained positive-reach assumption, $r_b(a,b)>0$. Since $b$ can be chosen from $\{a,b\}$ only when attention reaches the full binary menu, independence between preferences and stopping implies $\pi(b|\{a,b\})=r_b(a,b)\tau(\{\succ:b\succ a\})$. Hence $\tau(\{\succ:b\succ a\})=\pi(b|\{a,b\})/r_b(a,b)$ and is point identified.

If instead $b$ precedes $a$, the analogous argument identifies $\tau(\{\succ:a\succ b\})$ from $\{a,b\}$ and $\{a\}$. Since preferences are strict, $\tau(\{\succ:b\succ a\})=1-\tau(\{\succ:a\succ b\})$. Thus $\tau(\{\succ:b\succ a\})$ is point identified in either case.\qed

\subsection{Proof of Proposition 6}

Take any eligible $R\in\mathcal D$ with $R\supseteq S$ for which $\varphi(a|R)$ is defined and any eligible $T\in\mathcal D$ with $a\in T\subseteq S$ for which $\varphi(a|T)$ is defined. Whenever $\varphi(a|R)$ and $\varphi(a|T)$ are defined from the observed suffixes, Sequential Path Independence gives $\varphi(a|R)=\phi(a|R)$ and $\varphi(a|T)=\phi(a|T)$. Attention overload therefore implies $\varphi(a|R)\leq\phi(a|S)\leq\varphi(a|T)$. Taking the maximum over the eligible supersets $R$ and the minimum over the eligible subsets $T$ proves the result. If either collection is empty, the corresponding bound is omitted.\qed

\subsection{Proof of Proposition 7}

Let $\theta(a,S):=\tau(\{\succ\in\mathcal P:a\text{ is $\succ$-best in }S\})$. Whenever $f(a|T)$ is recovered from the data, H-LAO implies $f(a|T)=\theta(a,T)$. Because the full-attention choice rule generated by $\tau$ satisfies random-utility regularity, for $a\in T\subseteq S\subseteq R$ we have $f(a|T)\geq\theta(a,S)\geq f(a|R)$ whenever the displayed $f$ values are well-defined. Taking the minimum over the well-defined values $f(a|T)$ and the maximum over the well-defined values $f(a|R)$ gives the regularity bounds.

It remains to establish the additional list-based lower bound. Take any $R\supseteq S$ for which
\begin{align*}
\frac{\pi(a|R)-\Psi(a,S,R)}
{\varphi(a_{s_{|S|}}|R)}
\end{align*}
is well-defined. By the definition of $\Psi(a,S,R)$, this requires that $S$ be a prefix of $R$. Hence the prefixes of $R$ containing $a$ can be partitioned into those strictly contained in $S$, the prefix $S$ itself, and those strictly containing $S$. The H-LAO choice equation therefore gives
\begin{align*}
\pi(a|R)
&=
\underbrace{
\sum_{\substack{1\leq j\leq|R|:\\a\in R_{:j}\subsetneq S}}
f(a|R_{:j})\mu(R_{:j}|R)
}_{\Psi(a,S,R)}
+
\theta(a,S)\mu(S|R)\\
&\quad+
\sum_{\substack{1\leq j\leq|R|:\\R_{:j}\supsetneq S}}
\theta(a,R_{:j})\mu(R_{:j}|R).
\end{align*}

For every prefix $R_{:j}\supsetneq S$, random-utility regularity implies $\theta(a,R_{:j})\leq\theta(a,S)$. It follows that
\begin{align*}
\pi(a|R)
&\leq
\Psi(a,S,R)
+
\theta(a,S)
\sum_{\substack{1\leq j\leq|R|:\\R_{:j}\supseteq S}}
\mu(R_{:j}|R).
\end{align*}
Since $S$ is a prefix of $R$, the sum on the right is exactly the probability of reaching the last alternative of $S$ in $R$, which equals $\phi(a_{s_{|S|}}|R)=\varphi(a_{s_{|S|}}|R)$. Hence
\begin{align*}
\theta(a,S)
\geq
\frac{\pi(a|R)-\Psi(a,S,R)}
{\varphi(a_{s_{|S|}}|R)}.
\end{align*}

This holds for every $R\supseteq S$ for which the displayed expression is well-defined. Taking the maximum over such $R$ and combining it with the regularity bounds yields the stated result.\qed

\subsection{Observed-Domain and Sensitivity Analysis}

The full-domain, positive-engagement, independent H-LAO model remains our benchmark. This subsection builds nested sensitivity envelopes that isolate the identifying content of observed-menu coverage, positive reach, preference--stopping independence, and Sequential Path Independence. Removing these restrictions sequentially makes their contribution to identification transparent while preserving sharpness for each stated specification.

\subsubsection{Incomplete domains and zero reach}

We first allow $\pi(o^*|S)=1$ on observed menus, thereby covering zero reach beyond the positive-engagement setting of the recursive identification results. The next result gives a sharp representation on the observed domain $\mathcal D$. Full-domain extendibility is a stronger requirement that can be imposed when substantively appropriate.

Recall the observed list $a_1,\dots,a_{|X|}$. Write every observed menu as $S=\{a_{s_1},\dots,a_{s_{|S|}}\}$ in the inherited list order. We call $\mathcal D$ \textit{suffix closed} if $S_{j:}\in\mathcal D$ for every $S\in\mathcal D$ and $j=1,\dots,|S|$.

On a suffix-closed domain, define the recovered reach probabilities directly from outside choices:
\begin{align*}
\varphi(a_{s_j}|S)
=
\prod_{k=1}^j
\left[1-\pi(o^*|S_{k:})\right],
\qquad j=1,\dots,|S|.
\end{align*}
This product is well defined even when some reach probability is zero. Adopt the terminal convention $\varphi(a_{s_{|S|+1}}|S)=0$ and let
\begin{align*}
m_0(S)&=1-\varphi(a_{s_1}|S),\\
m_j(S)&=\varphi(a_{s_j}|S)-\varphi(a_{s_{j+1}}|S),
\qquad j=1,\dots,|S|.
\end{align*}
Thus, $m_j(S)$ is the recovered probability of considering exactly the prefix $S_{:j}$, while $m_0(S)=\pi(o^*|S)$ is the recovered probability of empty consideration.

Because each recovered reach is a product of factors in $[0,1]$, the sequence is weakly decreasing along every list. Hence $m_j(S)\geq0$ for every $S\in\mathcal D$ and $j=0,\dots,|S|$, and these masses add up to one automatically. We say that recovered attention is \textit{valid on $\mathcal D$} if
\begin{align*}
\varphi(a|S)\geq\varphi(a|R)
\qquad\text{whenever }a\in S\subseteq R\text{ and }S,R\in\mathcal D.
\end{align*}
This is attention overload on the observed domain. Sequential Path Independence and the validity of the within-menu prefix distribution hold by construction.

For $\tau\in\Delta(\mathcal P)$, define the choice probabilities generated by the recovered attention rule as
\begin{align*}
G_\tau(a|S)
=
\sum_{\succ\in\mathcal P}
\sum_{j=1}^{|S|}
m_j(S)
\Indicator\{a\in S_{:j},\ a\text{ is $\succ$-best in }S_{:j}\}
\tau(\succ).
\end{align*}
The incomplete-domain preference polytope is
\begin{align*}
\mathcal T_{\mathcal D}(\pi)
=
\left\{
\tau\in\Delta(\mathcal P):
G_\tau(a|S)=\pi(a|S)
\text{ for every }S\in\mathcal D\text{ and }a\in S
\right\}.
\end{align*}

\begin{prop}[Observed-Domain H-LAO]\label{prop:observed-domain-hlao}
Let $\mathcal D$ be suffix closed. The choice rule $\pi$ has an H-LAO representation on $\mathcal D$ if and only if recovered attention is valid on $\mathcal D$ and $\mathcal T_{\mathcal D}(\pi)$ is nonempty. Moreover, $\mathcal T_{\mathcal D}(\pi)$ is the sharp identified set for the preference distribution $\tau$. It is a compact polytope, and the sharp bounds on every preference event $A\subseteq\mathcal P$ are
\begin{align*}
\min_{\tau\in\mathcal T_{\mathcal D}(\pi)}
\sum_{\succ\in A}\tau(\succ)
\qquad\text{and}\qquad
\max_{\tau\in\mathcal T_{\mathcal D}(\pi)}
\sum_{\succ\in A}\tau(\succ).
\end{align*}
The result does not require positive reach.
\end{prop}

\begin{proof}
Suppose first that $(\phi,\tau)$ is an H-LAO representation on $\mathcal D$. For the first item of each observed menu,
\begin{align*}
\phi(a_{s_1}|S)=1-\pi(o^*|S).
\end{align*}
Sequential Path Independence and induction on $j$ give the product formula for $\varphi(a_{s_j}|S)$, including when one of its factors is zero. Hence $\varphi=\phi$ on the observed domain. The list-based attention rule therefore assigns probability $m_j(S)$ to $S_{:j}$ and probability $m_0(S)$ to the empty set. These probabilities are nonnegative, add to one, and satisfy attention overload. Independence between preferences and stopping gives $G_\tau(a|S)=\pi(a|S)$, so $\tau\in\mathcal T_{\mathcal D}(\pi)$.

Conversely, suppose recovered attention is valid and take $\tau\in\mathcal T_{\mathcal D}(\pi)$. Assign probability $m_0(S)$ to empty consideration and $m_j(S)$ to prefix $S_{:j}$. Their automatic nonnegativity and adding-up make this a list-based attention rule, while validity supplies attention overload. The product definition implies
\begin{align*}
\varphi(a_{s_j}|S)
=
\varphi(a_{s_{j-1}}|S)
\varphi(a_{s_j}|S_{j:}),
\end{align*}
so Sequential Path Independence holds. Draw $\succ$ independently according to $\tau$. The defining equations of $\mathcal T_{\mathcal D}(\pi)$ reproduce every inside choice probability, and $m_0(S)=\pi(o^*|S)$ reproduces outside choice. Thus every $\tau$ in the set is attainable, proving sharpness.

The set is the intersection of a finite-dimensional simplex with finitely many linear equalities. It is therefore a compact polytope. Preference-event probabilities are linear in $\tau$, so the displayed extrema are attained and are the sharp bounds.
\end{proof}

The observed-domain result complements the closed-form limited-data bounds in the main paper. Those bounds can be computed without enumerating rankings, while the polytope above exhausts the information about $\tau$ on any suffix-closed observed domain. Under full menu data and positive reach, it agrees with the standard random-utility polytope $\mathcal T(f)$. When terminal reach is zero, the recursive full-attention rule $f$ may be undefined, but the direct choice equations above remain valid and simply leave the unreached preference comparisons unrestricted.

For $p=|X|$, the direct formulation has $p!$ preference-probability variables, one adding-up constraint, and at most $\sum_{S\in\mathcal D}|S|$ choice equations. It is practical for small and moderate universes, while event-specific ranking generation extends its reach. The closed-form pairwise result below avoids ranking enumeration entirely and requires only singleton and binary menus. The dependence-robust formulation adds menu-specific joint variables in exchange for its greater flexibility; Section SA-5 reports target-specific implementation strategies and computational benchmarks.

\subsubsection{Illustration and implementation}

The replication files provide a four-alternative, suffix-closed example with $10$ of the $15$ nonempty menus and item-specific reach factors $(.86,.73,0,.64)$ in list order. The zero factor makes the recursive full-attention rule unavailable for some comparisons, while the direct polytope $\mathcal T_{\mathcal D}(\pi)$ remains well defined. For the event $d\succ a$, the sharp benchmark-independence bounds are $[.545,.780]$ and contain the generating value $.674$; allowing arbitrary preference--stopping dependence widens them to $[.219,1]$. The replication script constructs the choice rule and verifies feasibility, containment, and both sets of endpoints. Column generation certifies the two benchmark endpoints using only $5$ of the $24$ ranking columns in each restricted master problem.

Implementation should follow the target. Pairwise shares use the ranking-free formula in the main paper, and full-data rank probabilities use standard Block--Marschak sums. Ranking optimization is needed only for genuinely joint preference events or sharp incomplete-domain bounds. Section SA-5 gives the implementation hierarchy, solver certificates, and computational benchmarks.

We next relax the benchmark restrictions sequentially. The first envelope retains recovered stopping probabilities but relaxes preference--stopping independence; the second also removes Sequential Path Independence and treats prefix masses as latent.

\subsubsection{Preference--Stopping Dependence}

The benchmark H-LAO model makes preferences independent of the stopping position. We retain the recovered marginal stopping probabilities but allow arbitrary dependence within each observed menu. This extension is tailored to repeated cross sections: the marginal preference distribution is stable across menus, but the coupling between preferences and stopping may be menu specific. A panel model requiring one persistent joint preference--stopping type across all menus would impose additional restrictions.

For every $S\in\mathcal D$, $j=0,\dots,|S|$, and $\succ\in\mathcal P$, let $q_S(\succ,j)$ denote the joint probability of preference $\succ$ and stopping at $j$, where $j=0$ denotes empty consideration. Let $\mathcal Q_{\mathcal D}(\pi)$ be the collection of $(\tau,q)$ satisfying
\begin{align*}
&q_S(\succ,j)\geq0,\\
&\sum_{\succ\in\mathcal P}q_S(\succ,j)=m_j(S),\\
&\sum_{j=0}^{|S|}q_S(\succ,j)=\tau(\succ),\\
&\pi(a|S)=\sum_{j=1}^{|S|}\sum_{\succ\in\mathcal P}
\Indicator\{a\in S_{:j},\ a\text{ is $\succ$-best in }S_{:j}\}
q_S(\succ,j)
\end{align*}
for every relevant $S$, $j$, $\succ$, and $a\in S$. Define
\begin{align*}
\mathcal T_{\mathcal D}^{\mathrm{dep}}(\pi)
=
\{\tau:(\tau,q)\in\mathcal Q_{\mathcal D}(\pi)
\text{ for some }q\}.
\end{align*}

\begin{prop}[Dependence-Robust H-LAO]\label{prop:dependent-hlao}
Let $\mathcal D$ be suffix closed and suppose recovered attention is valid. Then $\mathcal T_{\mathcal D}^{\mathrm{dep}}(\pi)$ is the sharp identified set for the stable marginal preference distribution when preferences and stopping may be arbitrarily dependent within each observed menu. It is a compact polytope and contains the independent H-LAO identified set:
\begin{align*}
\mathcal T_{\mathcal D}(\pi)
\subseteq
\mathcal T_{\mathcal D}^{\mathrm{dep}}(\pi).
\end{align*}
Sharp bounds on preference events are obtained by linear programming over $\mathcal Q_{\mathcal D}(\pi)$. Positive reach is not required.
\end{prop}

\begin{proof}
Under a dependent representation, $q_S(\succ,j)$ is the joint distribution of the two latent types in menu $S$. Its stopping marginal must equal the identified prefix mass $m_j(S)$, its preference marginal must equal the stable distribution $\tau$, and utility maximization on each prefix gives the displayed choice equations. Thus every admissible representation belongs to $\mathcal Q_{\mathcal D}(\pi)$.

Conversely, take any feasible $(\tau,q)$. In each observed menu, draw $(\succ,j)$ according to $q_S$, consider the empty set when $j=0$ and prefix $S_{:j}$ otherwise, and select the $\succ$-best item in that prefix. The stopping marginal is the recovered list-based attention rule, so it satisfies Sequential Path Independence and attention overload. The preference marginal is $\tau$ in every menu, and the final constraints reproduce the observed choices. Hence every projected $\tau$ is attainable.

All defining constraints are linear and the variables are probabilities, so $\mathcal Q_{\mathcal D}(\pi)$ and its projection are compact polytopes. Under independence, $q_S(\succ,j)=\tau(\succ)m_j(S)$ satisfies the marginal constraints and reduces the choice equation to the definition of $\mathcal T_{\mathcal D}(\pi)$, proving the inclusion.
\end{proof}

The dependence-robust result gives a direct sensitivity analysis. Comparing event bounds over $\mathcal T_{\mathcal D}(\pi)$ and $\mathcal T_{\mathcal D}^{\mathrm{dep}}(\pi)$ isolates the identifying contribution of preference--stopping independence. If the dependence-robust set is empty, the observed list and marginal attention restrictions are rejected even after independence is removed. If it is nonempty while the independent set is empty, the data diagnose independence as the source of misspecification.

\subsubsection{Dropping Sequential Path Independence}

The preceding results use Sequential Path Independence to recover every prefix mass from outside-option probabilities on suffix menus. To assess this restriction, retain the observed list, prefix consideration, attention overload, and a stable marginal preference distribution, but do not impose Sequential Path Independence. This relaxation also permits arbitrary menu-specific dependence between preferences and stopping.

For every observed menu $S=\{a_{s_1},\dots,a_{s_{|S|}}\}$, let $m_j(S)$ be the latent probability of stopping at prefix $S_{:j}$, with $m_0(S)$ denoting empty consideration. For every $\succ\in\mathcal P$ and $j=0,\dots,|S|$, let $q_S(\succ,j)$ be the joint probability of preference $\succ$ and stopping at $j$. Define $\mathcal Q_{\mathcal D}^{\mathrm{noPI}}(\pi)$ by the following linear restrictions:
\begin{align}
&m_j(S)\geq0,\qquad m_0(S)=\pi(o^*|S),\qquad \sum_{j=0}^{|S|}m_j(S)=1,\label{eq:nopi-stopping}\\
&q_S(\succ,j)\geq0,\qquad \sum_{\succ\in\mathcal P}q_S(\succ,j)=m_j(S),\qquad \sum_{j=0}^{|S|}q_S(\succ,j)=\tau(\succ),\label{eq:nopi-marginals}\\
&\pi(a|S)=\sum_{j=1}^{|S|}\sum_{\succ\in\mathcal P}\Indicator\{a\in S_{:j},\ a\text{ is $\succ$-best in }S_{:j}\}q_S(\succ,j),\label{eq:nopi-choice}
\end{align}
for all relevant menus, positions, preferences, and inside alternatives. If $a_{s_j}=a_{r_k}\in S\subseteq R$, attention overload requires
\begin{align}
\sum_{\ell=j}^{|S|}m_\ell(S)
\geq
\sum_{\ell=k}^{|R|}m_\ell(R).
\label{eq:nopi-overload}
\end{align}
Indeed, the two sums are exactly the latent probabilities of reaching the common alternative in the smaller and larger menus. Let
\begin{align*}
\mathcal T_{\mathcal D}^{\mathrm{noPI}}(\pi)
=
\{\tau:(\tau,q,m)\in\mathcal Q_{\mathcal D}^{\mathrm{noPI}}(\pi)
\text{ for some }(q,m)\}.
\end{align*}

\begin{prop}[Path-Independence-Robust H-LAO]\label{prop:nopi-hlao}
For any observed-menu domain $\mathcal D$, $\mathcal T_{\mathcal D}^{\mathrm{noPI}}(\pi)$ is the sharp identified set for the stable marginal preference distribution when consideration sets are list prefixes, preferences and stopping may be arbitrarily dependent within menu, and Sequential Path Independence is not imposed. It is a compact polytope, does not require positive reach or suffix closure, and gives sharp preference-event bounds by linear programming. Whenever $\mathcal D$ is suffix closed and recovered attention is valid,
\begin{align*}
\mathcal T_{\mathcal D}(\pi)
\subseteq
\mathcal T_{\mathcal D}^{\mathrm{dep}}(\pi)
\subseteq
\mathcal T_{\mathcal D}^{\mathrm{noPI}}(\pi).
\end{align*}
\end{prop}

\begin{proof}
Consider first any representation in the stated class. Its menu-specific distribution over prefix stopping positions supplies nonnegative masses $m_j(S)$ satisfying the stopping and adding-up restrictions; the outside-option exclusion gives $m_0(S)=\pi(o^*|S)$. Let $q_S(\succ,j)$ be the joint distribution of preference and stopping in menu $S$. Stability of the preference marginal and the stopping marginal give the marginal restrictions. Utility maximization on each realized prefix gives the choice restrictions, and attention overload gives the reach inequalities. Every admissible representation therefore projects into $\mathcal T_{\mathcal D}^{\mathrm{noPI}}(\pi)$.

Conversely, take any feasible $(\tau,q,m)$. In each observed menu $S$, draw $(\succ,j)$ according to $q_S$, consider the empty set when $j=0$ and prefix $S_{:j}$ otherwise, and choose the $\succ$-best alternative in a nonempty prefix. The marginal stopping probabilities are $m_j(S)$, so the process is list based and its reach probabilities satisfy attention overload by the displayed reach inequalities. Its preference marginal is the same $\tau$ in every menu. The outside and inside choice equations reproduce $\pi$. Sequential Path Independence has not been used. Thus every projected $\tau$ is attainable, proving sharpness.

All constraints are linear and all variables lie in finite-dimensional probability simplices, so the feasible set and its projection are compact polytopes. Linear optimization therefore gives attained sharp bounds on every preference event. An independent H-LAO representation is a special case of a dependence-robust representation. Every dependence-robust representation satisfying Sequential Path Independence satisfies the preceding stopping, marginal, choice, and attention-overload constraints after its recovered prefix masses are treated as latent variables. This proves both inclusions.
\end{proof}

The path-independence-robust result converts the unobserved prefix masses into identified-set variables rather than recovering attention pointwise. Comparing the benchmark, dependence-robust, and no-Sequential-Path-Independence bounds therefore measures, in sequence, the identifying content of preference--stopping independence and Sequential Path Independence. The second relaxation naturally widens the bounds while delivering a sharp specification for arbitrary observed-menu domains without requiring suffix observations.

\subsubsection{Pairwise bounds under weakened restrictions}

The robustness envelopes can remain informative about selected preference events even when the complete ranking distribution is not point identified. The binary case makes this content transparent.

\begin{prop}[Robust Pairwise Preference Bounds]\label{prop:robust-pairwise-hlao}
Suppose $a$ precedes $b$ and write $\theta=\tau(\{\succ:b\succ a\})$.
\begin{enumerate}
\item If $\{a,b\}$ and $\{b\}$ are observed and Sequential Path Independence is maintained while preferences and stopping may be dependent, let $r=r_b(a,b)$ be the recovered probability of reaching $b$. Using these two menus alone, the sharp identified interval is
\begin{align*}
\pi(b|\{a,b\})
\leq
\theta
\leq
\pi(b|\{a,b\})+1-r.
\end{align*}
Its width is $1-r$; in particular, $r=1$ point identifies $\theta=\pi(b|\{a,b\})$ without preference--stopping independence.
\item If Sequential Path Independence is also removed and preferences and stopping may be dependent, the binary menu alone gives the sharp interval
\begin{align*}
\pi(b|\{a,b\})
\leq
\theta
\leq
1.
\end{align*}
If an exogenously assigned second list puts $b$ before $a$, pooling the two list strata gives the sharp interval
\begin{align*}
\pi(b|\{a,b\})
\leq
\theta
\leq
1-\pi^{\mathrm{rev}}(a|\{a,b\}).
\end{align*}
\end{enumerate}
Additional observed menus can only tighten these intervals.
\end{prop}

\begin{proof}
Let $E=\{\succ:b\succ a\}$ and let $J=2$ denote attention reaching the full binary prefix. Under Sequential Path Independence, $\Prob(J=2)=r$ is recovered. The later item $b$ is chosen exactly on $E\cap\{J=2\}$, so
\begin{align*}
\pi(b|\{a,b\})
=
\Prob(E\cap\{J=2\}).
\end{align*}
Consequently, $\theta=\Prob(E)\geq\pi(b|\{a,b\})$. Moreover,
\begin{align*}
1
\geq
\Prob(E\cup\{J=2\})
=
\theta+r-\pi(b|\{a,b\}),
\end{align*}
which gives $\theta\leq\pi(b|\{a,b\})+1-r$. To prove sharpness, fix any $\theta$ in this interval and assign masses
\begin{align*}
&\Prob(E,J=2)=\pi(b|\{a,b\}),
&&\Prob(E^c,J=2)=r-\pi(b|\{a,b\}),\\
&\Prob(E,J<2)=\theta-\pi(b|\{a,b\}),
&&\Prob(E^c,J<2)=1-r-\theta+\pi(b|\{a,b\}).
\end{align*}
All four masses are nonnegative. The two $J<2$ masses can be distributed across empty consideration and the singleton prefix to match their recovered stopping probabilities; choices at those prefixes do not depend on whether $E$ holds. This construction preserves the stable preference marginal, the recovered stopping marginal, and every choice probability in the two menus.

Without Sequential Path Independence, choosing the later item still implies $E$, and therefore $\theta\geq\pi(b|\{a,b\})$. Conversely, for any $\theta\in[\pi(b|\{a,b\}),1]$, assign probability $\pi(b|\{a,b\})$ to the full prefix and couple all of it with $E$. Assign the remaining preference mass across empty consideration and the singleton prefix, whose combined probability is $1-\pi(b|\{a,b\})$. This reproduces the binary choice probabilities and proves sharpness using that menu alone. In a reversed exogenous list, choosing the later item $a$ implies $E^c$, so $1-\theta\geq\pi^{\mathrm{rev}}(a|\{a,b\})$. Applying the preceding construction separately within each list stratum proves sharpness of the pooled interval.
\end{proof}

Under benchmark independence and positive reach, the point-identified share $\pi(b|\{a,b\})/r$ lies in the dependence-robust interval because $\pi(b|\{a,b\})=r\theta$. The result therefore quantifies the identifying contribution of independence: dropping it increases the width from zero to exactly $1-r$. Without Sequential Path Independence, a single ordering still supplies a sharp one-sided revealed-preference share, while exogenous reversal of the list supplies the opposite side.

\subsubsection{Exogenous list variation}

As a final sensitivity exercise, suppose exogenously assigned list strata $\ell=1,\dots,L$ share a common preference distribution $\tau$ but may have stratum-specific attention. Under benchmark independence, let $m_{\ell j}(S)$ be the recovered prefix masses and $S_{:j}^{\ell}$ the $j$th prefix under list $\ell$. Stack the inside-choice probabilities in $p$ and define
\begin{align*}
A_{(\ell,S,a),\succ}
=
\sum_{j=1}^{|S|}m_{\ell j}(S)
\Indicator\{a\in S_{:j}^{\ell},\ a\text{ is $\succ$-best in }S_{:j}^{\ell}\}.
\end{align*}

\begin{coro}[Multiple-list pooling]\label{coro: multiple list pooling}
The sharp pooled benchmark set is
\begin{align*}
\mathcal T^{\mathrm{pool}}
=\bigcap_{\ell=1}^L\mathcal T_{\mathcal D_\ell}(\pi_\ell)
=\{\tau\geq0:\boldsymbol 1'\tau=1,\ A\tau=p\}.
\end{align*}
The complete ranking distribution is globally point identified if and only if
\begin{align*}
\operatorname{rank}
\begin{pmatrix}\boldsymbol 1'\\ A\end{pmatrix}
=|\mathcal P|,
\qquad
\operatorname{rank}
\begin{pmatrix}\boldsymbol 1'\\ A\end{pmatrix}
\leq(|X|-2)2^{|X|-1}+2.
\end{align*}
Thus, when $|X|\geq4$, the rank bound rules out global identification of all $|X|!$ ranking probabilities from list variation alone. Each additional list nevertheless weakly tightens preference-event bounds and can identify selected targets. Under dependence or without Sequential Path Independence, the sharp pooled set is the intersection of the corresponding stratum-specific sensitivity sets.
\end{coro}

\begin{proof}
The pooled set is the intersection because each list stratum must share the same preference marginal and may use its own attention representation. The matrix form follows from the choice equations. Full column rank is equivalent to injectivity. Every row of $A$ is a linear combination of full-attention best-in-subset indicators; within each menu those indicators add to $\boldsymbol 1'$, giving the stated rank bound. Adding a list appends rows to $A$, so it cannot enlarge the identified set.
\end{proof}

\section{Econometric Methods}\label{section: econometric supplemental}

This section follows Section 4 of the main paper. It first proves Theorems 6--9 and Corollaries 4--6, then develops additional inference procedures for homogeneous AOM and H-LAO.

\subsection{Proof of Theorem 6}

We work conditionally on the realized menu assignments, as specified in the main paper. Thus, the menu-specific sample sizes are fixed and observations belonging to distinct menus are independent. We fix a preference $\succ$ throughout this proof. As in the main paper, let
\begin{align*}
    \mathcal I_\succ
    =\{(a,S,T):T,S\in\mathcal D,\ T\subset S,\ a\in T,\ U_\succeq(a)\cap T\neq T\}.
\end{align*}
denote the nonredundant comparison index, and let $\mathfrak c_\succ=|\mathcal I_\succ|$. To ease notation, we also assume that alternatives in a choice problem $S$ are ordered such that $a_1\prec a_2 \prec \cdots \prec a_{|S|}$. Now further fix a choice problem $S$. Recall that the choice probabilities and their estimates are
\begin{align*}
    \pi(a_j|S) = \Prob[y_i=a_j|Y_i=S],\ \text{and\ } \hat{\pi}(a_j|S) = \frac{1}{N_S}\sum_{i=1}^n \Indicator(y_i=a_j,Y_i=S).
\end{align*}
Standard deviation and standard error of the estimated choice probabilities are $\sigma(a_j|S)$ and $\hat{\sigma}(a_j|S)$, respectively,
\begin{align*}
    \sigma(a_j|S) = \sqrt{\frac{\pi(a_j|S)(1-\pi(a_j|S))}{N_S}},\ \text{and }\hat{\sigma}(a_j|S) = \sqrt{\frac{\hat{\pi}(a_j|S)(1-\hat{\pi}(a_j|S))}{N_S}}.
\end{align*}

As our testing procedure involves the probabilities of picking an alternative in an upper contour set, also let
\begin{align*}
    \pi(U_{\succeq}(a_j)|S) = \Prob[y_i\succeq a_j|Y_i=S],\ \text{and\ } \hat{\pi}(U_{\succeq}(a_j)|S) = \frac{1}{N_S}\sum_{i=1}^n \Indicator(y_i\succeq a_j,Y_i=S).
\end{align*}
Collect the preceding population probabilities and their estimates in the column vectors
\begin{align*}
    \bpi_S
    &=\Big(\pi(a_1|S),\dots,\pi(a_{|S|}|S),
    \pi(U_{\succeq}(a_1)|S),\dots,\pi(U_{\succeq}(a_{|S|})|S)\Big)',\\
    \hat{\bpi}_S
    &=\Big(\hat{\pi}(a_1|S),\dots,\hat{\pi}(a_{|S|}|S),
    \hat{\pi}(U_{\succeq}(a_1)|S),\dots,\hat{\pi}(U_{\succeq}(a_{|S|})|S)\Big)'.
\end{align*}
The upper-contour entries in $\hat{\bpi}_S$ are linear combinations of the individual-alternative probabilities. We retain them because they streamline the proof; implementation can work directly with the individual probabilities.

We first present the following lemma, which provides a strong approximation to the estimated choice probabilities.
As in the main paper, every logarithmic factor in the econometric arguments below is understood to be truncated below at one. This convention handles one-dimensional comparison sets, while the maintained restriction $N_S\geq2$ already excludes zero sample-size logarithms.

\begin{lem}\label{lemma: strong approx for one menu}
    In a possibly enlarged probability space, there exists a mean-zero Gaussian vector $\mathfrak{n}_S$ with the same covariance matrix as $\hat{\bpi}_S$, such that
    \begin{align*}
        \Prob\Big[ \frac{N_S}{\log(N_S)}\big\Vert \hat{\bpi}_S - \bpi_S - \mathfrak{n}_S \big\Vert_\infty > 2\Big(\frac{t}{\log(N_S)} + c_1\Big) \Big] \leq c_2\exp\{-c_3t\},
    \end{align*}
    where $c_1$--$c_3$ are universal constants.
\end{lem}

Now consider the collection $\mathcal{D}$ of menus that enter at least one comparison for testing $\succ$. These menus occur in nested pairs. Using any fixed ordering of $\mathcal D$, define $\bpi$ and $\hat{\bpi}$ as the column vectors obtained by stacking $\bpi_S$ and $\hat{\bpi}_S$, respectively, in that common order, and let $\mathfrak n$ stack the corresponding $\mathfrak n_S$ in the same order. Because every retained menu enters at least one comparison and each comparison uses two menus,
\begin{align*}
    \mathfrak c_{\mathcal D}\leq2\mathfrak c_\succ
\end{align*}
whenever $\mathfrak c_\succ\geq1$. This relation lets us replace logarithms of $\mathfrak c_{\mathcal D}$ by logarithms of $\mathfrak c_\succ$ in the final rates. We have the following strong approximation result.

\begin{lem}\label{lemma: strong approx for all menus}
    Under the setting of Lemma \ref{lemma: strong approx for one menu},
    \begin{align*}
        \Prob\Big[ \max_{S\in\mathcal{D} }\frac{N_S}{\log(N_S)}\big\Vert \hat{\bpi}_S - \bpi_S - \mathfrak{n}_S \big\Vert_\infty > c_1\Big(\log(\mathfrak{c}_{\mathcal{D}}) + 1\Big) \Big] \leq c_2\exp\Big\{-c_3\Big(\log\big(\min_{S\in\mathcal{D}}N_S\big)-1\Big)\log(\mathfrak{c}_{\mathcal{D}})\Big\},
    \end{align*}
    where
    \begin{align*}
        \mathfrak{c}_{\mathcal{D}} = \text{\# of choice problems in }\mathcal{D},
    \end{align*}
    $c_1$--$c_3$ are universal constants, and $c_3$ could be made arbitrarily large by appropriate choice of $c_1$.
\end{lem}

It is worth mentioning that we retained the sample sizes associated with each choice problem separately, instead of showing a bound on $\Vert \hat{\bpi} - \bpi - \mathfrak{n} \Vert_\infty$ that necessarily depends on the minimum sample size. This turns out to deliver sharper and more informative results.

We denote the nonredundant constraints implied by $\succ$-Regularity by the matrix $\bR_\succ$, whose rows are indexed by $\mathcal I_\succ$. The row associated with $(a,S,T)\in\mathcal I_\succ$ has $+1$ in the coordinate for $\pi(a|S)$, $-1$ in the coordinate for $\pi(U_\succeq(a)|T)$, and zeros elsewhere. Hence,
\begin{align*}
    [\bR_\succ\bpi]_{(a,S,T)}
    =\pi(a|S)-\pi(U_\succeq(a)|T)
    =D(a|S,T).
\end{align*}
The variances of $\bR_\succ\hat{\bpi}$ are collected in $\bsigma^2$, and its estimate is $\hat{\bsigma}^2$. Also recall from the main paper that elements of $\bsigma^2$ are denoted by $\sigma^2(a|S,T)$:
\begin{align*}
    \sigma^2(a|S,T) = \sigma^2(a|S) + \sigma^2(U_{\succeq}(a)|T) = \frac{\pi(a|S)(1-\pi(a|S))}{N_S} + \frac{\pi(U_{\succeq}(a)|T)(1-\pi(U_{\succeq}(a)|T))}{N_T}.
\end{align*}
We employ $\boldsymbol{\Omega}$ and $\hat{\boldsymbol{\Omega}}$ for the true and estimated correlation matrices of $\bR_\succ\hat{\bpi}$. Rows and columns of $\hat{\boldsymbol{\Omega}}$ corresponding to zero estimated standard errors are set to zero. Set $\bz = (\bR_\succ\mathfrak{n})\oslash \bsigma$, where $\oslash$ is the Hadamard (element-wise) division. To state the following result, define
\begin{align*}
   \mathfrak{c}_\sigma = \min_{(a,S,T)\in\mathcal I_\succ}\Bigg\{\min\Big\{ \frac{N_S}{\log(N_S)}, \frac{N_T}{\log(N_T)} \Big\} \cdot \max\Big\{ \sigma(a|S),\ \sigma(U_{\succeq}(a)|T)\Big\}\Bigg\},
\end{align*}
which, loosely speaking, corresponds to the root minimum effective sample size.

\begin{lem}\label{lemma: strong approximation for the comparisons}
    The centered and normalized constraints $\bR_\succ(\hat{\bpi} - \bpi)\oslash \bsigma$ satisfy
    \begin{align*}
        &\Prob\Big[ \big\Vert \bR_\succ(\hat{\bpi} - \bpi)\oslash \bsigma - \bz \big\Vert_\infty > c_1\frac{\log(\mathfrak{c}_{\mathcal{D}})}{ \mathfrak{c}_\sigma} \Big] \leq  c_2\exp\Big\{-c_3\Big(\log\big(\min_{S\in\mathcal{D}}N_S\big)-1\Big)\log(\mathfrak{c}_{\mathcal{D}})\Big\},
    \end{align*}
    where $c_1$--$c_3$ are universal constants, and $c_3$ could be made arbitrarily large by appropriate choice of $c_1$.
\end{lem}

The next step is to replace the infeasible standard error, for which we employ the lemma below. Division by $\hat{\bsigma}$ uses the main paper's coordinatewise convention: $0/0$ is zero and a nonzero numerator divided by zero is signed infinity. The standard-error concentration bound below controls the event that any such denominator vanishes.

\begin{lem}\label{lemma: taking hat off sigma}
    Let $\xi_{1},\xi_{2}> 0$ with $\xi_{2}\to 0$. Then,
    \begin{align*}
    &\Prob\Bigg[\Big\Vert (\bR_\succ\hat{\bpi} - \bR_\succ\bpi) \oslash \bsigma - (\bR_\succ\hat{\bpi} - \bR_\succ\bpi) \oslash \hat{\bsigma} \Big\Vert_\infty > \Big(\xi_{1} + c_4\log^{\frac{1}{2}}(\mathfrak{c}_{\succ}) + c_1\frac{\log(\mathfrak{c}_{\mathcal{D}})}{ \mathfrak{c}_\sigma}\Big)\cdot \xi_{2} \Bigg]\\
    &\quad\qquad\leq c_2\exp\Big\{-c_3\Big(\log\big(\min_{S\in\mathcal{D}}N_S\big)-1\Big)\log(\mathfrak{c}_{\mathcal{D}})\Big\} + \exp\Big\{-\frac{\xi_1^2}{2}\Big\} + c_5\exp\Big\{ -c_6 \mathfrak{c}_\sigma^2\xi_{2}^2 + \log(\mathfrak{c}_{\succ} ) \Big\},
    \end{align*}
    where $c_1$--$c_3$ are universal constants defined in Lemma \ref{lemma: strong approximation for the comparisons},
    \begin{align*}
        \mathfrak{c}_{\succ} = \text{\# of rows in }\bR_\succ,
    \end{align*}
    and $c_4$--$c_6$ are universal constants.
\end{lem}

As a byproduct of the discussion, we have

\begin{lem}\label{lemma: error in sigma}
    Let $\xi_{2}> 0$ with $\xi_{2}\to 0$. Then,
    \begin{align*}
        \Prob\Big[ \big\Vert\left(\hat{\bsigma} - \bsigma\right) \oslash \bsigma \big\Vert_\infty > \xi_{2} \Big] \leq c_5\exp\Big\{ -c_6  \mathfrak{c}_\sigma^2\xi_{2}^2 + \log(\mathfrak{c}_{\succ} ) \Big\},
    \end{align*}
    where $c_5$ and $c_6$ are universal constants defined in Lemma \ref{lemma: taking hat off sigma}.
\end{lem}

The proof of the theorem then follows from Lemmas \ref{lemma: strong approximation for the comparisons} and \ref{lemma: taking hat off sigma}. Specifically, we have that
   \begin{align*}
    &\Prob\Big[ \big\Vert \bR_\succ(\hat{\bpi} - \bpi)\oslash \hat{\bsigma} - \bz \big\Vert_\infty > c_1\frac{\log(\mathfrak{c}_{\mathcal{D}})}{ \mathfrak{c}_\sigma} + \Big(\xi_{1} + c_4\log^{\frac{1}{2}}(\mathfrak{c}_{\succ}) + c_1\frac{\log(\mathfrak{c}_{\mathcal{D}})}{ \mathfrak{c}_\sigma}\Big)\cdot \xi_{2} \Big]\\
    \leq&  c_2\exp\Big\{-c_3\Big(\log\big(\min_{S\in\mathcal{D}}N_S\big)-1\Big)\log(\mathfrak{c}_{\mathcal{D}})\Big\}  + \exp\Big\{-\frac{\xi_1^2}{2}\Big\} + c_5\exp\Big\{ -c_6 \mathfrak{c}_\sigma^2\xi_{2}^2 + \log(\mathfrak{c}_{\succ} ) \Big\}
\end{align*}
by applying the union bound. To simplify, we set
\begin{align*}
    \xi_2 \propto \frac{\log^{\frac{1}{2}}(\mathfrak{c}_{\succ})}{\mathfrak{c}_{\sigma}},
\end{align*}
and let $\xi_1$ be some large constant. This will imply the probabilistic bound on the Gaussian approximation
\begin{align*}
    \big\Vert \bR_\succ(\hat{\bpi} - \bpi)\oslash \hat{\bsigma} - \bz \big\Vert_\infty = \Op\left( \frac{\log(\mathfrak{c}_{\succ})}{\mathfrak{c}_{\sigma}} \right).
\end{align*}

\subsubsection{Proof of Lemma \ref{lemma: strong approx for one menu}}

Conditional on the menu assignments, let $u_{S,1},\dots,u_{S,N_S}$ be i.i.d. uniform random variables on $[0,1]$. Given the choice probabilities $\pi(a_j|S)$, define the disjoint intervals
\begin{align*}
    I_j = \left[ \sum_{j'<j}\pi(a_{j'}|S),\ \sum_{j'\leq j}\pi(a_{j'}|S) \right),
\end{align*}
with the right endpoint of the final interval included. Let $I_{j:}=\bigcup_{k\geq j}I_k$. The probability-integral construction gives
\begin{align*}
    \pi(a_j|S)=\Prob(u_{S,i}\in I_j),
    \qquad
    \pi(U_{\succeq}(a_j)|S)=\Prob(u_{S,i}\in I_{j:}),
\end{align*}
and the corresponding sample analogues are
\begin{align*}
    \hat{\pi}(a_j|S)
    &=\frac{1}{N_S}\sum_{i=1}^{N_S}\Indicator(u_{S,i}\in I_j),\\
    \hat{\pi}(U_{\succeq}(a_j)|S)
    &=\frac{1}{N_S}\sum_{i=1}^{N_S}\Indicator(u_{S,i}\in I_{j:}).
\end{align*}

Let $\mathcal{H}_S$ contain the indicator functions of the intervals $I_j$ and $I_{j:}$. Every member has total variation at most two. Indexing this finite class by distinct real numbers and setting the remaining index functions to zero verifies the hypotheses of the bounded-variation KMT corollary in \citet[Lemma SA26]{Cattaneo-Feng-Underwood_2024_JASA}; see also \citet[Remark 1]{cattaneo2025strong}. That lemma therefore supplies, on an enlarged probability space, a Gaussian process $G_S$ with the covariance of $\sqrt{N_S}(\hat\bpi_S-\bpi_S)$ and universal constants such that
\begin{align*}
    \Prob\left[
    \left\Vert\sqrt{N_S}(\hat\bpi_S-\bpi_S)-G_S\right\Vert_\infty
    >\frac{2t+c_1\log(N_S)}{\sqrt{N_S}}
    \right]
    \leq c_2\exp(-c_3t).
\end{align*}
Setting $\mathfrak n_S=G_S/\sqrt{N_S}$ and enlarging $c_1$ if needed gives the displayed bound in the lemma. The covariance of $\mathfrak n_S$ is exactly that of $\hat\bpi_S$.

\subsubsection{Proof of Lemma \ref{lemma: strong approx for all menus}}

Conditional on the menu assignments, choices from distinct menus are independent. We may therefore carry out the preceding construction for every $S$ on one product probability space, with independent auxiliary uniforms and Gaussian vectors across menus. Apply Lemma \ref{lemma: strong approx for one menu} with
\begin{align*}
    t_S=C\log(N_S)\log(\mathfrak{c}_{\mathcal D})
\end{align*}
and take a union bound over the $\mathfrak{c}_{\mathcal D}$ menus. After increasing the universal constants, the threshold is $c_1\{\log(\mathfrak{c}_{\mathcal D})+1\}$ and the union probability is bounded by
\begin{align*}
    c_2\exp\left[-c_3\left\{\log\left(\min_{S\in\mathcal D}N_S\right)-1\right\}
    \log(\mathfrak{c}_{\mathcal D})\right],
\end{align*}
as stated.

\subsubsection{Proof of Lemma \ref{lemma: strong approximation for the comparisons}}
To start, note that
\begin{align*}
    \bR_\succ(\hat{\bpi} - \bpi)\oslash \bsigma - \bz &= \bR_\succ\big(\hat{\bpi} - \bpi - \mathfrak{n}\big) \oslash \bsigma
    = \tilde{\bR}_\succ
    \begin{bmatrix}
        \frac{N_S}{\log(N_S)}\big( \hat{\bpi}_S - \bpi_S - \mathfrak{n}_S \big) \\
        \frac{N_T}{\log(N_T)}\big( \hat{\bpi}_T - \bpi_T - \mathfrak{n}_T \big) \\
        \vdots
    \end{bmatrix}.
\end{align*}
In other words, we rewrite the normalized constraints as the product of some matrix $\tilde{\bR}_\succ$ and the scaled choice probabilities. Since each row of $\bR_\succ$ consists of exactly one $+1$ and one $-1$, the rows of $\tilde{\bR}_\succ$ will contain two non-zero entries, taking the form
\begin{align*}
    \frac{\log(N_S)}{N_S}\frac{1}{\sigma(a|S,T)} \quad \text{and}\quad -\frac{\log(N_T)}{N_T}\frac{1}{\sigma(a|S,T)}.
\end{align*}
The next step is to bound the row-sum norm of $\tilde{\bR}_\succ$ (defined as the maximum of the $l_1$ norm of each row of a matrix). Specifically, consider a generic row that corresponds to a comparison $(a,S,T)\in\mathcal I_\succ$. Then the $l_1$ norm of this row is bounded by
\begin{align*}
    2 \Bigg[\min\Big( \frac{N_S}{\log(N_S)}, \frac{N_T}{\log(N_T)} \Big)\Bigg]^{-1}\Bigg[\max\Big( \frac{\pi(a|S)(1-\pi(a|S))}{N_S}, \frac{\pi(U_{\succeq}(a)|T)(1-\pi(U_{\succeq}(a)|T))}{N_T}\Big)\Bigg]^{-\frac{1}{2}}.
\end{align*}

To summarize, we have
\begin{align*}
    &\big\Vert\bR_\succ(\hat{\bpi} - \bpi)\oslash \bsigma - \bz\big\Vert_\infty \\
    &\leq 2 \max_{(a,S,T)\in\mathcal I_\succ}
    \Bigg[\min\Big( \frac{N_S}{\log(N_S)}, \frac{N_T}{\log(N_T)} \Big)\Bigg]^{-1}
    \Bigg[\max\Big( \sigma(a|S),\ \sigma(U_{\succeq}(a)|T) \Big)\Bigg]^{-1} \\
    &\qquad\times \max_{S\in\mathcal{D} }\frac{N_S}{\log(N_S)}
    \big\Vert \hat{\bpi}_S - \bpi_S - \mathfrak{n}_S \big\Vert_\infty,
\end{align*}
which implies the claim of the lemma.

\subsubsection{Proof of Lemma \ref{lemma: taking hat off sigma}}

Recall that individual components of $\bsigma^2$ are denoted by $\sigma^2(a|S,T)$. On the event that every required $\hat\sigma(a|S,T)$ is positive, simple algebra gives the following bound:
\begin{align*}
    &\Big\Vert (\bR_\succ\hat{\bpi} - \bR_\succ\bpi) \oslash \bsigma - (\bR_\succ\hat{\bpi} - \bR_\succ\bpi) \oslash \hat{\bsigma} \Big\Vert_\infty
    \leq \Big\Vert (\bR_\succ\hat{\bpi} - \bR_\succ\bpi) \oslash \bsigma  \Big\Vert_\infty
    \cdot
    \max_{(a,S,T)\in\mathcal I_\succ}\left|\frac{\sigma (a|S,T)}{\hat{\sigma}(a|S,T)} - 1\right|.
\end{align*}
We will control the two terms on the right-hand side separately.

Let $\xi_{1}>0$ be some constant which can depend on the sample size. Then by the triangle inequality, Lemma \ref{lemma: strong approximation for the comparisons}, maximal inequality for sub-Gaussian random variables, and the Borell-Tsirelson-Ibragimov-Sudakov inequality,
\begin{align*}
&\Prob\left[\Big\Vert (\bR_\succ\hat{\bpi} - \bR_\succ\bpi) \oslash \bsigma  \Big\Vert_\infty  > c_4\log^{\frac{1}{2}}(\mathfrak{c}_{\succ}) + \xi_{1} + c_1\frac{\log(\mathfrak{c}_{\mathcal{D}})}{ \mathfrak{c}_\sigma}\right] \\
&\qquad\leq \Prob\left[ \Big\Vert \bR_\succ(\hat{\bpi} - \bpi)\oslash \bsigma - \bz \Big\Vert_\infty > c_1\frac{\log(\mathfrak{c}_{\mathcal{D}})}{ \mathfrak{c}_\sigma} \right] + \Prob\Big[\Vert\bz\Vert_\infty > c_4\log^{\frac{1}{2}}(\mathfrak{c}_{\succ}) + \xi_{1}\Big] \\
&\qquad\leq
c_2\exp\Big\{-c_3\Big(\log\big(\min_{S\in\mathcal{D}}N_S\big)-1\Big)\log(\mathfrak{c}_{\mathcal{D}})\Big\} + \exp\Big\{-\frac{\xi_1^2}{2}\Big\},
\end{align*}
where $\bz$ and the constants $c_1$--$c_3$ are defined in Lemma \ref{lemma: strong approximation for the comparisons},
\begin{align*}
    \mathfrak{c}_{\succ} = \text{\# of rows in }\bR_\succ,
\end{align*}
and $c_4$ is a universal constant.

Next consider the standard error estimator, $\hat\sigma (a|S,T)$. We start with
\begin{align*}
\left|\hat\sigma (a|S,T)^2 - \sigma (a|S,T)^2\right| \leq \frac{1}{N_S}\left|\hat{\pi}(a|S) - \pi(a|S)\right| + \frac{1}{N_T}\left|\hat{\pi}(U_\succeq(a)|T) - \pi(U_\succeq(a)|T)\right|.
\end{align*}
Consider, for example, the first term on the right-hand side in the above. Using Bernstein's inequality, one has
\begin{align*}
\Prob\left[ \frac{1}{N_S\sigma (a|S,T)^2}\left|\hat{\pi}(a|S) - \pi(a|S)\right| > \xi_{2} \right]
 &\leq 2\exp\left\{ -\frac{1}{2}\frac{N_S^4\sigma (a|S,T)^4\xi_{2}^2}{N_S{\pi}(a|S)(1-{\pi}(a|S)) + \frac{1}{3}N_S^2\sigma (a|S,T)^2\xi_{2}} \right\}\\
 &= 2\exp\left\{ -\frac{1}{2}\frac{N_S^2\sigma (a|S,T)^4\xi_{2}^2}{\frac{1}{N_S}\pi(a|S)(1-\pi(a|S)) + \frac{1}{3}\sigma (a|S,T)^2\xi_{2}} \right\}\\
 &\leq 2\exp\left\{ -\frac{1}{2}\frac{N_S^2\sigma (a|S,T)^4\xi_{2}^2}{\sigma (a|S,T)^2 + \frac{1}{3}\sigma (a|S,T)^2\xi_{2}} \right\}\\
 &= 2\exp\left\{ -\frac{1}{2}\frac{N_S^2\sigma (a|S,T)^2\xi_{2}^2}{1 + \frac{1}{3}\xi_{2}} \right\}\\
&\leq 2\exp\left\{ -\frac{1}{4}N_S^2\sigma (a|S,T)^2\xi_{2}^2 \right\}
\leq 2\exp\left\{ -\frac{1}{4}  \mathfrak{c}_\sigma^2 \xi_{2}^2 \right\}.
\end{align*}
provided that $\xi_{2}\to 0$. Using the union bound, we deduce that
 \begin{align*}
 \Prob\left[ \max_{(a,S,T)\in\mathcal I_\succ}\frac{\left|\hat\sigma (a|S,T)^2 - \sigma (a|S,T)^2\right|}{\sigma (a|S,T)^2} \geq \xi_{2} \right] \leq 4\exp\left\{ -\frac{1}{16}  \mathfrak{c}_\sigma^2\xi_{2}^2 + \log(\mathfrak{c}_{\succ} )\right\},\qquad \text{provided that $\xi_{2}\to 0$},
 \end{align*}
 which also implies that
\begin{align*}
\Prob\left[ \max_{(a,S,T)\in\mathcal I_\succ}\frac{\left|\hat\sigma (a|S,T) - \sigma (a|S,T)\right|}{\sigma (a|S,T)} \geq \xi_{2} \right] \leq 4\exp\left\{ -\frac{1}{16}  \mathfrak{c}_\sigma^2\xi_{2}^2 + \log(\mathfrak{c}_{\succ} ) \right\},\qquad \text{provided that $\xi_{2}\to 0$}.
\end{align*}
If any required estimated standard error is zero, the maximum relative error in this display equals at least one. Hence, for $\xi_2<1$, the same probability bound also controls the complement of the event on which the preceding division is defined.
To control the reciprocal ratio in the first display of the proof, apply the preceding bound with $\xi_2/2$. On the complementary event, and for all sufficiently large sample sizes such that $\xi_2\leq1$,
\begin{align*}
\left|\frac{\sigma(a|S,T)}{\hat\sigma(a|S,T)}-1\right|
&=\frac{\left|\hat\sigma(a|S,T)/\sigma(a|S,T)-1\right|}
{\hat\sigma(a|S,T)/\sigma(a|S,T)}\\
&\leq\frac{\xi_2/2}{1-\xi_2/2}
\leq\xi_2.
\end{align*}
Replacing the numerical constants by universal constants $c_5$ and $c_6$ gives the probability bound stated in Lemma \ref{lemma: taking hat off sigma}. This closes the proof.

\subsubsection{Proof of Lemma \ref{lemma: error in sigma}}

This result follows directly from the proof of Lemma \ref{lemma: taking hat off sigma}

\subsection{Proof of Theorem 7}

Let
\begin{align*}
    \mathcal E_+
    =\left\{\min_{(a,S,T)\in\mathcal I_\succ}
    \hat\sigma(a|S,T)>0\right\}.
\end{align*}
All algebraic calculations involving division by $\hat{\bsigma}$ and the estimated correlation matrix are carried out on $\mathcal E_+$. On $\mathcal E_+^c$, we use the main paper's coordinatewise convention and bound its contribution to rejection probability by $\Prob(\mathcal E_+^c)$. Moreover, Lemma \ref{lemma: error in sigma} gives, for any $\xi_2<1$ used below,
\begin{align*}
    \Prob(\mathcal E_+^c)
    \leq c_5\exp\left\{-c_6\mathfrak c_\sigma^2\xi_2^2
    +\log(\mathfrak c_\succ)\right\}=o(1)
\end{align*}
for the choice of $\xi_2$ at the end of the proof, with a sufficiently large proportionality constant. This bound is no larger than the covariance-estimation remainder below and can be absorbed by increasing its universal constant. Thus bounds established on $\mathcal E_+$ extend to the full sample space without changing their rates.

The proof of Theorem 7 considers a sequence of approximations, and one step involves the comparison between Gaussian vectors $\hat{\bz}$ and $\bz$
\begin{align*}
    \hat{\bz}\ |\ \text{Data}\sim \mathcal{N}(0,\hat{\boldsymbol{\Omega}}),\quad \bz\sim \mathcal{N}(0,\boldsymbol{\Omega}).
\end{align*}
We have the following technical lemma on the estimation error in the correlation matrix.

\begin{lem}\label{lemma: correlation estimation}
    Let $\xi_{2}> 0$ with $\xi_{2}\to 0$. Then,
    \begin{align*}
        \Prob\Big[ \big\Vert \hat{\boldsymbol{\Omega}} - \boldsymbol{\Omega} \big\Vert_\infty > \xi_{2} \Big] \leq c_5\exp\Big\{ -c_6 \mathfrak{c}_\sigma^2\xi_{2}^2 + 2\log(\mathfrak{c}_{\succ} ) \Big\},
    \end{align*}
    where $c_5$ and $c_6$ are universal constants.
\end{lem}
We remark that the constants in the above result may be different from those in Lemmas \ref{lemma: taking hat off sigma} and \ref{lemma: error in sigma}. Since they are universal constants and do not depend on the underlying data-generating process, however, we decide to keep the same notation.

The following technical lemma helps establish distributional proximity in terms of the estimation error in correlation/covariance matrices.

\begin{lem}[Proposition 2.1 in \citealt*{Chernozhukov-Chetverikov-Kato-Koike_2022_AOS}]\label{lemma: gaussian comparison}
Let $\bx,\by\in\mathbb{R}^{\mathfrak{c}_{\succ}}$ be two mean-zero Gaussian random vectors with covariance matrices $\bSigma^{\bx}$ and $\bSigma^{\by}$, respectively. Further assume that the diagonal elements in $\bSigma^{\bx}$ are all one. Then,
\begin{align*}
    \sup_{\substack{A\subseteq \mathbb{R}^{\mathfrak{c}_{\succ}} \\ A\ \text{rectangular}}} \Big| \Prob[ \bx \in A ]-\Prob[ \by \in A ] \Big| \leq c_7\sqrt{\Vert\bSigma^{\bx}-\bSigma^{\by}\Vert_\infty}\log(\mathfrak{c}_{\succ}),
\end{align*}
where $c_7$ is a universal constant.
\end{lem}

Now we are ready to state the following result, which provides a feasible Gaussian approximation to $\bz$.

\begin{lem}\label{lemma: feasible gaussian approx from comparison}
Take $\xi_{2}> 0$ such that $\xi_{2}\to 0$. Then
\begin{align*}
    \Prob\left[
    \sup_{\substack{A\subseteq \mathbb{R}^{\mathfrak{c}_{\succ}} \\ A\ \text{rectangular}}}
    \left| \Prob[ \bz \in A ] - \Prob[ \hat{\bz} \in A | \text{Data}] \right|
    \leq c_7\xi_{2}^{\frac{1}{2}}\log(\mathfrak{c}_{\succ})
    \right]
    \geq 1 - c_5\exp\Big\{ -c_6  \mathfrak{c}_\sigma^2\xi_{2}^2 + 2\log(\mathfrak{c}_{\succ} ) \Big\},
\end{align*}
where the universal constants $c_5$--$c_7$ are defined in Lemmas \ref{lemma: correlation estimation} and \ref{lemma: gaussian comparison}.
\end{lem}

The results we have established so far allow making connections between the critical values. Write $\bz=(Z_1,\dots,Z_{\mathfrak c_\succ})'$ and $\hat\bz=(\hat Z_1,\dots,\hat Z_{\mathfrak c_\succ})'$. Define
\begin{align*}
    \hat{\mathsf{T}}^{\mathtt{G}}(\succ)
    &=\max\left\{0,\max_{1\leq\ell\leq\mathfrak c_\succ}\hat Z_\ell\right\},\quad
    \mathsf{T}^{\mathtt{G}}(\succ)
    =\max\left\{0,\max_{1\leq\ell\leq\mathfrak c_\succ}Z_\ell\right\},
\end{align*}
and the corresponding critical values
\begin{align*}
    \hat{\mathrm{cv}}(\alpha,\succ) &= \inf \{ t\geq 0:\ \Prob[\hat{\mathsf{T}}^{\mathtt{G}}(\succ)\leq t|\text{Data}]\geq 1-\alpha \}, \text{ and }
    {\mathrm{cv}}(\alpha,\succ) = \inf \{ t\geq 0:\ \Prob[\mathsf{T}^{\mathtt{G}}(\succ)\leq t]\geq 1-\alpha \}.
\end{align*}

\begin{lem}\label{lemma: feasible vs infeasible critical values}
Take $\xi_{2}>0$ such that $\xi_{2}\to0$, let $\delta_2=c_7\xi_2^{1/2}\log(\mathfrak{c}_\succ)$, and suppose
\begin{align*}
    0<\alpha-\delta_2<\alpha+\delta_2<\frac12.
\end{align*}
Then the critical values $\hat{\mathrm{cv}}(\alpha,\succ)$ and $\mathrm{cv}(\alpha,\succ)$ satisfy
\begin{align*}
    &\Prob\Big[\mathrm{cv}\big(\alpha + \delta_2,\succ\big)\leq \hat{\mathrm{cv}}(\alpha,\succ) \leq \mathrm{cv}\big(\alpha - \delta_2,\succ\big) \Big] \geq 1 - c_5\exp\Big\{ -c_6 \mathfrak{c}_\sigma^2\xi_{2}^2 + 2\log(\mathfrak{c}_{\succ} ) \Big\},
\end{align*}
where the universal constants $c_5$--$c_7$ are defined in Lemmas \ref{lemma: correlation estimation} and \ref{lemma: gaussian comparison}.
\end{lem}

To wrap up the proof of Theorem 7, we rely on a sequence of error bounds. Let
\begin{align*}
    \mathsf{T}(\succ)
    &=\max\left\{0,
      \max_{1\leq\ell\leq\mathfrak c_\succ}
      \left[(\bR_\succ\hat{\bpi})\oslash\hat{\bsigma}\right]_\ell
      \right\},\\
    \mathsf{T}^\circ(\succ)
    &=\max\left\{0,
      \max_{1\leq\ell\leq\mathfrak c_\succ}
      \left[\bR_\succ(\hat{\bpi}-\bpi)\oslash\hat{\bsigma}\right]_\ell
      \right\},\\
    \tilde{\mathsf{T}}^\circ(\succ)
    &=\max\left\{0,
      \max_{1\leq\ell\leq\mathfrak c_\succ}
      \left[\bR_\succ(\hat{\bpi}-\bpi)\oslash\bsigma\right]_\ell
      \right\}.
\end{align*}
To start, under $\mathsf{H}_0$ and after enlarging $c_5$ to absorb $\Prob(\mathcal E_+^c)$, we have
\begin{align*}
    \Prob\Big[ \mathsf{T}(\succ) > \hat{\mathrm{cv}}(\alpha,\succ) \Big]
    &\leq \Prob\Big[ \mathsf{T}^\circ(\succ) > \hat{\mathrm{cv}}(\alpha,\succ),\ \mathcal E_+ \Big] +\Prob(\mathcal E_+^c) \\
    &\leq
    \Prob\Big[ \mathsf{T}^\circ(\succ) > \mathrm{cv}\big(\alpha + c_7\xi_{2}^{\frac{1}{2}}\log(\mathfrak{c}_{\succ}),\succ\big) \Big]
    + c_5\exp\Big\{ -c_6 \mathfrak{c}_\sigma^2\xi_{2}^2 + 2\log(\mathfrak{c}_{\succ} ) \Big\}.
\end{align*}
Next, we apply Lemma \ref{lemma: taking hat off sigma} and obtain that
\begin{align*}
    &\Prob\Big[ \mathsf{T}^\circ(\succ) > \mathrm{cv}\big(\alpha + c_7\xi_{2}^{\frac{1}{2}}\log(\mathfrak{c}_{\succ}),\succ\big) \Big]  \\
    &\leq \Prob\Big[ \tilde{\mathsf{T}}^\circ(\succ) > \mathrm{cv}\big(\alpha + c_7\xi_{2}^{\frac{1}{2}}\log(\mathfrak{c}_{\succ}),\succ\big) - \Big(\xi_{1} + c_4\log^{\frac{1}{2}}(\mathfrak{c}_{\succ}) + c_1\frac{\log(\mathfrak{c}_{\mathcal{D}})}{ \mathfrak{c}_\sigma}\Big)\cdot \xi_{2} \Big] \\
    &+ c_2\exp\Big\{-c_3\Big(\log\big(\min_{S\in\mathcal{D}}N_S\big)-1\Big)\log(\mathfrak{c}_{\mathcal{D}})\Big\} + \exp\Big\{-\frac{\xi_1^2}{2}\Big\} + c_5\exp\Big\{ -c_6 \mathfrak{c}_\sigma^2\xi_{2}^2 + \log(\mathfrak{c}_{\succ} ) \Big\}.
\end{align*}

The final error bound in our analysis is due to Lemma \ref{lemma: strong approximation for the comparisons}, which implies that
\begin{align*}
    &\Prob\Big[ \tilde{\mathsf{T}}^\circ(\succ) > \mathrm{cv}\big(\alpha + c_7\xi_{2}^{\frac{1}{2}}\log(\mathfrak{c}_{\succ}),\succ\big) - \Big(\xi_{1} + c_4\log^{\frac{1}{2}}(\mathfrak{c}_{\succ}) + c_1\frac{\log(\mathfrak{c}_{\mathcal{D}})}{ \mathfrak{c}_\sigma}\Big)\cdot \xi_{2} \Big] \\
    &\leq \Prob\Big[\mathsf{T}^{\mathtt{G}}(\succ) > \mathrm{cv}\big(\alpha + c_7\xi_{2}^{\frac{1}{2}}\log(\mathfrak{c}_{\succ}),\succ\big) - \Big(\xi_{1} + c_4\log^{\frac{1}{2}}(\mathfrak{c}_{\succ}) + c_1\frac{\log(\mathfrak{c}_{\mathcal{D}})}{ \mathfrak{c}_\sigma}\Big)\cdot \xi_{2} - c_1\frac{\log(\mathfrak{c}_{\mathcal{D}})}{ \mathfrak{c}_\sigma}\Big]\\
    &\qquad + c_2\exp\Big\{-c_3\Big(\log\big(\min_{S\in\mathcal{D}}N_S\big)-1\Big)\log(\mathfrak{c}_{\mathcal{D}})\Big\}.
\end{align*}
Collecting all pieces, the size distortion of $\Prob[ \mathsf{T}(\succ) > \hat{\mathrm{cv}}(\alpha,\succ) ]$ is bounded as
\begin{align*}
    &\Prob[ \mathsf{T}(\succ) > \hat{\mathrm{cv}}(\alpha,\succ) ] \\
    &\leq \Prob\Big[\mathsf{T}^{\mathtt{G}}(\succ) > \mathrm{cv}\big(\alpha + c_7\xi_{2}^{\frac{1}{2}}\log(\mathfrak{c}_{\succ}),\succ\big) - \Big(\xi_{1} + c_4\log^{\frac{1}{2}}(\mathfrak{c}_{\succ}) + c_1\frac{\log(\mathfrak{c}_{\mathcal{D}})}{ \mathfrak{c}_\sigma}\Big)\cdot \xi_{2} - c_1\frac{\log(\mathfrak{c}_{\mathcal{D}})}{ \mathfrak{c}_\sigma}\Big]\\
    &\quad + c_2\exp\Big\{-c_3\Big(\log\big(\min_{S\in\mathcal{D}}N_S\big)-1\Big)\log(\mathfrak{c}_{\mathcal{D}})\Big\} + \exp\Big\{-\frac{\xi_1^2}{2}\Big\} + c_5\exp\Big\{ -c_6 \mathfrak{c}_\sigma^2\xi_{2}^2 + 2\log(\mathfrak{c}_{\succ} ) \Big\}.
\end{align*}
In the last step, we have absorbed a multiplicative constant 2 into the universal constant $c_2$ and $c_5$.

To proceed, we employ the following anti-concentration result for normal random vectors. The restriction $\alpha+\delta_2<1/2$ also resolves the truncation at zero: since any coordinate of $\bz$ is standard normal,
\begin{align*}
    \Prob\left[\max_{1\leq\ell\leq\mathfrak c_\succ}Z_\ell\leq0\right]\leq\frac12,
\end{align*}
so every oracle critical value used below is strictly positive. In a neighborhood of such a critical value, $\max\{0,\max_\ell Z_\ell\}$ and $\max_\ell Z_\ell$ coincide.

\begin{lem}[Lemma D.4 in the supplemental appendix to \citealt{Chernozhukov-Chetverikov-Kato_2019_RESTUD}]\label{lemma: anti concentration}
Let $\bz\in\mathbb{R}^{\mathfrak{c}_{\succ}}$ be a mean-zero normal random vector such that $\Var[Z_{\ell}]=1$ for all $1\leq \ell\leq \mathfrak{c}_{\succ}$. Then, for any $t\in\mathbb{R}$ and any $\epsilon>0$,
\begin{align*}
    \Prob\left[ \left|\max_{1\leq\ell\leq\mathfrak c_\succ}Z_\ell- t\right| \leq \epsilon \right] \leq 4\epsilon\Big(\sqrt{2\log(\mathfrak{c}_{\succ})} + 1\Big).
\end{align*}
\end{lem}

For brevity, define
\begin{align*}
    \eta_2
    =\left(\xi_1+c_4\log^{1/2}(\mathfrak c_\succ)
    +c_1\frac{\log(\mathfrak c_{\mathcal D})}{\mathfrak c_\sigma}\right)\xi_2
    +c_1\frac{\log(\mathfrak c_{\mathcal D})}{\mathfrak c_\sigma}.
\end{align*}
The lemma above immediately implies that
\begin{align*}
    &\mathrm{cv}\left(\alpha+\delta_2
    +4\eta_2\{\sqrt{2\log(\mathfrak c_\succ)}+1\},\succ\right)
    \leq
    \mathrm{cv}(\alpha+\delta_2,\succ)-\eta_2.
\end{align*}
A direct consequence is that
\begin{align*}
    &\Prob\left[\mathsf{T}^{\mathtt G}(\succ)
    >\mathrm{cv}(\alpha+\delta_2,\succ)-\eta_2\right]
    \leq\alpha+\delta_2
    +4\eta_2\{\sqrt{2\log(\mathfrak c_\succ)}+1\}.
\end{align*}
Therefore, by possibly enlarging the constants,
\begin{align*}
    &\Prob[ \mathsf{T}(\succ) > \hat{\mathrm{cv}}(\alpha,\succ) ] \leq \alpha \\
    &\quad + c_2\exp\Big\{-c_3\Big(\log\big(\min_{S\in\mathcal{D}}N_S\big)-1\Big)\log(\mathfrak{c}_{\mathcal{D}})\Big\} + \exp\Big\{-\frac{\xi_1^2}{2}\Big\} + c_5\exp\Big\{ -c_6 \mathfrak{c}_\sigma^2\xi_{2}^2 + 2\log(\mathfrak{c}_{\succ} ) \Big\}\\
    &\quad+ c_7\xi_{2}^{\frac{1}{2}}\log(\mathfrak{c}_{\succ})
    + c_1\xi_{2}\Big(\xi_{1} + \log^{\frac{1}{2}}(\mathfrak{c}_{\succ})\Big) \log^{\frac{1}{2}}(\mathfrak{c}_{\succ})\\
    &\quad+ c_1\frac{\log(\mathfrak c_{\mathcal D})\log^{1/2}(\mathfrak c_\succ)}{\mathfrak c_\sigma}.
\end{align*}

Under the growth conditions of Theorem 7, all probability-level shifts above are smaller than $1/2-\alpha$ for sufficiently large samples. Now set
\begin{align*}
    \xi_1 \propto \log^{\frac{1}{2}}(\mathfrak{c}_{\sigma}),\quad \text{and}\quad \xi_2 \propto \frac{\log^{\frac{1}{2}}(\mathfrak{c}_{\succ})\log^{\frac{1}{2}}(\mathfrak{c}_{\sigma})}{\mathfrak{c}_\sigma},
\end{align*}
Using $\mathfrak c_{\mathcal D}\leq2\mathfrak c_\succ$, the last term in the preceding display is of smaller order than the rate below under the stated growth conditions. Thus the chosen values give the desired bound
\begin{align*}
    \Prob[ \mathsf{T}(\succ) > \hat{\mathrm{cv}}(\alpha,\succ) ] \leq \alpha + c \frac{\log^{\frac{5}{4}}(\mathfrak{c}_{\succ})\log^{\frac{1}{4}}(\mathfrak{c}_{\sigma})}{\mathfrak{c}_\sigma^{\frac{1}{2}}}
\end{align*}
for some constant $c$.

\subsubsection{Proof of Lemma \ref{lemma: correlation estimation}}

Each restriction compares probabilities from two menus. Two restrictions, indexed by $\ell$ and $\ell'$, therefore involve at most four menus, denoted by $T\subset S$ and $T'\subset S'$. If $\{S,T\}\cap\{S',T'\}=\emptyset$, conditional independence across menus makes both the population covariance and its estimate zero. It remains to control pairs of restrictions that share at least one menu.

To start, we first decompose the difference as
\begin{align*}
&\left| \frac{\hat{\sigma}(a|S,T;a'|S',T')}{\hat{\sigma}(a|S,T)\hat{\sigma}(a'|S',T')} - \frac{\sigma (a|S,T;a'|S',T')}{\sigma (a|S,T)\sigma (a'|S',T')} \right| \\
 &\leq \left| \frac{\hat{\sigma}(a|S,T;a'|S',T')-\sigma (a|S,T;a'|S',T')}{\sigma (a|S,T)\sigma (a'|S',T')}  \right| + \left| \frac{\hat{\sigma}(a|S,T;a'|S',T')}{\hat{\sigma}(a|S,T)\hat{\sigma}(a'|S',T')} \right|\cdot\left| \frac{\hat{\sigma}(a|S,T)\hat{\sigma}(a'|S',T')}{\sigma (a|S,T)\sigma (a'|S',T')} - 1 \right|\\
 &\leq \left| \frac{\hat{\sigma}(a|S,T;a'|S',T')-\sigma (a|S,T;a'|S',T')}{\sigma (a|S,T)\sigma (a'|S',T')}  \right| + \left| \frac{\hat{\sigma}(a|S,T)\hat{\sigma}(a'|S',T')}{\sigma (a|S,T)\sigma (a'|S',T')} - 1 \right|\\
&\leq \left| \frac{\hat{\sigma}(a|S,T;a'|S',T')-\sigma (a|S,T;a'|S',T')}{\sigma (a|S,T)\sigma (a'|S',T')}  \right| + \left| \frac{\hat{\sigma}(a|S,T)}{\sigma (a|S,T)} - 1 \right| + \left| \frac{\hat{\sigma}(a'|S',T')}{\sigma (a'|S',T')} - 1 \right| \\
&\qquad\qquad\qquad+ \left| \frac{\hat{\sigma}(a|S,T)}{\sigma (a|S,T)} - 1 \right| \cdot \left| \frac{\hat{\sigma}(a'|S',T')}{\sigma (a'|S',T')} - 1 \right|.
\end{align*}
Therefore,
\begin{align*}
&\max_{(a,S,T)\in\mathcal I_\succ}\max_{(a',S',T')\in\mathcal I_\succ} \left| \frac{\hat{\sigma}(a|S,T;a'|S',T')}{\hat{\sigma}(a|S,T)\hat{\sigma}(a'|S',T')} - \frac{\sigma (a|S,T;a'|S',T')}{\sigma (a|S,T)\sigma (a'|S',T')} \right| \\
&\leq
\max_{(a,S,T)\in\mathcal I_\succ}\max_{(a',S',T')\in\mathcal I_\succ}\left| \frac{\hat{\sigma}(a|S,T;a'|S',T')-\sigma (a|S,T;a'|S',T')}{\sigma (a|S,T)\sigma (a'|S',T')}  \right| \\
&\qquad + 2\max_{(a,S,T)\in\mathcal I_\succ}  \left| \frac{\hat{\sigma}(a|S,T)}{\sigma (a|S,T)} - 1 \right|+ \max_{(a,S,T)\in\mathcal I_\succ}  \left| \frac{\hat{\sigma}(a|S,T)}{\sigma (a|S,T)} - 1 \right|^2.
\end{align*}
We already have an error bound for the last two terms from Lemma \ref{lemma: error in sigma}, and hence it suffices to study the first term in the above display.

Here $\sigma(a|S,T;a'|S',T')$ denotes a covariance, whereas $\sigma(a|S,T)$ denotes a standard deviation. Up to interchanging the two restrictions, there are four overlap configurations: (i) $S=S'$ and $T\neq T'$; (ii) $T=T'$ and $S\neq S'$; (iii) $T=S'$; and (iv) $S=S'$ and $T=T'$. Menus not identified by these equalities are distinct.

\noindent \textbf{Case (i)}. The covariance $\sigma (a|S,T;a'|S,T')$ can take two forms: $\frac{1}{N_S} \pi(a|S)(1 - \pi(a|S))$ or $-\frac{1}{N_S}\pi(a|S)\pi(a'|S)$.
Nevertheless, the following bound applies
\begin{align*}
\left|\hat{\sigma}(a|S,T;a'|S,T') - \sigma (a|S,T;a'|S,T')\right| &\leq
\frac{1}{N_S} \left| \hat{\pi}(a|S) - \pi(a|S) \right| + \frac{1}{N_S} \left| \hat{\pi}(a'|S) - \pi(a'|S) \right|.
\end{align*}
Next we apply Bernstein's inequality, which gives
\begin{align*}
&\ \Prob\left[ \frac{1}{N_S\sigma (a|S,T)\sigma (a'|S,T')}\left|\hat{\pi}(a|S) - \pi(a|S)\right| \geq \xi_{2} \right] \\
 &\leq 2\exp\left\{ -\frac{1}{2}\frac{N_S^4\sigma (a|S,T)^2\sigma (a'|S,T')^2\xi_{2}^2}{N_S\pi(a|S)(1-\pi(a|S)) + \frac{1}{3}N_S^2\sigma (a|S,T)\sigma (a'|S,T')\xi_{2}} \right\}\\
 &= 2\exp\left\{ -\frac{1}{2}\frac{N_S^2\sigma (a|S,T)^2\sigma (a'|S,T')^2\xi_{2}^2}{\frac{1}{N_S}\pi(a|S)(1-\pi(a|S)) + \frac{1}{3}\sigma (a|S,T)\sigma (a'|S,T')\xi_{2}} \right\}\\
 &\leq 2\exp\left\{ -\frac{1}{2}\frac{N_S^2\sigma (a|S,T)^2\sigma (a'|S,T')^2\xi_{2}^2}{\sigma (a|S,T)^2 + \frac{1}{3}\sigma (a|S,T)\sigma (a'|S,T')\xi_{2}} \right\}
 \\
 &\leq 2\exp\left\{ -\frac{1}{2}\frac{N_S^2\sigma (a|S,T)^2\sigma (a'|S,T')^2\xi_{2}^2}{(\sigma (a|S,T)\vee\sigma (a'|S,T'))^2 + \frac{1}{3}(\sigma (a|S,T)\vee\sigma (a'|S,T'))^2\xi_{2}} \right\}\\
 &= 2\exp\left\{ -\frac{1}{2}\frac{N_S^2(\sigma (a|S,T)\wedge\sigma (a'|S,T'))^2\xi_{2}^2}{1 + \frac{1}{3}\xi_{2}} \right\}\\
&\leq 2\exp\left\{ -\frac{1}{4}N_S^2(\sigma (a|S,T)\wedge\sigma (a'|S,T'))^2\xi_{2}^2 \right\} \leq 2\exp\left\{ -\frac{1}{4}  \mathfrak{c}_\sigma^2\xi_{2}^2 \right\}.
\end{align*}
provided that $\xi_{2}\to 0$. Therefore, we have
\begin{align*}
\Prob\left[ \frac{1}{\sigma (a|S,T)\sigma (a'|S,T')}\left|\hat{\sigma}(a|S,T;a'|S,T') - \sigma (a|S,T;a'|S,T')\right| \geq \xi_{2} \right] \leq 4\exp\left\{ -\frac{1}{16}  \mathfrak{c}_\sigma^2\xi_{2}^2 \right\}.
\end{align*}

\noindent \textbf{Case (ii)}. The covariance $\sigma (a|S,T;a'|S',T)$ can be conveniently written as
\begin{align*}
\sigma (a|S,T;a'|S',T) &=  \frac{1}{N_{T}}\Big(\pi(U_{\succeq}(a)|T)\wedge\pi(U_{\succeq}(a')|T)-\pi(U_{\succeq}(a)|T)\pi(U_{\succeq}(a')|T)\Big).
\end{align*}
Then, we have the bound
\begin{align*}
    &\left|\hat{\sigma}(a|S,T;a'|S',T) - \sigma (a|S,T;a'|S',T)\right|\\
    &\leq
    \frac{1}{N_{T}} \left| \hat{\pi}(U_{\succeq}(a)|T) - \pi(U_{\succeq}(a)|T) \right| + \frac{1}{N_{T}} \left| \hat{\pi}(U_{\succeq}(a')|T) - \pi(U_{\succeq}(a')|T) \right|.
\end{align*}
Again by using the union bound and Bernstein's inequality, one has
\begin{align*}
\Prob\left[ \frac{1}{\sigma (a|S,T)\sigma (a'|S',T)}\left|\hat{\sigma}(a|S,T;a'|S',T) - \sigma (a|S,T;a'|S',T)\right| \geq \xi_{2} \right] \leq 4\exp\left\{ -\frac{1}{16}  \mathfrak{c}_\sigma^2\xi_{2}^2 \right\}.
\end{align*}

\noindent \textbf{Case (iii)}. Here $S'=T$. The covariance $\sigma(a|S,T;a'|T,T')$ can take two forms:
\begin{align*}
\underbrace{\frac{1}{N_{T}} \pi(U_{\succeq}(a)|T)\pi(a'|T)}_{\textstyle \text{if } a'\not\in U_{\succeq}(a)}\qquad  \text{or}\qquad \underbrace{-\frac{1}{N_T}\Big(\pi(a'|T)-\pi(a'|T)\pi(U_{\succeq}(a)|T)\Big)}_{\textstyle \text{if } a'\in U_{\succeq}(a)}.
\end{align*}
In either case, we have
\begin{align*}
\left|\hat{\sigma}(a|S,T;a'|T,T') - \sigma (a|S,T;a'|T,T')\right| &\leq
\frac{1}{N_{T}} \left| \hat{\pi}(U_{\succeq}(a)|T) - \pi(U_{\succeq}(a)|T) \right| + \frac{1}{N_{T}} \left| \hat{\pi}(a'|T) - \pi(a'|T) \right|.
\end{align*}
Again by using the union bound and Bernstein's inequality, one has
\begin{align*}
\Prob\left[ \frac{\left|\hat{\sigma}(a|S,T;a'|T,T') - \sigma (a|S,T;a'|T,T')\right|}{\sigma (a|S,T)\sigma (a'|T,T')} \geq \xi_{2} \right] \leq 4\exp\left\{ -\frac{1}{16} \mathfrak{c}_\sigma^2\xi_{2}^2 \right\}.
\end{align*}

\noindent \textbf{Case (iv)}. The covariance $\sigma (a|S,T;a'|S',T')$ can take two forms:
\begin{align*}
\sigma (a|S,T;a'|S,T) &= \underbrace{\frac{1}{N_S} \pi(a|S)(1 - \pi(a|S)) + \frac{1}{N_{T}}\pi(U_{\succeq}(a)|T)\Big(1-\pi(U_{\succeq}(a)|T)\Big)}_{\textstyle \text{if $a=a'$}},\\
\sigma (a|S,T;a'|S,T) &= \underbrace{-\frac{1}{N_S} \pi(a|S)\pi(a'|S) + \frac{1}{N_{T}}\Big(\pi(U_{\succeq}(a)|T)\wedge\pi(U_{\succeq}(a')|T)-\pi(U_{\succeq}(a)|T)\pi(U_{\succeq}(a')|T)\Big)}_{\textstyle \text{if $a\neq a'$}}.
\end{align*}
Then,
\begin{align*}
    &\left|\hat{\sigma}(a|S,T;a'|S,T) - \sigma (a|S,T;a'|S,T)\right| \\
    &\leq \frac{|\hat\pi(a|S)-\pi(a|S)|}{N_S}
    +\frac{|\hat\pi(a'|S)-\pi(a'|S)|}{N_S}\\
    &\qquad+\frac{|\hat\pi(U_\succeq(a)|T)-\pi(U_\succeq(a)|T)|}{N_T}
    +\frac{|\hat\pi(U_\succeq(a')|T)-\pi(U_\succeq(a')|T)|}{N_T}.
\end{align*}
Again, by using the union bound and Bernstein's inequality,
\begin{align*}
\Prob\left[ \frac{\left|\hat{\sigma}(a|S,T;a'|S,T) - \sigma (a|S,T;a'|S,T)\right|}{\sigma (a|S,T)\sigma (a'|S,T)} \geq \xi_{2} \right] \leq 8\exp\left\{ -\frac{1}{64}  \mathfrak{c}_\sigma^2\xi_{2}^2 \right\}.
\end{align*}
Combining the four cases and taking a union bound over at most $\mathfrak{c}_{\succ}^2$ pairs gives, after changing universal constants,
\begin{align*}
\Prob\left[
\max_{\ell,\ell'}
\frac{|\hat\sigma_{\ell\ell'}-\sigma_{\ell\ell'}|}{\sigma_\ell\sigma_{\ell'}}
>\xi_2
\right]
\leq c_5\exp\left\{-c_6\mathfrak c_\sigma^2\xi_2^2+2\log(\mathfrak c_\succ)\right\}.
\end{align*}
Applying Lemma \ref{lemma: error in sigma} to the remaining standard-error terms in the decomposition above, and absorbing their constants, proves the stated bound for $\Vert \hat{\boldsymbol{\Omega}}-\boldsymbol{\Omega}\Vert_\infty$.

\subsubsection{Proof of Lemma \ref{lemma: feasible gaussian approx from comparison}}

This follows directly from Lemmas \ref{lemma: correlation estimation} and \ref{lemma: gaussian comparison}.

\subsubsection{Proof of Lemma \ref{lemma: feasible vs infeasible critical values}}

Work on the event in Lemma \ref{lemma: feasible gaussian approx from comparison}. The level restrictions in Lemma \ref{lemma: feasible vs infeasible critical values} imply that the critical values in this argument are positive. At nonnegative cutoffs, the distribution functions of the truncated maxima correspond to rectangular events for the untruncated Gaussian vectors, so the uniform comparison applies. Therefore,
\begin{align*}
\Prob\Big[\hat{\mathsf{T}}^{\mathtt{G}}(\succ)>  \mathrm{cv}\big(\alpha - c_7\xi_{2}^{\frac{1}{2}}\log(\mathfrak{c}_{\succ}),\succ\big)\Big| \text{Data}\Big]
&\leq \Prob\Big[\mathsf{T}^{\mathtt{G}}(\succ)> \mathrm{cv}\big(\alpha - c_7\xi_{2}^{\frac{1}{2}}\log(\mathfrak{c}_{\succ}),\succ\big)\Big] + c_7\xi_{2}^{\frac{1}{2}}\log(\mathfrak{c}_{\succ})\\
&\leq \alpha - c_7\xi_{2}^{\frac{1}{2}}\log(\mathfrak{c}_{\succ}) + c_7\xi_{2}^{\frac{1}{2}}\log(\mathfrak{c}_{\succ}) = \alpha.
\end{align*}
Therefore, on this event,
\begin{align*}
    \hat{\mathrm{cv}}(\alpha,\succ) \leq \mathrm{cv}\big(\alpha - c_7\xi_{2}^{\frac{1}{2}}\log(\mathfrak{c}_{\succ}),\succ\big).
\end{align*}
Similarly, for every $0\leq t<\mathrm{cv}(\alpha+\delta_2,\succ)$, the same uniform comparison gives
\begin{align*}
    \Prob\big[\hat{\mathsf{T}}^{\mathtt{G}}(\succ)\leq t\mid\text{Data}\big]
    &\leq\Prob\big[\mathsf{T}^{\mathtt{G}}(\succ)\leq t\big]+\delta_2
    <1-\alpha.
\end{align*}
Here the oracle probability is marginal, not conditional on the data. By the definition of the feasible critical value, this implies
\begin{align*}
    \hat{\mathrm{cv}}(\alpha,\succ) \geq \mathrm{cv}\big(\alpha + c_7\xi_{2}^{\frac{1}{2}}\log(\mathfrak{c}_{\succ}),\succ\big)
\end{align*}
under the same event.

\subsection{Proof of Corollary 4}

Corollary 4 restricts the comparison index in Theorem 7 to $(a,S,\{a,b\})$ with $S\in\mathcal D_{ab}$. The standardized statistic and its feasible Gaussian analogue are therefore coordinate restrictions of the corresponding objects in Theorem 7. The coupling, covariance-estimation, and Gaussian-comparison arguments apply unchanged with $\mathfrak c_\succ=|\mathcal D_{ab}|$ and with $\mathfrak c_\sigma$ specialized to the binary-screening comparisons, yielding the first displayed size-distortion bound.

It remains to prove the sharper fixed-pair statement. Let $\bz_{ab}\sim\mathcal N(0,\boldsymbol\Omega_{ab})$ and $\widehat{\bz}_{ab}\mid\mathrm{Data}\sim\mathcal N(0,\widehat{\boldsymbol\Omega}_{ab})$ be the corresponding oracle and feasible Gaussian vectors for the binary-screening comparisons indexed by $S\in\mathcal D_{ab}$. Set
\begin{align*}
    \rho_{ab}=\lambda_{\min}(\boldsymbol\Omega_{ab}),
    \qquad
    \Delta_{ab}=\|\widehat{\boldsymbol\Omega}_{ab}-\boldsymbol\Omega_{ab}\|_\infty.
\end{align*}
Lemma \ref{lemma: binary comparison covariance} and Corollary \ref{coro: binary comparison nondegenerate} below imply that the balance condition gives $\rho_{ab}\geq1/(1+C^2)$. The correlation-estimation argument in Lemma \ref{lemma: correlation estimation}, restricted to these $|\mathcal D_{ab}|$ coordinates, gives
\begin{align*}
    \Prob(\Delta_{ab}>\xi_2)
    \leq c_5\exp\{-c_6\mathfrak c_\sigma^2\xi_2^2+2\log(|\mathcal D_{ab}|)\}.
\end{align*}
On the event $\Delta_{ab}\leq\xi_2<e^{-1}$, Theorem 2.3 of \citet{lopes2022central} and the eigenvalue bound yield
\begin{align*}
    \sup_{A\ \mathrm{rectangular}}
    \left|\Prob(\bz_{ab}\in A)
    -\Prob(\widehat{\bz}_{ab}\in A\mid\mathrm{Data})\right|
    \leq c(1+C^2)\log(|\mathcal D_{ab}|)\log(1/\xi_2)\xi_2.
\end{align*}
The result permits $\widehat{\boldsymbol\Omega}_{ab}$ to be singular; only the oracle correlation matrix must satisfy the displayed eigenvalue bound.

Let $L_{ab}=\max\{\log(|\mathcal D_{ab}|\mathfrak c_\sigma),1\}$ and choose $\xi_1=A L_{ab}^{1/2}$ and $\xi_2=A L_{ab}^{1/2}/\mathfrak c_\sigma$ for a sufficiently large universal constant $A$. Repeat the proof of Theorem 7 with the preceding Lopes bound in place of Lemma \ref{lemma: gaussian comparison}. Because the menus entering a fixed-pair screen consist of the binary menu and at most one larger menu per comparison, $\mathfrak c_{\mathcal D}\leq|\mathcal D_{ab}|+1$. The Gaussian-comparison contribution is bounded by
\begin{align*}
    c(1+C^2)\frac{L_{ab}^{5/2}}{\mathfrak c_\sigma},
\end{align*}
while the studentization and anti-concentration contribution and the strong-coupling contribution are each of order at most $L_{ab}^{3/2}/\mathfrak c_\sigma$. With $A$ sufficiently large, the exponential remainder probabilities are also of smaller order. The stated condition makes the resulting level shift vanish and gives
\begin{align*}
    \Prob\big[\mathsf T(ab)>\widehat{\mathrm{cv}}(\alpha,ab)\big]
    \leq\alpha+c(1+C^2)
    \frac{\log^{5/2}(|\mathcal D_{ab}|\mathfrak c_\sigma)}{\mathfrak c_\sigma}.
\end{align*}
This argument concerns a fixed ordered pair $(a,b)$. Applying the same result to a statistic that stacks multiple pairs would additionally require a lower bound for the minimum eigenvalue of the joint correlation matrix.\qed

\subsection{Proof of Theorem 8}

If an estimated standard error is zero, the lower confidence bound is zero by convention and coverage is immediate. We therefore work on the complementary event. We start with three lemmas, the first two giving a Gaussian coupling and the third handling uncertainty in the standard error.

\begin{lem}\label{lemma: strong approx for all supersets of S}
    In a possibly enlarged probability space, there exists a mean-zero Gaussian vector $\mathfrak{n}_R$ for each $R\supseteq S$ with the same covariance matrix as $\hat{\bpi}_R$, such that
    \begin{align*}
        &\Prob\Bigg[
        \max_{\substack{R\supseteq S\\R\in\mathcal D}}
        \frac{N_R}{\log(N_R)}
        \|\hat\bpi_R-\bpi_R-\mathfrak n_R\|_\infty
        >c_1\{\log(\mathfrak c_{\supseteq S})+1\}\Bigg]\\
        &\quad\leq c_2\exp\Bigg[-c_3
        \left\{\log\left(\min_{\substack{R\supseteq S\\R\in\mathcal D}}N_R\right)-1\right\}
        \log(\mathfrak c_{\supseteq S})\Bigg],
    \end{align*}
    where
    \begin{align*}
        \mathfrak{c}_{\supseteq S} = \text{\# of supersets of }S,
    \end{align*}
    $c_1$--$c_3$ are universal constants, and $c_3$ could be made arbitrarily large by appropriate choice of $c_1$.
\end{lem}

\begin{lem}\label{lemma: strong approximation with normalization}
    Under the setting of Lemma \ref{lemma: strong approx for all supersets of S}, the centered and normalized choice probabilities satisfy
    \begin{align*}
        &\Prob\Big[ \max_{R\supseteq S, R\in\mathcal{D}}\Big| \frac{\hat{\pi}(a|R) - \pi(a|R)}{\sigma(a|R)} - \mathsf{Z}_{a,S}(R) \Big| > c_1\frac{\log(\mathfrak{c}_{\supseteq S})}{ \mathfrak{c}_\sigma} \Big]\\
        &\qquad\leq c_2\exp\Big\{-c_3\Big(\log\big(\min_{R\supseteq S, R\in\mathcal{D}}N_R\big)-1\Big)\log(\mathfrak{c}_{\supseteq S})\Big\},
    \end{align*}
    where $\mathsf{Z}_{a,S}(R)$ are independent standard Gaussian random variables,
    \begin{align*}
        \mathfrak{c}_\sigma = \min_{R\supseteq S,R\in\mathcal{D}} \frac{\sqrt{N_R \pi(a|R)(1-\pi(a|R))}}{\log(N_R)},
    \end{align*}
    and $c_1$--$c_3$ are universal constants such that $c_3$ could be made arbitrarily large by appropriate choice of $c_1$.
\end{lem}

\begin{lem}\label{lemma: taking hat off sigma - sigma(a|R)}
    Let $\xi_{1},\xi_{2}> 0$ with $\xi_{2}\to 0$. Then,
    \begin{align*}
    &\Prob\Bigg[ \max_{R\supseteq S, R\in\mathcal{D}}\Big| \frac{\hat{\pi}(a|R) - \pi(a|R)}{\sigma(a|R)} - \frac{\hat{\pi}(a|R) - \pi(a|R)}{\hat{\sigma}(a|R)} \Big|
    > \Big(\xi_{1} + c_4\log^{\frac{1}{2}}(\mathfrak{c}_{\supseteq S}) + c_1\frac{\log(\mathfrak{c}_{\supseteq S})}{ \mathfrak{c}_\sigma}\Big)\cdot \xi_{2} \Bigg]\\
    &\quad\qquad\leq c_2\exp\Big\{-c_3\Big(\log\big(\min_{R\supseteq S, R\in\mathcal{D}}N_R\big)-1\Big)\log(\mathfrak{c}_{\supseteq S})\Big\} + \exp\Big\{-\frac{\xi_1^2}{2}\Big\} + c_5\exp\Big\{ -c_6 \mathfrak{c}_\sigma^2\xi_{2}^2 + \log(\mathfrak{c}_{\supseteq S} ) \Big\},
    \end{align*}
    where $c_1$--$c_3$ are universal constants defined in Lemma \ref{lemma: strong approximation with normalization}, and $c_4$--$c_6$ are additional universal constants.
\end{lem}

The rest of the proof then follows from the observation that $\phi(a|S)\geq \max_{R\supseteq S,R\in\mathcal{D}}\ChoiProb(a|R)$, and therefore
\begin{align*}
&\mathbb{P}\Big[ \phi(a|S) < \underline{\widehat{\phi}}(a|S) \Big]\\
&\leq \mathbb{P}\Big[ \exists R\supseteq S,\ R\in\mathcal{D}: \hat{\ChoiProb}(a|R) - \mathrm{cv}(\alpha,\underline{\phi}(a|S))\cdot\hat{\sigma}(a|R) > \ChoiProb(a|R) \Big]\\
&=\mathbb{P}\Big[ \max_{R\supseteq S, R\in\mathcal{D}}\frac{\hat{\pi}(a|R) - \pi(a|R)}{\hat{\sigma}(a|R)} > \mathrm{cv}(\alpha,\underline{\phi}(a|S)) \Big] \\
&\leq \mathbb{P}\Bigg[
\max_{\substack{R\supseteq S\\R\in\mathcal D}}\mathsf{Z}_{a,S}(R)
>\mathrm{cv}(\alpha,\underline\phi(a|S))
-c_1\frac{\log(\mathfrak c_{\supseteq S})}{\mathfrak c_\sigma}\\
&\hspace{31mm}
-\left\{\xi_1+c_4\log^{1/2}(\mathfrak c_{\supseteq S})
+c_1\frac{\log(\mathfrak c_{\supseteq S})}{\mathfrak c_\sigma}\right\}\xi_2
\Bigg]\\
&\quad+c_2\exp\Bigg[-c_3
\left\{\log\left(\min_{\substack{R\supseteq S\\R\in\mathcal D}}N_R\right)-1\right\}
\log(\mathfrak c_{\supseteq S})\Bigg]\\
&\quad+\exp(-\xi_1^2/2)
+c_5\exp\{-c_6\mathfrak c_\sigma^2\xi_2^2+\log(\mathfrak c_{\supseteq S})\}.
\end{align*}
Apply the anti-concentration result from Lemma \ref{lemma: anti concentration}, which gives
\begin{align*}
&\mathbb{P}\Big[ \max_{R\supseteq S, R\in\mathcal{D}}\mathsf{Z}_{a,S}(R) > \mathrm{cv}(\alpha,\underline{\phi}(a|S)) - c_1\frac{\log(\mathfrak{c}_{\supseteq S})}{ \mathfrak{c}_\sigma} - \Big(\xi_{1} + c_4\log^{\frac{1}{2}}(\mathfrak{c}_{\supseteq S}) + c_1\frac{\log(\mathfrak{c}_{\supseteq S})}{ \mathfrak{c}_\sigma}\Big)\cdot \xi_{2} \Big] \\
&\leq \alpha + 4\Big(\xi_{1} + c_4\log^{\frac{1}{2}}(\mathfrak{c}_{\supseteq S}) + c_1\frac{\log(\mathfrak{c}_{\supseteq S})}{ \mathfrak{c}_\sigma}\Big)\xi_{2}\Big(\sqrt{2\log(\mathfrak{c}_{\supseteq S})} + 1\Big) + 4c_1\frac{\log(\mathfrak{c}_{\supseteq S})}{ \mathfrak{c}_\sigma}\Big(\sqrt{2\log(\mathfrak{c}_{\supseteq S})} + 1\Big).
\end{align*}
Collecting pieces, we have
\begin{align*}
&\mathbb{P}\Big[ \phi(a|S) < \underline{\widehat{\phi}}(a|S) \Big]\\
&\leq \alpha + 4\Big(\xi_{1} + c_4\log^{\frac{1}{2}}(\mathfrak{c}_{\supseteq S}) + c_1\frac{\log(\mathfrak{c}_{\supseteq S})}{ \mathfrak{c}_\sigma}\Big)\xi_{2}\Big(\sqrt{2\log(\mathfrak{c}_{\supseteq S})} + 1\Big) + 4c_1\frac{\log(\mathfrak{c}_{\supseteq S})}{ \mathfrak{c}_\sigma}\Big(\sqrt{2\log(\mathfrak{c}_{\supseteq S})} + 1\Big)\\
&\qquad + c_2\exp\Big\{-c_3\Big(\log\big(\min_{R\supseteq S, R\in\mathcal{D}}N_R\big)-1\Big)\log(\mathfrak{c}_{\supseteq S})\Big\} + \exp\Big\{-\frac{\xi_1^2}{2}\Big\} + c_5\exp\Big\{ -c_6 \mathfrak{c}_\sigma^2\xi_{2}^2 + \log(\mathfrak{c}_{\supseteq S} ) \Big\}.
\end{align*}
Therefore, by possibly enlarging the constants,
\begin{align*}
&\mathbb{P}\Big[ \phi(a|S) < \underline{\widehat{\phi}}(a|S) \Big]\\
    &\leq \alpha + c_1\Big(\xi_{1} + \log^{\frac{1}{2}}(\mathfrak{c}_{\supseteq S}) + \frac{\log(\mathfrak{c}_{\supseteq S})}{ \mathfrak{c}_\sigma}\Big)\xi_{2}\log^{\frac{1}{2}}(\mathfrak{c}_{\supseteq S}) \\
    &\qquad +c_1\frac{\log^{3/2}(\mathfrak c_{\supseteq S})}{\mathfrak c_\sigma}\\
    &\qquad + c_2\exp\Big\{-c_3\Big(\log\big(\min_{R\supseteq S, R\in\mathcal{D}}N_R\big)-1\Big)\log(\mathfrak{c}_{\supseteq S})\Big\} + \exp\Big\{-\frac{\xi_1^2}{2}\Big\} + c_5\exp\Big\{ -c_6 \mathfrak{c}_\sigma^2\xi_{2}^2 + \log(\mathfrak{c}_{\supseteq S} ) \Big\}.
\end{align*}

Now set
\begin{align*}
    \xi_1 \propto \log^{\frac{1}{2}}(\mathfrak{c}_{\sigma}),\quad \text{and}\quad \xi_2 \propto \frac{\log^{\frac{1}{2}}(\mathfrak{c}_{\supseteq S})\log^{\frac{1}{2}}(\mathfrak{c}_{\sigma})}{\mathfrak{c}_\sigma},
\end{align*}
The added strong-coupling term is dominated by the displayed rate because $\log(\mathfrak c_\sigma)$ diverges. Thus these choices give the desired bound
\begin{align*}
    \mathbb{P}\Big[ \phi(a|S) < \underline{\widehat{\phi}}(a|S) \Big] \leq \alpha + c \frac{\log^{\frac{3}{2}}(\mathfrak{c}_{\supseteq S})\log(\mathfrak{c}_{\sigma})}{\mathfrak{c}_\sigma}
\end{align*}
for some constant $c$.

\subsubsection{Proof of Lemma \ref{lemma: strong approx for all supersets of S}}

Apply Lemma \ref{lemma: strong approx for one menu} to the collection
\begin{align*}
    \mathcal D_S=\{R\in\mathcal D:R\supseteq S\}.
\end{align*}
Conditional on the menu assignments, the samples indexed by distinct $R$ are independent, so the menu-level couplings can be placed on one product probability space. Taking
\begin{align*}
    t_R=C\log(N_R)\log(\mathfrak c_{\supseteq S})
\end{align*}
and applying a union bound over the $\mathfrak c_{\supseteq S}$ menus yields exactly the probability and threshold displayed in the lemma, by the same calculation as in the proof of Lemma \ref{lemma: strong approx for all menus}.

\subsubsection{Proof of Lemma \ref{lemma: strong approximation with normalization}}

Let $\mathfrak n_R(a)$ denote the coordinate of $\mathfrak n_R$ corresponding to $\hat\pi(a|R)$ and set
\begin{align*}
    \mathsf{Z}_{a,S}(R)=\frac{\mathfrak n_R(a)}{\sigma(a|R)}.
\end{align*}
The variables $\mathsf{Z}_{a,S}(R)$ are independent standard normal variables because the menu-level Gaussian vectors are independent and $\Var(\mathfrak n_R(a))=\sigma^2(a|R)$. On the coupling event from the preceding lemma,
\begin{align*}
&\max_{R\in\mathcal D_S}
\left|\frac{\hat\pi(a|R)-\pi(a|R)}{\sigma(a|R)}-\mathsf{Z}_{a,S}(R)\right|\\
&\quad\leq
\max_{R\in\mathcal D_S}
\frac{\log(N_R)}{N_R\sigma(a|R)}
\max_{R\in\mathcal D_S}
\frac{N_R}{\log(N_R)}
\left\Vert\hat\bpi_R-\bpi_R-\mathfrak n_R\right\Vert_\infty\\
&\quad\leq
\frac{c_1\log(\mathfrak c_{\supseteq S})}{\mathfrak c_\sigma},
\end{align*}
where $N_R\sigma(a|R)=\sqrt{N_R\pi(a|R)(1-\pi(a|R))}$. The probability bound is inherited from the preceding lemma.

\subsubsection{Proof of Lemma \ref{lemma: taking hat off sigma - sigma(a|R)}}

For every $R\in\mathcal D_S$,
\begin{align*}
    |\hat\sigma^2(a|R)-\sigma^2(a|R)|
    \leq\frac{|\hat\pi(a|R)-\pi(a|R)|}{N_R}.
\end{align*}
Bernstein's inequality and a union bound therefore give
\begin{align*}
\Prob\left[
\max_{R\in\mathcal D_S}
\left|\frac{\hat\sigma(a|R)}{\sigma(a|R)}-1\right|>\xi_2
\right]
\leq c_5\exp\left\{-c_6\mathfrak c_\sigma^2\xi_2^2
+\log(\mathfrak c_{\supseteq S})\right\}.
\end{align*}
On the complementary event, for $\xi_2$ sufficiently small,
\begin{align*}
\left|\frac{\hat\pi(a|R)-\pi(a|R)}{\sigma(a|R)}
-\frac{\hat\pi(a|R)-\pi(a|R)}{\hat\sigma(a|R)}\right|
\leq2\xi_2
\left|\frac{\hat\pi(a|R)-\pi(a|R)}{\sigma(a|R)}\right|.
\end{align*}
Lemma \ref{lemma: strong approximation with normalization} and the Gaussian maximal inequality imply, except on events with probabilities given by the first two terms in the lemma's bound,
\begin{align*}
\max_{R\in\mathcal D_S}
\left|\frac{\hat\pi(a|R)-\pi(a|R)}{\sigma(a|R)}\right|
\leq \xi_1+c_4\log^{1/2}(\mathfrak c_{\supseteq S})
+c_1\frac{\log(\mathfrak c_{\supseteq S})}{\mathfrak c_\sigma}.
\end{align*}
Combining the three events and absorbing the factor two into the universal constants proves the result.

\subsection{Proof of Theorem 9}

The homogeneous-AOM procedures test moment inequalities associated with a candidate preference. H-LAO instead generally produces an identified set for a distribution of preferences. For the benchmark and dependence-robust models imposing Sequential Path Independence, $\mathcal D$ is suffix closed, as required by Proposition \ref{prop:observed-domain-hlao}. Proposition \ref{prop:nopi-hlao} does not require this closure. We use projection: construct a simultaneous confidence region for the primitive choice rule, retain every latent structure compatible with some choice rule in that region, and then project onto the preference or attention object of interest. This approach remains valid when reach is weak because it never divides by an estimated reach probability.

Let $p_0$ be the true choice rule and let $\mathcal C_{1-\alpha}$ be any confidence region satisfying
\begin{align*}
    \Prob[p_0\in\mathcal C_{1-\alpha}]\geq1-\alpha.
\end{align*}
For a candidate choice rule $p$, use the convention that $\mathcal T_{\mathcal D}(p)$ and $\mathcal T_{\mathcal D}^{\mathrm{dep}}(p)$ are empty when recovered attention is invalid. Define
\begin{align*}
    \widehat{\mathcal T}_{1-\alpha}^{\mathrm{ind}}
    &=\bigcup_{p\in\mathcal C_{1-\alpha}}
    \mathcal T_{\mathcal D}(p),\\
    \widehat{\mathcal T}_{1-\alpha}^{\mathrm{dep}}
    &=\bigcup_{p\in\mathcal C_{1-\alpha}}
    \mathcal T_{\mathcal D}^{\mathrm{dep}}(p),\\
    \widehat{\mathcal T}_{1-\alpha}^{\mathrm{noPI}}
    &=\bigcup_{p\in\mathcal C_{1-\alpha}}
    \mathcal T_{\mathcal D}^{\mathrm{noPI}}(p).
\end{align*}

On the event $p_0\in\mathcal C_{1-\alpha}$, sharp identification gives
\begin{align*}
    \tau_0\in\mathcal T_{\mathcal D}(p_0)
    \subseteq\widehat{\mathcal T}_{1-\alpha}^{\mathrm{ind}}
\end{align*}
under independence, and the same argument applies to the dependence-robust and no-SPI sharp sets. Because inclusion of $\tau_0$ is one common event, every projection of $\tau_0$ is covered simultaneously. Lemma \ref{lemma: simultaneous probability bands} supplies the finite-sample probability bound for the stated choice of $\mathcal C_{1-\alpha}$.

The same construction gives confidence regions for attention. For every candidate $p$, the product formula above recovers all reach probabilities without division. Taking their union over $p\in\mathcal C_{1-\alpha}$ therefore covers the true attention frequencies. On portions of the parameter space with positive terminal reach, one may also recursively compute $f$ for every candidate $p$ and project the resulting collection. Near zero reach, retaining the undivided choice equations is preferable because $f$ is then weakly or partially identified.\qed

\subsection{Proof of Corollary 5}

For a pair in which $a$ precedes $b$, abbreviate
\begin{align*}
    \underline y_{ba}&=\underline p(b|\{a,b\}),
    &\overline y_{ba}&=\overline p(b|\{a,b\}),\\
    \underline r_b(a,b)
    &=\left[1-\overline p(o^*|\{a,b\})\right]
      \left[1-\overline p(o^*|\{b\})\right],
    &\overline r_b(a,b)
    &=\left[1-\underline p(o^*|\{a,b\})\right]
      \left[1-\underline p(o^*|\{b\})\right].
\end{align*}
These are the exact extrema of the terminal reach over the rectangular cell bands. Define
\begin{align*}
    \widehat C^{\mathrm H}_{ba}
    =\left\{\theta\in[0,1]:
    [\underline r_b(a,b)\theta,\overline r_b(a,b)\theta]
    \cap[\underline y_{ba},\overline y_{ba}]\neq\emptyset
    \right\}.
\end{align*}

On the simultaneous event in Lemma \ref{lemma: simultaneous probability bands}, the true inside-choice probability belongs to $[\underline y_{ba},\overline y_{ba}]$ and the true terminal reach belongs to $[\underline r_b(a,b),\overline r_b(a,b)]$ for every pair. The Pairwise Preference Shares corollary in the main paper gives the undivided moment
\begin{align*}
    \pi(b|\{a,b\})
    =r_b(a,b)\Prob_\tau(b\succ a),
\end{align*}
including when $r_b(a,b)=0$. Hence every true pairwise share belongs to its displayed confidence set on the same event. The interval formula follows because $[\underline r\theta,\overline r\theta]$ intersects $[\underline y,\overline y]$ if and only if $\overline r\theta\geq\underline y$ and $\underline r\theta\leq\overline y$, with the stated zero-reach conventions.

Corollary 5 of the main paper is an Anderson--Rubin-type inversion: it tests the undivided structural equation rather than first estimating a ratio. If reach is bounded away from zero, both reach endpoints are eventually positive and the interval contracts at the rate of the primitive probability bands. If reach is zero, the data correctly leave the pairwise preference share unrestricted. Simultaneous correlated-Gaussian or bootstrap bands for the primitive probabilities can replace the Hoeffding bands to improve power while preserving the same inversion logic.\qed

\subsection{Proof of Corollary 6}

The finite-sample interval above obtains its uniform coverage from rectangular cell bands. Studentizing the same undivided structural equation provides a sharper complementary procedure under positive or weak but estimable reach, without introducing the unstable ratio estimator as the object being approximated.

Fix a pair in which $a$ precedes $b$ and abbreviate
\begin{align*}
    y_{ba}&=\pi(b|\{a,b\}),&
    e_{ab}&=\pi(o^*|\{a,b\}),&
    e_b&=\pi(o^*|\{b\}).
\end{align*}
For a candidate pairwise share $\theta\in[0,1]$, define
\begin{align*}
    g_{ba}(\theta)
    =y_{ba}-\theta(1-e_{ab})(1-e_b).
\end{align*}
Under independent H-LAO, the true share $\theta_{ba}=\Prob_\tau(b\succ a)$ satisfies $g_{ba}(\theta_{ba})=0$, including when terminal reach is zero. Let $N_{ab}=N_{\{a,b\}}$ and $N_b=N_{\{b\}}$. Conditional independence of menu subsamples and the multinomial covariance between the $b$ and outside-option cells in the binary menu give the delta-method variance
\begin{align*}
V_{ba}(\theta)
&=\frac{1}{N_{ab}}\Big[
y_{ba}(1-y_{ba})
+\theta^2(1-e_b)^2e_{ab}(1-e_{ab})
-2\theta(1-e_b)y_{ba}e_{ab}
\Big]\\
&\quad+
\frac{\theta^2(1-e_{ab})^2e_b(1-e_b)}{N_b}.
\end{align*}
Let $\widehat g_{ba}(\theta)$ and $\widehat V_{ba}(\theta)$ replace the three probabilities by their empirical counterparts. For a finite target collection $\mathcal B$ of ordered pairs, set
\begin{align*}
    c_{\mathcal B}(\alpha)
    =\Phi^{-1}\left(1-\frac{\alpha}{2|\mathcal B|}\right)
\end{align*}
and define
\begin{align*}
    \widehat C^{\mathrm S}_{ba}
    =\left\{\theta\in[0,1]:
    \widehat g_{ba}(\theta)^2
    \leq c_{\mathcal B}(\alpha)^2\widehat V_{ba}(\theta)
    \right\}.
\end{align*}
If $\widehat V_{ba}(\theta)=0$ for every $\theta\in[0,1]$, set $\widehat C^{\mathrm S}_{ba}=[0,1]$ by convention. Otherwise, the quadratic inequality includes any zero-variance candidate at which $\widehat g_{ba}(\theta)=0$. This is the quadratic inversion used in the software.

For a generic choice $W_{ab}$ from $\{a,b\}$ and $W_b$ from $\{b\}$, define the centered influence terms
\begin{align*}
    \psi_{ab}(W_{ab};\theta)
    &=\Indicator\{W_{ab}=b\}-y_{ba}
      +\theta(1-e_b)\big(\Indicator\{W_{ab}=o^*\}-e_{ab}\big),\\
    \psi_b(W_b;\theta)
    &=\theta(1-e_{ab})\big(\Indicator\{W_b=o^*\}-e_b\big).
\end{align*}
Their menu-sample averages have total variance $V_{ba}(\theta)$. The exact expansion around the population probabilities is
\begin{align*}
    \widehat g_{ba}(\theta)-g_{ba}(\theta)
    &=\frac{1}{N_{ab}}\sum_{i:Y_i=\{a,b\}}
      \psi_{ab}(y_i;\theta)
      +\frac{1}{N_b}\sum_{i:Y_i=\{b\}}\psi_b(y_i;\theta)\\
    &\quad-\theta(\widehat e_{ab}-e_{ab})(\widehat e_b-e_b).
\end{align*}

For each $(b,a)\in\mathcal B$, Corollary 6 of the main paper assumes
\begin{align*}
    N_{ab},N_b\to\infty,\qquad
    V_{ba}(\theta_{ba})>0,\qquad
    \frac{\widehat V_{ba}(\theta_{ba})}{V_{ba}(\theta_{ba})}
    \to_{\mathbb P}1,
\end{align*}
as well as
\begin{align*}
    &\frac{
      \mathbb E|\psi_{ab}(W_{ab};\theta_{ba})|^3/N_{ab}^2
      +\mathbb E|\psi_b(W_b;\theta_{ba})|^3/N_b^2}
      {V_{ba}(\theta_{ba})^{3/2}}
      \longrightarrow0,\\
    &\frac{
      \sqrt{e_{ab}(1-e_{ab})e_b(1-e_b)/(N_{ab}N_b)}}
      {\sqrt{V_{ba}(\theta_{ba})}}
      \longrightarrow0.
\end{align*}

At the true share, $g_{ba}(\theta_{ba})=0$. The first displayed condition is Lyapunov's condition for the two independent menu-sample sums in the exact expansion, so their sum divided by $\sqrt{V_{ba}(\theta_{ba})}$ converges to a standard normal distribution. The second condition and the independence of the two menu samples make the product remainder negligible relative to the same standard deviation. Variance consistency and Slutsky's theorem therefore imply
\begin{align*}
    \frac{\widehat g_{ba}(\theta_{ba})}
    {\sqrt{\widehat V_{ba}(\theta_{ba})}}
    \rightsquigarrow\mathcal N(0,1)
\end{align*}
for every pair. Variance consistency makes $\widehat V_{ba}(\theta_{ba})$ positive with probability tending to one; on this event, the quadratic inequality is equivalent to the studentized inequality. Hence the probability that a given true share is excluded is at most $\alpha/|\mathcal B|+o(1)$. The union bound over the fixed collection proves the simultaneous result. No division by terminal reach occurs.

The displayed variance follows from the gradient
\begin{align*}
    \left(1,\ \theta(1-e_b),\ \theta(1-e_{ab})\right)
\end{align*}
with respect to $(y_{ba},e_{ab},e_b)$, together with
\begin{align*}
    \Cov\big(\widehat y_{ba},\widehat e_{ab}\big)
    =-\frac{y_{ba}e_{ab}}{N_{ab}}.
\end{align*}
Primitive sufficient conditions include diverging effective counts in the relevant binary and singleton cells and the absence of asymptotic cancellation in $V_{ba}(\theta_{ba})$. These conditions allow reach to shrink, but they exclude a boundary at which every contribution to the moment variance vanishes.

The conditions of Corollary 6 of the main paper permit terminal reach to drift toward zero provided the undivided moment remains studentizable. At an exactly degenerate boundary, the finite-sample set $\widehat C^{\mathrm H}_{ba}$ remains valid. Bonferroni calibration avoids estimating cross-pair correlations and is the procedure covered by that corollary and used in the simulations.\qed

\subsection{Alternative Inference Procedures for Homogeneous AOM}\label{section: proposed econometric improvements}

This subsection develops valid alternatives to the feasible correlated-Gaussian procedure in the main paper. We first give a finite-sample projection method based on simultaneous confidence bands for the primitive choice probabilities. We then verify nondegeneracy for binary screening, develop a universal critical value that avoids covariance estimation, and explain how generalized moment selection and an alpha-split rule can be combined with these procedures. Throughout, we retain the notation and conditional sampling framework used above.

\subsubsection{Finite-Sample Projection from Choice-Probability Bands}

The Gaussian procedures exploit covariance information and are therefore our preferred methods when power is important. A useful comparison-free benchmark follows from concentration of the menu-level empirical probabilities themselves. Let $\mathcal Y(S)$ denote the observed outcome set: $\mathcal Y(S)=S$ for the homogeneous-AOM data and $\mathcal Y(S)=S\cup\{o^*\}$ when the H-LAO outside option is observed. Thus, for H-LAO, the choice-data framework in the main paper is extended only by allowing the observed outcome to equal $o^*$. For $\alpha\in(0,1)$, let
\begin{align*}
    d_{\mathcal D}=\sum_{S\in\mathcal D}|\mathcal Y(S)|
\end{align*}
be the number of menu--outcome cells. For $N_S\geq1$, define
\begin{align*}
    \epsilon_S(\alpha)
    =\sqrt{\frac{\log(2d_{\mathcal D}/\alpha)}{2N_S}},
\end{align*}
and, for every $a\in\mathcal Y(S)$,
\begin{align*}
    \underline p(a|S)
    &=\max\{0,\widehat\pi(a|S)-\epsilon_S(\alpha)\},\\
    \overline p(a|S)
    &=\min\{1,\widehat\pi(a|S)+\epsilon_S(\alpha)\}.
\end{align*}
Let $\mathcal C^{\mathrm H}_{1-\alpha}$ contain all candidate choice rules $p$ satisfying
\begin{align*}
    \underline p(a|S)&\leq p(a|S)\leq\overline p(a|S),\\
    \sum_{a\in\mathcal Y(S)}p(a|S)&=1
\end{align*}
for every observed menu and outcome. The superscript $\mathrm H$ refers to the Hoeffding construction, not to preference heterogeneity.

\begin{lem}[Simultaneous menu-probability region]\label{lemma: simultaneous probability bands}
Under the choice-data sampling framework of the main paper, conditionally on the realized menu assignments,
\begin{align*}
    \Prob\big[\pi\in\mathcal C^{\mathrm H}_{1-\alpha}\big]
    \geq1-\alpha.
\end{align*}
The result is finite sample, allows zero cell probabilities, and imposes no covariance nondegeneracy condition.
\end{lem}

\begin{proof}
For each menu--outcome cell, Hoeffding's inequality gives
\begin{align*}
    \Prob\left(
    |\widehat\pi(a|S)-\pi(a|S)|>\epsilon_S(\alpha)
    \right)
    \leq2\exp\{-2N_S\epsilon_S^2(\alpha)\}
    =\frac{\alpha}{d_{\mathcal D}}.
\end{align*}
A union bound over the $d_{\mathcal D}$ cells proves simultaneous coverage. The true probabilities satisfy the adding-up restrictions automatically.
\end{proof}

For homogeneous AOM, define the finite-sample projection confidence set
\begin{align*}
    \widehat{\mathcal P}^{\mathrm H}_{1-\alpha}
    =\left\{\succ:\text{$p$ satisfies $\succ$-Regularity for some }
    p\in\mathcal C^{\mathrm H}_{1-\alpha}\right\}.
\end{align*}

\begin{prop}[Finite-sample inference for homogeneous AOM]\label{prop: finite sample homogeneous projection}
Let $\mathcal P_\pi$ be the sharp identified set of preferences under homogeneous AOM. Then
\begin{align*}
    \Prob\left[
    \mathcal P_\pi\subseteq
    \widehat{\mathcal P}^{\mathrm H}_{1-\alpha}
    \right]
    \geq1-\alpha.
\end{align*}
For a fixed $\succ$, membership in $\widehat{\mathcal P}^{\mathrm H}_{1-\alpha}$ is a linear feasibility problem. Projecting the same feasible set through the lower and upper formulas in Theorem 2 of the main paper gives simultaneous finite-sample-valid outer intervals for every attention frequency associated with $\succ$.
\end{prop}

\begin{proof}
On the event $\pi\in\mathcal C^{\mathrm H}_{1-\alpha}$, every $\succ\in\mathcal P_\pi$ is feasible in the definition of $\widehat{\mathcal P}^{\mathrm H}_{1-\alpha}$ by taking $p=\pi$. Lemma \ref{lemma: simultaneous probability bands} gives the first claim. The simplex, band, and $\succ$-Regularity restrictions are linear in $p$. On the same simultaneous event, Theorem 2 of the main paper places every rationalizing attention frequency between its lower and upper formulas evaluated at the true $\pi$; optimizing those formulas over the feasible candidate probabilities therefore preserves coverage for all coordinates.
\end{proof}

This construction provides a degeneracy-robust, finite-sample complement to the high-dimensional Gaussian method and avoids both Gaussian approximation and covariance comparison. Rectangular concentration bands prioritize coverage over power, making them especially useful as a finite-sample benchmark and as the common primitive confidence region for H-LAO inference below.

\subsubsection{Feasible Correlated-Gaussian Critical Values}

The default procedure in the main paper uses the feasible correlated-Gaussian critical value $\widehat{\mathrm{cv}}(\alpha,\succ)$. This critical value simulates the distribution of
\begin{align*}
    \widehat{\mathsf{T}}^{\mathtt{G}}(\succ)
    =\max\left\{0,\max_{1\leq\ell\leq\mathfrak c_\succ}\hat Z_\ell\right\},
\end{align*}
where $\hat{\bz}|\text{Data}\sim\mathcal{N}(0,\hat{\boldsymbol{\Omega}})$. Its advantage is power: it uses the estimated dependence structure among the inequality comparisons, and hence can be much less conservative than a universal bound based only on the number of inequalities.

The proof of Theorem 7 passes from $\bz\sim\mathcal{N}(0,\boldsymbol{\Omega})$ to $\hat{\bz}\sim\mathcal{N}(0,\hat{\boldsymbol{\Omega}})$ using Lemma \ref{lemma: gaussian comparison}. Its fully general treatment accommodates singular or nearly singular covariance matrices and produces the extra square-root term in the final size-distortion bound. The refinements below sharpen that rate under nondegeneracy or avoid the comparison step altogether.

\subsubsection{Binary-Comparison Nondegeneracy}

The binary-comparison statistic has a special covariance structure. Fix two alternatives $a$ and $b$, and consider the comparisons
\begin{align*}
    D(a|S,b)=\pi(a|S)-\pi(a|b),\qquad S\in\mathcal{D}_{ab}.
\end{align*}
The only shared sampling component across different $S$ is the binary comparison $\widehat{\pi}(a|b)$. This produces a diagonal-plus-rank-one correlation matrix.

\begin{lem}[Binary-comparison covariance]\label{lemma: binary comparison covariance}
Let $\boldsymbol{\Omega}_{ab}$ be the correlation matrix of
\begin{align*}
    \left\{
    \frac{\widehat{D}(a|S,b)-D(a|S,b)}{\sigma(a|S,b)}:\ S\in\mathcal{D}_{ab}
    \right\}.
\end{align*}
Then
\begin{align*}
    \boldsymbol{\Omega}_{ab}
    =
    \operatorname{diag}\left(
    \frac{\sigma^2(a|S)}{\sigma^2(a|S,b)}
    \right)_{S\in\mathcal{D}_{ab}}
    + vv',
    \qquad
    v_S=\frac{\sigma(a|b)}{\sigma(a|S,b)}.
\end{align*}
Consequently,
\begin{align*}
    \lambda_{\min}(\boldsymbol{\Omega}_{ab})
    \geq
    \min_{S\in\mathcal{D}_{ab}}
    \frac{\sigma^2(a|S)}{\sigma^2(a|S,b)}.
\end{align*}
\end{lem}

\begin{proof}
Conditional on the menu sample sizes, the estimators from different choice problems are independent. Let $Z_S$ and $Z_b$ be mean-zero Gaussian random variables with variances $\sigma^2(a|S)$ and $\sigma^2(a|b)$, respectively, with $Z_b$ independent of all $Z_S$ and the $Z_S$ mutually independent across $S$. The oracle Gaussian fluctuation for the comparison $D(a|S,b)$ can be represented as $Z_S-Z_b$.

For $S\neq T$,
\begin{align*}
    \Cov(Z_S-Z_b,Z_T-Z_b)=\sigma^2(a|b),
\end{align*}
while
\begin{align*}
    \Var(Z_S-Z_b)=\sigma^2(a|S)+\sigma^2(a|b)=\sigma^2(a|S,b).
\end{align*}
After normalizing by $\sigma(a|S,b)$, the stated diagonal-plus-rank-one representation follows. Since $vv'$ is positive semidefinite, Weyl's inequality gives the lower bound on the minimum eigenvalue.
\end{proof}

\begin{coro}[Balanced standard errors imply nondegeneracy]\label{coro: binary comparison nondegenerate}
If there exists a constant $C<\infty$ such that
\begin{align*}
    C^{-1}\leq \frac{\sigma(a|S)}{\sigma(a|b)}\leq C
    \qquad\text{for all }S\in\mathcal{D}_{ab},
\end{align*}
then
\begin{align*}
    \lambda_{\min}(\boldsymbol{\Omega}_{ab})\geq \frac{1}{1+C^2}.
\end{align*}
\end{coro}

\begin{proof}
The ratio
\begin{align*}
    \frac{\sigma^2(a|S)}{\sigma^2(a|S,b)}
    =
    \frac{\sigma^2(a|S)}{\sigma^2(a|S)+\sigma^2(a|b)}
\end{align*}
is bounded below by $1/(1+C^2)$ under the stated balance condition. The result follows from Lemma \ref{lemma: binary comparison covariance}.
\end{proof}

This result verifies the key nondegeneracy condition needed for sharper Gaussian comparison in binary screening. Under the balance condition, one may replace Lemma \ref{lemma: gaussian comparison} with Theorem 2.3 of \citet{lopes2022central}, whose covariance-error dependence is nearly linear. The sharper fixed-pair bound in Corollary 4 makes this improvement formal. It does not assert the same eigenvalue bound for a joint statistic stacking multiple ordered pairs.

\subsubsection{Universal Critical Values Without Gaussian Comparison}

A comparison-free alternative avoids estimating the correlation matrix altogether. Define
\begin{align*}
    c_B(\alpha,\mathfrak c_\succ)
    =\Phi^{-1}(1-\alpha/\mathfrak c_\succ).
\end{align*}
The universal test rejects $\mathsf{H}_0$ when
\begin{align*}
    \mathsf{T}(\succ)>c_B(\alpha,\mathfrak c_\succ).
\end{align*}
The critical value depends only on the number of comparisons. It is typically more conservative than $\widehat{\mathrm{cv}}(\alpha,\succ)$, but it does not require the feasible Gaussian vector $\hat{\bz}$ or Lemma \ref{lemma: gaussian comparison}.

\begin{lem}[Universal critical values]\label{lemma: universal critical values}
Fix $\alpha\in(0,1/2)$ and a preference $\succ$ with $\mathfrak{c}_{\succ}\geq1$. Suppose that, under $\mathsf{H}_0$,
\begin{align*}
    \Prob\left[
    \left\Vert \bR_\succ(\hat{\bpi}-\bpi)\oslash\hat{\bsigma}-\bz\right\Vert_\infty>\eta
    \right]\leq \rho,
\end{align*}
where $\bz$ is the oracle Gaussian vector with unit variances. Then
\begin{align*}
    \Prob\left[\mathsf{T}(\succ)>c_B(\alpha,\mathfrak c_\succ)\right]
    \leq
    \alpha+\rho+4\eta\left(\sqrt{2\log(\mathfrak c_\succ)}+1\right).
\end{align*}
\end{lem}

\begin{proof}
Under $\mathsf{H}_0$, the population component $\bR_\succ\bpi\oslash\hat{\bsigma}$ is weakly negative coordinatewise. Therefore, on the event
\begin{align*}
    \left\Vert \bR_\succ(\hat{\bpi}-\bpi)\oslash\hat{\bsigma}-\bz\right\Vert_\infty\leq\eta,
\end{align*}
the rejection event implies
\begin{align*}
    \max_{1\leq\ell\leq\mathfrak c_\succ}Z_\ell
    >c_B(\alpha,\mathfrak c_\succ)-\eta.
\end{align*}
Thus,
\begin{align*}
    \Prob\left[\mathsf{T}(\succ)>c_B(\alpha,\mathfrak c_\succ)\right]
    \leq
    \Prob\left[\max_{1\leq\ell\leq\mathfrak c_\succ}Z_\ell
    >c_B(\alpha,\mathfrak c_\succ)-\eta\right]+\rho.
\end{align*}
The first term is bounded by
\begin{align*}
    \Prob\left[\max_{1\leq\ell\leq\mathfrak c_\succ}Z_\ell
    >c_B(\alpha,\mathfrak c_\succ)\right]
    +
    \Prob\left[\left|\max_{1\leq\ell\leq\mathfrak c_\succ}Z_\ell
    -c_B(\alpha,\mathfrak c_\succ)\right|\leq\eta\right].
\end{align*}
By the union bound,
\begin{align*}
    \Prob\left[\max_{1\leq\ell\leq\mathfrak c_\succ}Z_\ell
    >c_B(\alpha,\mathfrak c_\succ)\right]
    \leq \mathfrak c_\succ\{1-\Phi(c_B(\alpha,\mathfrak c_\succ))\}=\alpha.
\end{align*}
The anti-concentration bound in Lemma \ref{lemma: anti concentration} controls the second term and gives the stated result.
\end{proof}

Taking $\eta$ and $\rho$ from the strong approximation proof of Theorem 6 yields a valid test whose excess size is governed by the strong-coupling error, up to the anti-concentration factor. The improvement is conceptual and theoretical: the extra square-root loss from estimated-covariance Gaussian comparison is optional. The tradeoff is power, because $c_B(\alpha,\mathfrak c_\succ)$ protects against the worst case using only the number of inequalities.

\subsubsection{Generalized Moment Selection}\label{subsubsection: gms supplemental}

The statistic $\mathsf{T}(\succ)$ can involve many inequalities that are far inside the null. These slack inequalities increase the dimension of the max statistic and can make the critical value conservative. Generalized moment selection addresses this by using the data to focus the critical-value calculation on inequalities that are close to binding.

The follow-up simulation pipeline records two critical values from the same Gaussian draws. The first is the zero-centered least-favorable critical value, labeled \texttt{LF} in the software, which retains every inequality and directly implements Theorem 7 of the main paper. The second is the generalized moment-selection (GMS) refinement in equation (4.4) of \citet{Andrews-Soares_2010_ECMA}. If $N$ is the total sample size, the implementation sets \texttt{MNRatioGMS}$=1/\log(N)$ and multiplies each estimated moment entering the Gaussian recentering by $\sqrt{\texttt{MNRatioGMS}}=1/\sqrt{\log(N)}$, truncating positive recentering terms at zero. Thus, binding and local-to-binding inequalities remain protected while deeply slack inequalities receive a negative recentering.

Moment selection can reduce the effective contribution of slack inequalities to the max statistic and thereby improve power. It does not remove the Gaussian-comparison step. The all-inequalities \texttt{LF} procedure is the formal Theorem 7 baseline; GMS is reported as an additional numerical refinement whose validity relies on the safeguards in \citet{Andrews-Soares_2010_ECMA}.

\subsubsection{Hybrid Critical Values}

The feasible correlated-Gaussian and universal critical values can also be combined by splitting the nominal level.

\begin{lem}[Alpha-split hybrid critical values]\label{lemma: hybrid critical values}
Suppose two critical values $c_1$ and $c_2$ satisfy, under $\mathsf{H}_0$,
\begin{align*}
    \Prob[\mathsf{T}(\succ)>c_k]\leq \alpha_k+\delta_k,
    \qquad k=1,2.
\end{align*}
Then
\begin{align*}
    \Prob[\mathsf{T}(\succ)>\min\{c_1,c_2\}]
    \leq
    \alpha_1+\alpha_2+\delta_1+\delta_2.
\end{align*}
\end{lem}

\begin{proof}
The event $\{\mathsf{T}(\succ)>\min\{c_1,c_2\}\}$ is contained in
\begin{align*}
    \{\mathsf{T}(\succ)>c_1\}\cup\{\mathsf{T}(\succ)>c_2\}.
\end{align*}
The result follows from the union bound.
\end{proof}

For example, take $c_1=\widehat{\mathrm{cv}}(\alpha_G,\succ)$ and $c_2=c_B(\alpha_B,\mathfrak{c}_{\succ})$ with $\alpha_G+\alpha_B=\alpha$. The hybrid rule rejects when
\begin{align*}
    \mathsf{T}(\succ)>\min\{\widehat{\mathrm{cv}}(\alpha_G,\succ),c_B(\alpha_B,\mathfrak{c}_{\succ})\}.
\end{align*}
This rule is valid up to the sum of the two approximation errors. It allows the correlated-Gaussian test to carry most of the power while preserving a comparison-free route to rejection.

\subsection{Alternative Inference Procedures for Heterogeneous H-LAO}\label{section: hlao inference}

This subsection develops additional H-LAO inference procedures beyond those stated in the main paper. It first gives finite-sample linear-programming procedures under preference--stopping dependence and without Sequential Path Independence, and then discusses how projection and Gaussian procedures can be combined.

The exact independent projection preserves Sequential Path Independence and preference--stopping independence through polynomial joint feasibility equations. The dependence-robust projection is likewise nonlinear when the exact product formula for reach is imposed while $p$ varies. The linear outer construction below delivers transparent computation and finite-sample coverage; its relaxation disappears as the primitive probability bands contract.

\subsubsection{A Dependence-Robust Linear-Programming Procedure}

For larger preference events, an LP implementation is available after forming simultaneous outer bands for the recovered reach probabilities. For $S=\{a_{s_1},\dots,a_{s_{|S|}}\}$, write $S_{k:}=\{a_{s_k},\dots,a_{s_{|S|}}\}$ and define
\begin{align*}
    \underline\varphi_j(S)
    &=\prod_{k=1}^j
      \left[1-\overline p(o^*|S_{k:})\right],\\
    \overline\varphi_j(S)
    &=\prod_{k=1}^j
      \left[1-\underline p(o^*|S_{k:})\right].
\end{align*}
On the event in Lemma \ref{lemma: simultaneous probability bands}, every true recovered reach probability lies in the corresponding interval.

Consider variables $v_j(S)$, $m_j(S)$, $p(a|S)$, $\tau(\succ)$, and $q_S(\succ,j)$. Let $\widehat{\mathcal Q}^{\mathrm{LP}}_{1-\alpha}$ be defined by
\begin{align*}
    &\underline\varphi_j(S)\leq v_j(S)\leq\overline\varphi_j(S),
      \qquad v_{|S|+1}(S)=0,\\
    &m_0(S)=1-v_1(S),
      \qquad m_j(S)=v_j(S)-v_{j+1}(S)\geq0,\\
    &v_j(S)\geq v_k(R)
      \quad\text{whenever }a_{s_j}=a_{r_k}\in S\subseteq R,\\
    &q_S(\succ,j)\geq0,
      \qquad \sum_{\succ\in\mathcal P}q_S(\succ,j)=m_j(S),\\
    &\sum_{j=0}^{|S|}q_S(\succ,j)=\tau(\succ),
      \qquad \sum_{\succ\in\mathcal P}\tau(\succ)=1,\\
    &\underline p(a|S)\leq p(a|S)\leq\overline p(a|S),\\
    &p(a|S)=\sum_{j=1}^{|S|}\sum_{\succ\in\mathcal P}
      \Indicator\{a\in T_j(S),\ a\text{ is $\succ$-best in }T_j(S)\}
      q_S(\succ,j)
\end{align*}
for all relevant menus, positions, preferences, and inside alternatives. The third line imposes attention overload, matching the same alternative across the two inherited list positions.

\begin{prop}[Finite-sample LP inference for dependence-robust H-LAO]\label{prop: dependence robust lp inference}
Suppose the dependence-robust H-LAO model in Proposition \ref{prop:dependent-hlao} holds with preference distribution $\tau_0$. Then, with probability at least $1-\alpha$, the true latent structure is feasible in $\widehat{\mathcal Q}^{\mathrm{LP}}_{1-\alpha}$. Consequently, for every collection of preference events $A$, the intervals obtained by minimizing and maximizing
\begin{align*}
    \sum_{\succ\in A}\tau(\succ)
\end{align*}
over $\widehat{\mathcal Q}^{\mathrm{LP}}_{1-\alpha}$ cover all corresponding true event probabilities simultaneously. Each endpoint is the solution to a linear program. The conclusion also covers the independent H-LAO model, viewed as a subclass of the dependence-robust model.
\end{prop}

\begin{proof}
On the simultaneous probability-band event, set $v_j(S)$ equal to the true recovered reach, $m_j(S)$ equal to the true prefix-stopping probability, $p=\pi$, and $(\tau,q)$ equal to the true marginal preference distribution and menu-specific preference--stopping couplings. The product bounds contain every $v_j(S)$. Attention validity gives the monotonicity and attention-overload constraints, while Proposition \ref{prop:dependent-hlao} gives the remaining marginal and choice equations. Thus the true structure is feasible. Lemma \ref{lemma: simultaneous probability bands} bounds the probability of this common event, and linear projection gives simultaneous coverage of all event probabilities.
\end{proof}

The LP uses interval restrictions on reach rather than imposing the estimated nonlinear product equations exactly. It is therefore a finite-sample outer procedure. As every $N_S$ diverges, the primitive bands and the induced reach bands contract, so this relaxation disappears. Under benchmark independence, adding $q_S(\succ,j)=\tau(\succ)m_j(S)$ recovers the exact restriction but makes the program bilinear. The explicit pairwise procedure above avoids that difficulty for the most empirically useful preference margins; exact projection or global polynomial optimization remains available for more general independent-H-LAO events.

\subsubsection{A Linear Projection Procedure without Sequential Path Independence}

The robustness envelope in Proposition \ref{prop:nopi-hlao} permits a simpler exact projection over the primitive probability region because its prefix masses are latent and need not satisfy nonlinear product equalities. On any observed-menu domain $\mathcal D$, let $\widehat{\mathcal Q}^{\mathrm{noPI}}_{1-\alpha}$ contain variables $p(a|S)$ for $a\in S\cup\{o^*\}$, $m_j(S)$, $q_S(\succ,j)$, and $\tau(\succ)$ satisfying
\begin{align*}
    &\underline p(a|S)\leq p(a|S)\leq\overline p(a|S),
      \qquad \sum_{a\in S\cup\{o^*\}}p(a|S)=1,\\
    &m_j(S)\geq0,\qquad m_0(S)=p(o^*|S),
      \qquad \sum_{j=0}^{|S|}m_j(S)=1,\\
    &q_S(\succ,j)\geq0,\qquad
      \sum_{\succ\in\mathcal P}q_S(\succ,j)=m_j(S),
      \qquad \sum_{j=0}^{|S|}q_S(\succ,j)=\tau(\succ),\\
    &p(a|S)=\sum_{j=1}^{|S|}\sum_{\succ\in\mathcal P}
      \Indicator\{a\in T_j(S),\ a\text{ is $\succ$-best in }T_j(S)\}
      q_S(\succ,j),
\end{align*}
for all relevant indices, together with
\begin{align*}
    \sum_{\ell=j}^{|S|}m_\ell(S)
    \geq
    \sum_{\ell=k}^{|R|}m_\ell(R)
    \qquad
    \text{whenever }a_{s_j}=a_{r_k}\in S\subseteq R.
\end{align*}

\begin{prop}[Finite-sample LP inference without Sequential Path Independence]\label{prop: nopi lp inference}
Suppose the model in Proposition \ref{prop:nopi-hlao} holds with marginal preference distribution $\tau_0$. Then, with probability at least $1-\alpha$, its true choice rule and latent structure are feasible in $\widehat{\mathcal Q}^{\mathrm{noPI}}_{1-\alpha}$. Consequently, minimizing and maximizing any collection of preference-event probabilities over this set gives simultaneous finite-sample confidence bounds. Every endpoint is a linear program, no suffix-closure or positive-reach condition is required, and the procedure exactly projects the rectangular cell region with menu-specific adding-up onto the path-independence-robust identified set.
\end{prop}

\begin{proof}
On the simultaneous-band event, set $p=\pi$ and set $(\tau,q,m)$ equal to the true stable preference marginal, menu-specific preference--stopping couplings, and prefix masses. The primitive bands and adding-up restrictions contain $p$. Proposition \ref{prop:nopi-hlao} gives all remaining equalities and inequalities, so the true structure is feasible. The event has probability at least $1-\alpha$ by Lemma \ref{lemma: simultaneous probability bands}. Because the displayed constraints are exactly those defining $\mathcal T_{\mathcal D}^{\mathrm{noPI}}(p)$ for some $p$ in the primitive region, optimizing a linear event probability performs the stated projection without further relaxation.
\end{proof}

\subsubsection{Relation to the Gaussian Procedures}

We implement a hybrid primitive-probability region as follows. Let $K=d_{\mathcal D}$ be the number of observed menu--outcome cells and set the screening threshold to $b=5$. For a row-i.i.d. cell $(a,S)$, write $X_{aS}=N_S\widehat\pi(a|S)$ and let $\widehat s_{aS}$ denote its estimated standard error. The cell is regular when $X_{aS}\geq b$, $N_S-X_{aS}\geq b$, and $\widehat s_{aS}>\sqrt{\varepsilon_{\mathrm{mach}}}$; it is boundary or degenerate when $\widehat s_{aS}\leq\sqrt{\varepsilon_{\mathrm{mach}}}$, and sparse otherwise. Let $\widehat{\mathcal R}$ and $\widehat{\mathcal F}$ collect the regular and fallback cells. If both sets are nonempty, assign $\alpha_{\mathrm G}=\alpha_{\mathrm F}=\alpha/2$; if only one is nonempty, it receives the full error budget $\alpha$.

For $k\in\widehat{\mathcal R}$, let $\widehat c_{\mathrm G}$ be the conditional $(1-\alpha_{\mathrm G})$ quantile of
\begin{align*}
    \max_{k\in\widehat{\mathcal R}}
    \frac{|Z_k|}{\widehat s_k},
    \qquad
    Z\mid\text{data}\sim\mathcal N(0,\widehat V),
\end{align*}
where $\widehat V$ is the plug-in block-multinomial covariance matrix. The regular-cell interval is $[\widehat\pi_k-\widehat c_{\mathrm G}\widehat s_k,\widehat\pi_k+\widehat c_{\mathrm G}\widehat s_k]$, truncated to $[0,1]$. For $k=(a,S)\in\widehat{\mathcal F}$, put $\eta=\alpha_{\mathrm F}/K$. Write the two-sided Clopper--Pearson interval as $I^{\mathrm{CP}}_{aS}=[L_{aS},U_{aS}]$, where its lower endpoint is
\begin{align*}
    L_{aS}
    =\begin{cases}
      0, & X_{aS}=0,\\
      B^{-1}_{X_{aS},\,N_S-X_{aS}+1}(\eta/2), & X_{aS}>0,
    \end{cases}
\end{align*}
and its upper endpoint is
\begin{align*}
    U_{aS}
    =\begin{cases}
      1, & X_{aS}=N_S,\\
      B^{-1}_{X_{aS}+1,\,N_S-X_{aS}}(1-\eta/2), & X_{aS}<N_S,
    \end{cases}
\end{align*}
where $B^{-1}_{u,v}$ is the quantile function of a $\mathrm{Beta}(u,v)$ variable. The hybrid region contains every vector whose cells lie in their assigned intervals and whose probabilities add to one within each menu.

The fallback component has simultaneous finite-sample coverage at least $1-\alpha_{\mathrm F}$ by a union bound. The Gaussian component, and hence a hybrid region containing regular cells, is an asymptotic procedure whose coverage is subject to Gaussian-approximation, covariance-estimation, and simulation errors. If no Gaussian cell is used, the exact-binomial region is finite-sample valid. This coverage distinction concerns the primitive region; projection through the model-feasible sets preserves the coverage status of the region supplied to it.

The finite-sample bands emphasize robustness, whereas the correlated-Gaussian machinery in the main paper emphasizes power in high-dimensional problems. They can be combined. Under terminal reach bounded away from zero, the pairwise ratio is a smooth function of $\pi(b|\{a,b\})$, $\pi(o^*|\{a,b\})$, and $\pi(o^*|\{b\})$. Under weaker reach, Corollary 6 of the main paper applies the Gaussian approximation to the undivided moment instead. In both cases, simultaneous Gaussian or multiplier-bootstrap inference can use the dependence across pairs induced by shared menus. At degenerate zero reach, finite-sample Hoeffding or exact-binomial projection remains valid. Thus, the support condition determines the most informative implementation, not the validity of the underlying identified-set analysis.

The studentized moments also suggest a covariance-aware Gaussian-max calibration: stack the moments, estimate their candidate-dependent correlation matrix from the menu-level multinomial covariance blocks, and simulate the corresponding quantile. A fully uniform theory additionally entails joint approximation, correlation consistency, quantile continuity, simulation-error control, and uniform inversion over the candidate set. Our formal result and simulations therefore use the theorem-backed Bonferroni calibration, while the projection method already exploits covariance-aware Gaussian bands for the primitive probabilities.

\subsection{Cluster-Robust Inference for Repeated Choices}\label{section: cluster robust inference}

The preceding procedures extend to repeated-choice data in which observations are independent across clusters but may be arbitrarily dependent within a cluster. Let $g=1,\ldots,G$ index clusters, let $n_{gS}$ be cluster $g$'s number of observations from menu $S$, and let $N_S=\sum_g n_{gS}$. For observation $i$, write $S_i$ for the menu and $y_i$ for the chosen alternative. For every observed menu--outcome cell $(a,S)$, define the cluster influence contribution
\begin{align*}
    \widehat\psi_{g,aS}
    =\frac{1}{N_S}\sum_{i:g_i=g,\,S_i=S}
    \left\{\Indicator(y_i=a)-\widehat\pi(a|S)\right\}.
\end{align*}
After stacking all cells, the cluster covariance estimator and its multiplier analogue are
\begin{align*}
    \widehat V_{\mathrm{cl}}
    &=\frac{G}{G-1}\sum_{g=1}^G
      \widehat\psi_g\widehat\psi_g',
    &
    \widehat Z^*
    &=\sqrt{\frac{G}{G-1}}
      \sum_{g=1}^G\xi_g\widehat\psi_g,
      \qquad \xi_g\stackrel{\mathrm{iid}}{\sim}\mathcal N(0,1).
\end{align*}
Under the cluster-level analogues of the moment, covariance-consistency, and Gaussian-approximation conditions used above, the common multiplier maximum yields asymptotically valid simultaneous primitive-probability bands as the number of independent clusters grows. The same draws can instead be applied directly to the AOM inequalities and regular H-LAO diagnostic moments. Additional subject-level statistics, such as elicited preference reports, can be stacked with the choice cells when constructing confidence intervals for differences. Thus one joint cluster process supports the AOM ranking tests, H-LAO specification diagnostics, pairwise preference intervals, and projection or linear-programming bounds.

For clustered data, let $X_{g,aS}$ be cluster $g$'s number of choices of $a$ from $S$, and define
\begin{align*}
    G^+_{aS}&=\sum_{g=1}^G\Indicator\{X_{g,aS}>0\},
    &
    G^-_{aS}&=\sum_{g=1}^G\Indicator\{X_{g,aS}<n_{gS}\}.
\end{align*}
A cell is regular when $G^+_{aS}\geq b$, $G^-_{aS}\geq b$, and its cluster standard error exceeds $\sqrt{\varepsilon_{\mathrm{mach}}}$; it is boundary or degenerate when that standard error does not exceed the cutoff, and sparse otherwise. For regular cells, the multiplier analogue $\widehat Z^*$ replaces $Z$ in the Gaussian maximum above. Fallback cells use the cluster-Hoeffding half-width
\begin{align*}
    \left\{
    \frac{\log(2K/\alpha_{\mathrm F})}{2}
    \sum_{g=1}^G\left(\frac{n_{gS}}{N_S}\right)^2
    \right\}^{1/2}.
\end{align*}
This bound treats clusters as independent bounded contributions and remains valid under unrestricted dependence among observations within a cluster. The same error-budget rule applies: multiplier and cluster-Hoeffding cells each receive $\alpha/2$ when both occur, and the sole component receives $\alpha$ otherwise. Accordingly, the cluster-Hoeffding component is finite-sample valid, while a region containing multiplier cells has the asymptotic coverage status of the multiplier approximation. The resulting joint primitive region is propagated through the same AOM and H-LAO transformations and optimization problems as in the independent-row analysis.

\section{Implementation and Computational Scalability}\label{section: computational implementation}

This section collects the implementation details that are common across empirical designs. The numerical benchmarks vary the size of the universe and menu domain while holding the population choice system fixed and treating it as observed. They therefore measure the cost of constructing and solving the paper's optimization problems, rather than statistical performance under a particular simulation design. Section SA-6 separately reports finite-sample behavior and execution times for the Monte Carlo designs used in the paper.

\subsection{Homogeneous AOM}

For a fixed candidate preference, the econometric procedures in Section 4 of the main paper construct and evaluate the $\mathfrak c_\succ$ observed $\succ$-Regularity inequalities. For each observed pair $T\subset S$ with $|T|=\ell\geq2$, the comparison associated with the $\succ$-worst alternative in $T$ reduces to $\pi(a|S)\leq1$ and is omitted; singleton $T$ contributes no nonredundant comparison. Hence, with complete menu data,
\begin{align*}
    \mathfrak c_\succ
    =\sum_{k=3}^{|X|}\binom{|X|}{k}
    \sum_{\ell=2}^{k-1}\binom{k}{\ell}(\ell-1).
\end{align*}
For $|X|=6$, this formula gives 664 comparisons, whereas retaining every $a\in T\subset S$ gives 1,266 regularity rows.
The subset relations depend only on the menu domain and should be constructed once, stored sparsely, and reused across candidate preferences, bootstrap or Gaussian draws, and specification tests. On incomplete domains, only observed pairs $T\subset S$ enter, so the relevant count can be much smaller than the complete-data expression.

When the objective is to recover the entire compatible-preference set, Corollary 2 of the main paper replaces explicit enumeration of $|X|!$ rankings with a mixed-integer linear program. Its directed variables $z_{ba}$ indicate $b\succ a$. Pairwise totality and transitivity make $z$ a strict linear order, and the remaining constraints impose $\succ$-Regularity. The formulation uses $|X|(|X|-1)$ directed binary variables, of which $\binom{|X|}{2}$ are independent after totality, cubically many ordering constraints, and one row for each observed regularity comparison. Because the ranking is an optimization variable, the direct mixed-integer formulation retains a row for every $a\in T\subset S$. Rows that become vacuous under the selected ranking do not change the feasible set, but retaining them avoids having to determine the worst alternative before solving. This is why its total constraint count exceeds $\mathfrak c_\succ$. Binary and nonbinary revealed-preference screens can fix coordinates before optimization. Model compatibility requires one feasibility solve; for each unresolved pair, a solve imposing the reverse comparison determines whether that comparison occurs in any compatible preference. Thus the number of mixed-integer queries grows quadratically rather than factorially in $|X|$.

The implementation can be tightened further without changing the identified set. One direction of each totality equality can be eliminated, and transitivity rows can be added lazily whenever an integer solution contains a directed cycle. The number of nested-menu inequalities remains the main complete-data bottleneck, so screening and sparse observed domains are complementary to the mixed-integer representation rather than substitutes for it. Solver status is part of the reported result: status-checked infeasibility establishes incompatibility, while an unresolved numerical status is not interpreted as rejection of the model.

The accompanying replication benchmark uses the public \texttt{aomIdentify} interface on deterministic, model-compatible choice systems. For each universe size it records the number of observed menus, $\mathfrak c_\succ$, the factorial ranking count, binary variables, constraints, mixed-integer solves, construction time, solve time, and total elapsed time. Separate feasibility-only and all-pairs calls isolate the cost of the initial model check from the additional revealed-preference queries. This design replaces machine-specific timing anecdotes with output that can be regenerated under the reported package version, solver, hardware, and numerical tolerance.

\begin{table}[t]
\centering
\caption{Deterministic computational benchmark for homogeneous AOM}
\label{tab:computation-aom}
\resizebox{\textwidth}{!}{%
\begin{tabular}{rrrrrrrrr}
\toprule
$|X|$ & $|\mathcal D|$ & $\mathfrak c_\succ$ & $|X|!$ & Binary & Constraints & Build & Feasibility & All pairs \\
\midrule
3 & 7 & 3 & 6 & 6 & 24 & $<0.01$ & $<0.01$ & $<0.01$ \\
4 & 15 & 26 & 24 & 12 & 106 & $<0.01$ & $<0.01$ & 0.01 \\
5 & 31 & 145 & 120 & 20 & 395 & 0.02 & 0.02 & 0.03 \\
6 & 63 & 664 & 720 & 30 & 1,401 & 0.09 & 0.09 & 0.13 \\
7 & 127 & 2,723 & 5,040 & 42 & 4,886 & 0.43 & 0.43 & 0.60 \\
\bottomrule
\end{tabular}%
}
\begin{flushleft}
\footnotesize Notes: Times are median elapsed seconds over 3 runs. The column $\mathfrak c_\succ$ counts the nonredundant fixed-preference inequalities used for inference. Constraints counts the direct mixed-integer rows retained by the software, including totality, transitivity, and algebraically redundant regularity rows. Build is the time spent validating the population and constructing this system in the feasibility-only call. Feasibility includes system construction and one solver call. All pairs includes system construction, the feasibility solve, and one reverse-comparison query for every unordered pair. The deterministic choice system is generated by full attention and the ordering $1\succ2\succ\cdots\succ|X|$ on the complete menu domain.
\end{flushleft}
\end{table}

Table \ref{tab:computation-aom} shows the combinatorial growth of the complete-data restriction system without a factorial search over rankings. From three to seven alternatives, $\mathfrak c_\succ$ rises from $3$ to $2{,}723$ and the mixed-integer system from $24$ to $4{,}886$ rows. On the reported machine, constructing and solving the seven-alternative feasibility problem takes about half a second, and adding all $\binom{7}{2}$ reverse-comparison queries raises the total to less than one second. Because performance depends on instance difficulty and the observed domain, we report the package version, solver, hardware, tolerance, benchmark design, and scaling quantities needed to reproduce and interpret these timings.

\subsection{Heterogeneous H-LAO}

H-LAO has a target-specific computational hierarchy. Pairwise preference shares use Proposition 5 of the main paper and require only singleton and binary menus. Alternative-specific rank probabilities use Block--Marschak sums once $f$ is recovered, and full-attention agreement is computed directly as $\sum_{a\in S}\phi(a|S)f(a|S)$. None of these calculations enumerates rankings. General joint preference events and sharp incomplete-domain bounds use the preference polytope in Proposition \ref{prop:observed-domain-hlao}. Under benchmark independence, its direct formulation has $|X|!$ ranking-probability variables, one adding-up equation, and at most $\sum_{S\in\mathcal D}|S|$ choice equations. The dependence-robust and no-SPI formulations are larger because they add menu-specific preference--stopping variables or latent prefix masses.

For a direct benchmark solve, one column is constructed for each ranking. Its entries equal one in the adding-up row and, in each observed choice row, the recovered probability of the prefixes on which that ranking selects the corresponding alternative. Columns with identical choice-equation entries and the same target value are observationally equivalent and can be collapsed before optimization. This compression can be substantial on sparse menu domains even though the formal ranking count remains $|X|!$.

When the ranking matrix is too large to materialize, the benchmark-independent event LP admits column generation. A Phase-I restricted master with artificial variables first produces a feasible collection of rankings. The master dual assigns a shadow value to the adding-up and observed-choice equations. Mixed-integer pricing then minimizes the reduced cost over strict linear orders, using pairwise-order and prefix-winner variables with transitivity constraints. A negatively priced ranking is added to the master; otherwise the current endpoint is optimal up to the reported tolerance. The same procedure with the sign of the target reversed yields the other endpoint. Generated rankings are reused across endpoints and related events.

Column generation avoids storing irrelevant ranking columns but does not remove worst-case factorial complexity: pricing itself is a linear-ordering problem. A numerical endpoint is therefore reported as certified only when the master and pricing solvers terminate successfully, primal and dual residuals satisfy the requested tolerance, the final reduced cost is no smaller than minus that tolerance, and the resulting optimality-gap bound is acceptable. The \texttt{ramchoice} diagnostics record these quantities, the number of generated columns, pricing dimensions, iteration count, and solver statuses. Exhaustive four-alternative tests reproduce both endpoints. In a six-alternative software test, a cap of $100$ forces column generation and only $18$ of the possible $720$ ranking columns are needed. Dependence-robust and no-SPI event bounds currently use direct enumeration because introducing a ranking creates a block of menu-specific variables rather than one master column.

The replication benchmark compares direct enumeration and certified column generation on the same deterministic H-LAO choice systems. It records $|X|!$, observed menus and choice equations, enumerated and active ranking columns, pricing variables and constraints, iterations, certification diagnostics, and elapsed time. These general scalability results are distinct from the H-LAO Monte Carlo runtimes in Section SA-6, which include estimation and inference for specific data-generating processes.

\begin{table}[t]
\centering
\caption{Deterministic computational benchmark for H-LAO}
\label{tab:computation-hlao}
\resizebox{\textwidth}{!}{%
\begin{tabular}{rrrrrrrrrc}
\toprule
$|X|$ & $|\mathcal D|$ & Choice eq. & $|X|!$ & Enumerated & Active & Pricing binary & Enumeration & Column generation & Certified \\
\midrule
3 & 3 & 6 & 6 & 6 & 5 & 16 & 0.01 & 0.01 & Yes \\
4 & 4 & 10 & 24 & 24 & 9 & 32 & 0.01 & 0.02 & Yes \\
5 & 5 & 15 & 120 & 120 & 15 & 55 & 0.01 & 0.05 & Yes \\
6 & 6 & 21 & 720 & 720 & 18 & 86 & 0.03 & 0.11 & Yes \\
7 & 7 & 28 & 5,040 & 5,040 & 28 & 126 & 0.14 & 0.35 & Yes \\
\bottomrule
\end{tabular}%
}
\begin{flushleft}
\footnotesize Notes: Times are median elapsed seconds over 3 runs. The deterministic suffix-domain system has outside-option probability $0.2$ and selects the first listed alternative with probability $0.8$. Both methods bound the event that alternative 1 is top ranked. Active is the number of ranking columns retained by column generation. Certified requires successful master and pricing statuses and all residual, reduced-cost, and gap checks at the reported tolerance.
\end{flushleft}
\end{table}

Table \ref{tab:computation-hlao} illustrates the tradeoff. At seven alternatives, direct enumeration materializes all $5{,}040$ ranking columns, whereas column generation certifies the same event bounds using $28$. Direct enumeration is faster at the small sizes in this benchmark because repeated mixed-integer pricing has a fixed cost. Column generation's advantage is instead that memory and master-problem dimensions depend on the rankings that matter for the target rather than on $|X|!$, which becomes decisive once the full ranking matrix is too large to construct or store.

\section{Simulation Evidence}\label{section: simulation evidence}

This section provides the simulation designs and complete results for the procedures developed in the paper. The first subsection studies homogeneous AOM, reporting finite-sample size and power under alternative menu supports, size at a boundary null, and design-specific computation time. The second evaluates H-LAO estimation, confidence sets, specification tests, robustness to departures from the baseline assumptions, subject-clustered inference, and design-specific computation time.

\subsection{Homogeneous AOM}

The universe contains six alternatives. The data-generating preference is $a_1\succ a_2\succ\cdots\succ a_6$. We use the logit attention rule of \citet{Brady-Rehbeck_2016_ECMA},
\begin{align*}
    \mu(T|S)=\frac{|T|^\varsigma}{\sum_{T'\subseteq S}|T'|^\varsigma},
\end{align*}
where $|T|$ is the cardinality of $T$. We set $\varsigma=2$. Direct calculation verifies attention overload: attention frequency depends only on menu size and equals $0.833$, $0.750$, $0.700$, $0.667$, and $0.643$ for $|S|=2,3,4,5,6$, respectively. We compute the implied choice rule from Definition 2 of the main paper and generate independent menu-level samples with $N_S\in\{50,100,200,500\}$ observations per menu.

We test the four hypothesized preferences
\begin{align*}
    \succ_1&:\ a_1\succ a_2\succ a_3\succ a_4\succ a_5\succ a_6,
    &\succ_2&:\ a_2\succ a_3\succ a_4\succ a_5\succ a_6\succ a_1,\\
    \succ_3&:\ a_1\succ a_2\succ a_6\succ a_5\succ a_4\succ a_3,
    &\succ_4&:\ a_1\succ a_6\succ a_5\succ a_4\succ a_3\succ a_2.
\end{align*}
The data-generating ordering $\succ_1$ is compatible with AOM, so its rejection probability measures size. The other three orderings are incompatible under complete data, although some remain compatible on restricted menu domains. Table \ref{tab:simulation-aom-generated} reports rejection probabilities from 2,000 Monte Carlo replications and 2,000 critical-value draws per replication at nominal level $0.05$. The first four result columns use the all-inequalities zero-centered critical value covered directly by Theorem 7 of the main paper; the last four apply the GMS refinement described in Section \ref{subsubsection: gms supplemental}.

The complete-data design uses all $57$ menus of size at least two and contains $664$ nonredundant inequalities. None is violated by the data-generating ordering. The other three orderings violate $90$, $6$, and $23$ inequalities, respectively. Their rejection probabilities constitute a fixed-alternative power analysis. Power is highest against $\succ_2$, which places the true top alternative last and violates many inequalities. The other alternatives are closer to the identified set and are correspondingly harder to reject.

The remaining menu supports omit selected menu sizes. The number of inequalities consequently ranges from $664$ with complete data to $15$ with only binary menus and the grand menu. A dagger in the tables marks an ordering that satisfies every population inequality on the indicated domain. Both procedures are conservative for these compatible orderings. When the retained menus expose a violation, rejection probabilities generally increase with sample size; when they do not, the corresponding ordering is correctly retained.

The least-favorable procedure provides substantial size protection in this baseline design: its largest rejection probability for a compatible ordering is $0.009$. GMS reduces the contribution of deeply slack inequalities to the effective Gaussian maximum and raises power while keeping rejection probabilities at most $0.035$ for compatible orderings. With complete data and $N_S=200$, GMS raises rejection of the most separated false ordering from $0.208$ to $0.478$; at $N_S=500$, the corresponding probabilities are $0.680$ and $0.910$. With only binary menus and the grand menu, GMS raises rejection of the sole incompatible ordering from $0.179$ to $0.523$ at $N_S=200$ and from $0.555$ to $0.872$ at $N_S=500$. These comparisons quantify the practical power gain from moment selection alongside the theorem-backed least-favorable baseline.

\begin{table}[t]
\centering
\caption{Finite-sample size and power for homogeneous AOM}
\label{tab:simulation-aom-generated}
\resizebox{\textwidth}{!}{%
\begin{tabular}{lrrcccccccc}
\toprule
& & & \multicolumn{4}{c}{All inequalities (LF)} & \multicolumn{4}{c}{GMS refinement} \\
\cmidrule(lr){4-7} \cmidrule(lr){8-11}
$\{|S|:S\in\mathcal D\}$ & $N_S$ & $\mathfrak c_\succ$ & $\succ_1$ & $\succ_2$ & $\succ_3$ & $\succ_4$ & $\succ_1$ & $\succ_2$ & $\succ_3$ & $\succ_4$ \\
\midrule
2--6 & 50 & 664 & 0.001$^{\dagger}$ & 0.029 & 0.009 & 0.029 & 0.017$^{\dagger}$ & 0.191 & 0.035 & 0.070 \\
2--6 & 100 & 664 & 0.002$^{\dagger}$ & 0.087 & 0.017 & 0.054 & 0.011$^{\dagger}$ & 0.279 & 0.055 & 0.139 \\
2--6 & 200 & 664 & 0.000$^{\dagger}$ & 0.208 & 0.030 & 0.120 & 0.003$^{\dagger}$ & 0.478 & 0.128 & 0.351 \\
2--6 & 500 & 664 & 0.000$^{\dagger}$ & 0.680 & 0.197 & 0.569 & 0.002$^{\dagger}$ & 0.910 & 0.533 & 0.880 \\
\cdashline{1-11}
3--6 & 50 & 439 & 0.001$^{\dagger}$ & 0.015 & 0.002 & 0.003 & 0.021$^{\dagger}$ & 0.119 & 0.020 & 0.019 \\
3--6 & 100 & 439 & 0.002$^{\dagger}$ & 0.038 & 0.003 & 0.011 & 0.006$^{\dagger}$ & 0.140 & 0.011 & 0.033 \\
3--6 & 200 & 439 & 0.001$^{\dagger}$ & 0.043 & 0.005 & 0.029 & 0.005$^{\dagger}$ & 0.175 & 0.023 & 0.089 \\
3--6 & 500 & 439 & 0.000$^{\dagger}$ & 0.104 & 0.019 & 0.159 & 0.001$^{\dagger}$ & 0.304 & 0.070 & 0.400 \\
\cdashline{1-11}
4--6 & 50 & 159 & 0.002$^{\dagger}$ & 0.018 & 0.008$^{\dagger}$ & 0.007 & 0.035$^{\dagger}$ & 0.095 & 0.032$^{\dagger}$ & 0.026 \\
4--6 & 100 & 159 & 0.003$^{\dagger}$ & 0.017 & 0.005$^{\dagger}$ & 0.006 & 0.025$^{\dagger}$ & 0.094 & 0.024$^{\dagger}$ & 0.021 \\
4--6 & 200 & 159 & 0.001$^{\dagger}$ & 0.028 & 0.001$^{\dagger}$ & 0.001 & 0.009$^{\dagger}$ & 0.111 & 0.009$^{\dagger}$ & 0.015 \\
4--6 & 500 & 159 & 0.000$^{\dagger}$ & 0.034 & 0.000$^{\dagger}$ & 0.005 & 0.003$^{\dagger}$ & 0.136 & 0.003$^{\dagger}$ & 0.042 \\
\cdashline{1-11}
5--6 & 50 & 24 & 0.004$^{\dagger}$ & 0.010 & 0.008$^{\dagger}$ & 0.009$^{\dagger}$ & 0.032$^{\dagger}$ & 0.057 & 0.032$^{\dagger}$ & 0.029$^{\dagger}$ \\
5--6 & 100 & 24 & 0.002$^{\dagger}$ & 0.019 & 0.004$^{\dagger}$ & 0.004$^{\dagger}$ & 0.026$^{\dagger}$ & 0.083 & 0.026$^{\dagger}$ & 0.022$^{\dagger}$ \\
5--6 & 200 & 24 & 0.002$^{\dagger}$ & 0.015 & 0.002$^{\dagger}$ & 0.002$^{\dagger}$ & 0.013$^{\dagger}$ & 0.067 & 0.013$^{\dagger}$ & 0.011$^{\dagger}$ \\
5--6 & 500 & 24 & 0.001$^{\dagger}$ & 0.024 & 0.001$^{\dagger}$ & 0.001$^{\dagger}$ & 0.005$^{\dagger}$ & 0.082 & 0.005$^{\dagger}$ & 0.004$^{\dagger}$ \\
\cdashline{1-11}
2, 3, 4, 6 & 50 & 370 & 0.000$^{\dagger}$ & 0.026 & 0.011 & 0.032 & 0.011$^{\dagger}$ & 0.162 & 0.044 & 0.088 \\
2, 3, 4, 6 & 100 & 370 & 0.000$^{\dagger}$ & 0.086 & 0.019 & 0.057 & 0.003$^{\dagger}$ & 0.275 & 0.073 & 0.178 \\
2, 3, 4, 6 & 200 & 370 & 0.000$^{\dagger}$ & 0.215 & 0.050 & 0.148 & 0.002$^{\dagger}$ & 0.507 & 0.175 & 0.404 \\
2, 3, 4, 6 & 500 & 370 & 0.000$^{\dagger}$ & 0.665 & 0.206 & 0.549 & 0.001$^{\dagger}$ & 0.909 & 0.584 & 0.881 \\
\cdashline{1-11}
2, 3, 6 & 50 & 115 & 0.001$^{\dagger}$ & 0.040 & 0.013 & 0.035 & 0.006$^{\dagger}$ & 0.139 & 0.051 & 0.094 \\
2, 3, 6 & 100 & 115 & 0.000$^{\dagger}$ & 0.085 & 0.026 & 0.058 & 0.000$^{\dagger}$ & 0.233 & 0.097 & 0.182 \\
2, 3, 6 & 200 & 115 & 0.000$^{\dagger}$ & 0.174 & 0.049 & 0.111 & 0.001$^{\dagger}$ & 0.447 & 0.225 & 0.374 \\
2, 3, 6 & 500 & 115 & 0.000$^{\dagger}$ & 0.545 & 0.205 & 0.409 & 0.000$^{\dagger}$ & 0.865 & 0.621 & 0.821 \\
\cdashline{1-11}
2, 6 & 50 & 15 & 0.000$^{\dagger}$ & 0.040 & 0.000$^{\dagger}$ & 0.000$^{\dagger}$ & 0.000$^{\dagger}$ & 0.180 & 0.000$^{\dagger}$ & 0.001$^{\dagger}$ \\
2, 6 & 100 & 15 & 0.000$^{\dagger}$ & 0.087 & 0.000$^{\dagger}$ & 0.000$^{\dagger}$ & 0.000$^{\dagger}$ & 0.302 & 0.000$^{\dagger}$ & 0.000$^{\dagger}$ \\
2, 6 & 200 & 15 & 0.000$^{\dagger}$ & 0.179 & 0.000$^{\dagger}$ & 0.000$^{\dagger}$ & 0.000$^{\dagger}$ & 0.523 & 0.000$^{\dagger}$ & 0.000$^{\dagger}$ \\
2, 6 & 500 & 15 & 0.000$^{\dagger}$ & 0.555 & 0.000$^{\dagger}$ & 0.000$^{\dagger}$ & 0.000$^{\dagger}$ & 0.872 & 0.000$^{\dagger}$ & 0.000$^{\dagger}$ \\
\bottomrule
\end{tabular}
}%
\par
\begin{minipage}{0.94\textwidth}
\footnotesize Notes: Entries are Monte Carlo rejection probabilities at the five-percent level based on 2,000 replications and 2,000 critical-value draws. LF is the all-inequalities feasible Gaussian test corresponding to Theorem 7; GMS is the Andrews--Soares refinement with \texttt{MNRatioGMS}$=1/\log(N)$. A dagger marks a preference ordering that satisfies the population inequalities on the indicated menu support. Daggered entries measure size; the remaining entries measure power against fixed alternatives. The quantity $\mathfrak c_\succ$ is the number of $\succ$-Regularity inequalities. All menus of the listed sizes are observed. The four rankings are defined in the text. Each observed menu has the reported sample size.
\end{minipage}
\end{table}

The baseline null lies in the interior of the population inequality region. To evaluate size when many inequalities bind, we use a second data-generating process based on a nondegenerate capacity-one mixture of search priorities. All four reported orderings satisfy the population inequalities with maximum slack zero; the numbers of binding inequalities for $\succ_1$ through $\succ_4$ are $65$, $115$, $186$, and $301$. Thus every entry in Table \ref{tab:simulation-aom-boundary-generated} is a boundary-null rejection probability.

\begin{table}[t]
\centering
\caption{Boundary-null size for homogeneous AOM}
\label{tab:simulation-aom-boundary-generated}
\begin{tabular}{rcccccccc}
\toprule
& \multicolumn{4}{c}{All inequalities (LF)} & \multicolumn{4}{c}{GMS refinement} \\
\cmidrule(lr){2-5} \cmidrule(lr){6-9}
$N_S$ & $\succ_1$ & $\succ_2$ & $\succ_3$ & $\succ_4$ & $\succ_1$ & $\succ_2$ & $\succ_3$ & $\succ_4$ \\
\midrule
50 & 0.011 & 0.010 & 0.018 & 0.022 & 0.060 & 0.045 & 0.066 & 0.067 \\
100 & 0.009 & 0.009 & 0.020 & 0.031 & 0.065 & 0.059 & 0.067 & 0.068 \\
200 & 0.009 & 0.012 & 0.018 & 0.029 & 0.068 & 0.064 & 0.067 & 0.071 \\
500 & 0.003 & 0.009 & 0.013 & 0.021 & 0.065 & 0.071 & 0.060 & 0.067 \\
\bottomrule
\end{tabular}
\begin{minipage}{0.91\textwidth}
\footnotesize Notes: Entries are rejection probabilities at the five-percent level based on 2,000 replications and 2,000 critical-value draws. The DGP is a nondegenerate capacity-one priority mixture for which all four reported orderings satisfy the population inequalities with maximum slack zero. The numbers of binding inequalities for $\succ_1$ through $\succ_4$ are 65, 115, 186, 301 respectively. Thus every entry measures size at a boundary null rather than at the strictly interior null used in the baseline size-and-power design.
\end{minipage}
\end{table}

At the boundary, the least-favorable procedure rejects with probability between $0.003$ and $0.031$, while GMS rejection probabilities range from $0.045$ to $0.071$. GMS therefore concentrates rejection closer to the nominal level, modestly exceeding it in some cells while delivering the power gains documented above. The experiment makes this size--power tradeoff transparent: the all-inequalities procedure remains the theorem-backed primary method, and GMS provides a useful power refinement when greater boundary sensitivity is acceptable.

Table \ref{tab:simulation-aom-runtime} reports end-to-end execution time at $N_S=500$. Average time per replication ranges from $0.17$ seconds with $24$ inequalities to $2.31$ seconds with all $664$ inequalities. LF and GMS take nearly identical time because both evaluate the same restriction system and use the same number of Gaussian draws. Absolute times are machine dependent, but the comparison shows that the reported power refinement has essentially no additional computational cost in these designs.

\begin{table}[t]
\centering
\caption{Average homogeneous-AOM execution time at $N_S=500$}
\label{tab:simulation-aom-runtime}
\begin{tabular}{lrrr}
\toprule
$\{|S|:S\in\mathcal D\}$ & $\mathfrak c_\succ$ & LF & GMS \\
\midrule
2--6 & 664 & 2.31 & 2.30 \\
3--6 & 439 & 1.55 & 1.53 \\
4--6 & 159 & 0.51 & 0.51 \\
5--6 & 24 & 0.17 & 0.17 \\
2, 3, 4, 6 & 370 & 2.21 & 2.19 \\
2, 3, 6 & 115 & 0.86 & 0.86 \\
2, 6 & 15 & 0.25 & 0.25 \\
\bottomrule
\end{tabular}
\begin{minipage}{0.80\textwidth}
\footnotesize Notes: Entries are average wall-clock seconds per Monte Carlo replication. Each method is run end to end and jointly tests the four reported rankings using the same simulated sample and common Gaussian draws. A call includes data processing, construction and evaluation of the restrictions, and 2,000 critical-value draws. LF and GMS are run consecutively on the same compute node within each replication; absolute times remain machine dependent.
\end{minipage}
\end{table}

\clearpage
\subsection{Heterogeneous H-LAO}

We use a separate Monte Carlo experiment to evaluate the H-LAO estimators, confidence sets, specification diagnostics, and robustness envelopes. The experiment varies the dimensions that govern identification and finite-sample performance: the marginal preference distribution, the stopping rule, terminal reach, observed menu support, dependence between preferences and stopping, and Sequential Path Independence. Designs H01--H08 and H11 have four alternatives; H09 and H10 have six. For each configuration, we generate independent menu-level samples with $N_S\in\{50,100,200,500\}$ observations per observed menu. Table \ref{tab:hlao-simulation-design} records the eleven configurations.

\begin{table}[t]
\centering
\caption{H-LAO Monte Carlo configurations}
\label{tab:hlao-simulation-design}
\resizebox{\textwidth}{!}{%
\begin{tabular}{lcllllll}
\toprule
Design & $|X|$ & Preference distribution & Stopping rule & Reach & Menu support & Dependence & SPI \\
\midrule
H01 & 4 & Diffuse      & Geometric             & High & Full            & Independent & Yes \\
H02 & 4 & Concentrated & Geometric             & High & Full            & Independent & Yes \\
H03 & 4 & Diffuse      & Rank dependent        & High & Prefix rich     & Independent & Yes \\
H04 & 4 & Diffuse      & Alternative dependent & High & Prefix rich     & Independent & Yes \\
H05 & 4 & Diffuse      & Geometric             & Low  & Full            & Independent & Yes \\
H06 & 4 & Diffuse      & Rank dependent        & Low  & Prefix rich     & Independent & Yes \\
H07 & 4 & Concentrated & Alternative dependent & Low  & Prefix rich     & Dependent & Yes \\
H08 & 4 & Diffuse      & Rank dependent        & Zero & Prefix rich     & Independent & Yes \\
H09 & 6 & Diffuse      & Geometric             & High & Sparse pairwise & Independent & Yes \\
H10 & 6 & Concentrated & Rank dependent        & Low  & Sparse pairwise & Independent & Yes \\
H11 & 4 & Late concentrated & Menu dependent & Moderate & Full & Independent & No \\
\bottomrule
\end{tabular}%
}
\begin{flushleft}
\footnotesize Notes: Full support contains every nonempty menu. Prefix-rich support contains the menus required by the recursive reach construction for the selected lists. Sparse-pairwise support contains singleton and binary menus and targets the ranking-free pairwise procedure. In H07, preferences and stopping are coupled while their marginal distributions are held fixed. H11 satisfies prefix consideration and attention overload but violates Sequential Path Independence (SPI).
\end{flushleft}
\end{table}

The diffuse preference distribution assigns positive mass to every strict ordering, while the concentrated distribution places three quarters of its mass on one ordering and spreads the remainder over all orderings. Geometric stopping uses a common continuation probability. Rank-dependent stopping varies continuation with list position, and alternative-dependent stopping varies it with the identity of the item reached. The high-, low-, and zero-reach designs respectively evaluate stable recursion, weak-denominator inference, and identified-set inference at prefixes where point recovery is unavailable. H07 couples preferences and stopping while preserving their marginal distributions, thereby separating the dependence-robust analysis from the benchmark independence interpretation.

H11 isolates Sequential Path Independence. It places $0.95$ of preference mass on the reverse-list ordering and distributes the remaining $0.05$ according to the diffuse design. For the global list $1,2,3,4$, let $h=(0.92,0.82,0.72,0.62)$ and $d_k=1-0.10(k-1)$. If alternative $a$ appears in menu $S$, its latent reach probability is
\begin{align*}
    \phi(a|S)=h_a d_{|S|}.
\end{align*}
The resulting reach probabilities decrease along every observed list and weakly decrease when the menu expands, so the implied prefix masses are nonnegative and attention overload holds. Sequential Path Independence fails already at $S=\{1,2\}$:
\begin{align*}
\phi(2|\{1,2\})=0.82(0.90)=0.738
\neq \phi(1|\{1,2\})\phi(2|\{2\})
=0.828(0.82)=0.67896.
\end{align*}
The failure is observable through a particularly simple population certificate. The SPI recovery assigns masses $0.172$, $0.14904$, and $0.67896$ to $\emptyset$, $\{1\}$, and $\{1,2\}$, respectively. Every decision maker who stops at $\{1\}$ chooses alternative $1$, regardless of any dependence between preference and stopping, so either SPI polytope requires
\begin{align*}
\pi(1|\{1,2\})\geq0.14904.
\end{align*}
The H11 population instead has $\pi(1|\{1,2\})=0.1071558$, a violation of about $0.0418842$. Equivalently, the independent SPI recursion produces $f(1|\{1,2\})=-0.0616888$. This one inequality certifies that both the independent and dependence-robust SPI polytopes are empty. By contrast, the true no-SPI prefix masses $0.172$, $0.09$, and $0.738$, coupled independently with the stated preference distribution, reproduce the population choices and furnish an explicit feasible point of the no-SPI polytope.

For point estimation, we report bias and root mean squared error for recovered reach probabilities and, when the observed support and SPI permit the recursion, the full-attention rule $f$. For inference under independence and SPI, we report familywise coverage---the probability that every valid pairwise target is covered in a replication---and average width for three simultaneous pairwise procedures: finite-sample Hoeffding projection, hybrid Gaussian/exact-binomial projection, and Bonferroni studentization of the undivided moments. The last method uses the exact quadratic inversion implemented in the software; its reported width is the total length if the Fieller set is disconnected, and it returns $[0,1]$ at an exactly degenerate moment.

For four-alternative designs, the preference event is that alternative $4$ is preferred to alternative $1$. We compute dependence-robust projection intervals under SPI and no-SPI projection intervals for the targeted H01, H05, H07, and H11 configurations. Their population widths quantify how relaxing preference--stopping independence and then SPI changes identification. H07 is generated outside the independent benchmark but inside both robust models. H11 is generated outside both SPI models but inside the no-SPI model. All projection procedures begin from simultaneous menu-level multinomial bands. We compare finite-sample Hoeffding bands with the hybrid region defined in Section SA-4.10. The latter uses Gaussian calibration with the plug-in menu-level multinomial covariance structure for regular cells and exact Clopper--Pearson bounds for boundary and sparse cells, with the common error-budget allocation stated above.

The specification experiment uses four alternatives, complete menu data, high reach, and a preference distribution on the boundary of a Block--Marschak inequality. It contains a correctly specified null and local or fixed violations of either Recovered Attention Overload or Block--Marschak non-negativity at $N_S\in\{100,200,500\}$. The local attention-overload and Block--Marschak perturbations are $0.75/\sqrt{N_S}$ and $3/\sqrt{N_S}$, respectively; the fixed perturbations are $0.08$ and $0.30$. We compare three simultaneous diagnostics: a finite-sample Hoeffding outer test, a hybrid Gaussian/exact-binomial outer test, and the direct delta-Gaussian test available in this regular positive-reach design. The reported rejection probabilities use the targeted restriction under each alternative and the omnibus test under the null.
The experiment uses 2,000 Monte Carlo replications and 2,000 critical-value draws at nominal level $0.05$. The complete design has 44 menu-iid H-LAO estimation and inference tasks, four subject-clustered inference tasks, and 15 specification-diagnostic tasks.

\begin{table}[t]
\centering
\caption{H-LAO point estimation}
\label{tab:simulation-hlao-estimation}
\begin{tabular}{lrrrrr}
\toprule
& & \multicolumn{2}{c}{Reach probabilities} & \multicolumn{2}{c}{Full-attention rule} \\
\cmidrule(lr){3-4} \cmidrule(lr){5-6}
Design & $N_S$ & Bias & RMSE & Bias & RMSE \\
\midrule
H01 & 100 & -0.000 & 0.036 & -0.000 & 0.059 \\
H01 & 500 & 0.000 & 0.016 & -0.000 & 0.026 \\
\addlinespace[2pt]
H02 & 100 & -0.000 & 0.036 & -0.000 & 0.036 \\
H02 & 500 & -0.000 & 0.016 & -0.000 & 0.016 \\
\addlinespace[2pt]
H03 & 100 & -0.000 & 0.034 & -- & -- \\
H03 & 500 & -0.000 & 0.015 & -- & -- \\
\addlinespace[2pt]
H04 & 100 & 0.000 & 0.036 & -- & -- \\
H04 & 500 & -0.000 & 0.016 & -- & -- \\
\addlinespace[2pt]
H05 & 100 & -0.000 & 0.041 & -0.000 & 0.260 \\
H05 & 500 & 0.000 & 0.018 & -0.000 & 0.111 \\
\addlinespace[2pt]
H06 & 100 & -0.001 & 0.043 & -- & -- \\
H06 & 500 & 0.000 & 0.019 & -- & -- \\
\addlinespace[2pt]
H07 & 100 & -0.000 & 0.042 & -- & -- \\
H07 & 500 & 0.000 & 0.019 & -- & -- \\
\addlinespace[2pt]
H08 & 100 & 0.000 & 0.019 & -- & -- \\
H08 & 500 & 0.000 & 0.008 & -- & -- \\
\addlinespace[2pt]
H09 & 100 & -0.000 & 0.034 & -- & -- \\
H09 & 500 & 0.000 & 0.015 & -- & -- \\
\addlinespace[2pt]
H10 & 100 & 0.000 & 0.046 & -- & -- \\
H10 & 500 & 0.000 & 0.020 & -- & -- \\
\bottomrule
\end{tabular}
\begin{minipage}{0.88\textwidth}
\footnotesize Notes: Bias and RMSE are averaged across replications and the estimands available in each design. Dashes indicate that the menu support does not identify the recursive full-attention rule.
\end{minipage}
\end{table}

Table \ref{tab:simulation-hlao-estimation} shows that reach and full-attention estimates are essentially unbiased in every design where the target is identified. Reach RMSE ranges from $0.019$ to $0.046$ at $N_S=100$ and from $0.008$ to $0.020$ at $N_S=500$. Full-attention recovery is less precise under weak reach, as expected: in H05 its RMSE falls from $0.260$ to $0.111$, compared with $0.059$ to $0.026$ in the otherwise similar high-reach H01 design.

\begin{table}[t]
\centering
\caption{H-LAO pairwise confidence sets}
\label{tab:simulation-hlao-inference}
\begingroup
\renewcommand{\arraystretch}{0.90}
\resizebox{\textwidth}{!}{%
\begin{tabular}{lrrrrrrr}
\toprule
& & \multicolumn{2}{c}{Hoeffding projection} & \multicolumn{2}{c}{Gaussian projection} & \multicolumn{2}{c}{Studentized moment} \\
\cmidrule(lr){3-4} \cmidrule(lr){5-6} \cmidrule(lr){7-8}
Design & $N_S$ & Familywise & Width & Familywise & Width & Familywise & Width \\
\midrule
H01 & 50 & 1.000 & 0.871 & 1.000 & 0.821 & 0.926 & 0.456 \\
H01 & 100 & 1.000 & 0.789 & 1.000 & 0.626 & 0.936 & 0.321 \\
H01 & 200 & 1.000 & 0.656 & 1.000 & 0.434 & 0.945 & 0.226 \\
H01 & 500 & 1.000 & 0.416 & 1.000 & 0.272 & 0.955 & 0.143 \\
\cdashline{1-8}
H02 & 50 & 1.000 & 0.924 & 1.000 & 0.522 & 0.787 & 0.256 \\
H02 & 100 & 1.000 & 0.595 & 1.000 & 0.314 & 0.866 & 0.195 \\
H02 & 200 & 1.000 & 0.409 & 0.994 & 0.211 & 0.903 & 0.138 \\
H02 & 500 & 1.000 & 0.268 & 0.998 & 0.132 & 0.932 & 0.088 \\
\cdashline{1-8}
H03 & 50 & 1.000 & 0.836 & 0.999 & 0.780 & 0.938 & 0.404 \\
H03 & 100 & 1.000 & 0.755 & 1.000 & 0.589 & 0.941 & 0.285 \\
H03 & 200 & 1.000 & 0.607 & 1.000 & 0.385 & 0.938 & 0.201 \\
H03 & 500 & 1.000 & 0.380 & 1.000 & 0.241 & 0.960 & 0.127 \\
\cdashline{1-8}
H04 & 50 & 1.000 & 0.894 & 1.000 & 0.834 & 0.933 & 0.500 \\
H04 & 100 & 1.000 & 0.814 & 1.000 & 0.688 & 0.941 & 0.351 \\
H04 & 200 & 1.000 & 0.694 & 1.000 & 0.481 & 0.952 & 0.247 \\
H04 & 500 & 1.000 & 0.444 & 1.000 & 0.300 & 0.946 & 0.155 \\
\cdashline{1-8}
H05 & 50 & 1.000 & 1.000 & 1.000 & 0.995 & 0.887 & 0.876 \\
H05 & 100 & 1.000 & 0.999 & 1.000 & 0.964 & 0.921 & 0.705 \\
H05 & 200 & 1.000 & 0.988 & 1.000 & 0.877 & 0.938 & 0.510 \\
H05 & 500 & 1.000 & 0.890 & 1.000 & 0.687 & 0.951 & 0.317 \\
\cdashline{1-8}
H06 & 50 & 1.000 & 0.998 & 1.000 & 0.978 & 0.926 & 0.751 \\
H06 & 100 & 1.000 & 0.986 & 1.000 & 0.906 & 0.945 & 0.556 \\
H06 & 200 & 1.000 & 0.922 & 1.000 & 0.793 & 0.941 & 0.387 \\
H06 & 500 & 1.000 & 0.784 & 1.000 & 0.557 & 0.951 & 0.241 \\
\cdashline{1-8}
H08 & 50 & 1.000 & 1.000 & 1.000 & 1.000 & 1.000 & 1.000 \\
H08 & 100 & 1.000 & 1.000 & 1.000 & 1.000 & 1.000 & 1.000 \\
H08 & 200 & 1.000 & 1.000 & 1.000 & 1.000 & 1.000 & 1.000 \\
H08 & 500 & 1.000 & 1.000 & 1.000 & 1.000 & 1.000 & 1.000 \\
\cdashline{1-8}
H09 & 50 & 1.000 & 0.872 & 1.000 & 0.822 & 0.914 & 0.510 \\
H09 & 100 & 1.000 & 0.791 & 1.000 & 0.639 & 0.938 & 0.359 \\
H09 & 200 & 1.000 & 0.668 & 1.000 & 0.440 & 0.939 & 0.252 \\
H09 & 500 & 1.000 & 0.422 & 1.000 & 0.276 & 0.952 & 0.159 \\
\cdashline{1-8}
H10 & 50 & 1.000 & 1.000 & 1.000 & 0.996 & 0.150 & 0.407 \\
H10 & 100 & 1.000 & 1.000 & 1.000 & 0.809 & 0.566 & 0.312 \\
H10 & 200 & 1.000 & 0.915 & 1.000 & 0.426 & 0.740 & 0.243 \\
H10 & 500 & 1.000 & 0.538 & 0.998 & 0.244 & 0.866 & 0.156 \\
\bottomrule
\end{tabular}
}
\endgroup
\begin{minipage}{0.91\textwidth}
\footnotesize Notes: Familywise coverage is the probability that all valid pairwise targets are covered in a replication; width is averaged over pairwise targets and replications. H07 is omitted because pairwise point identification requires preference--stopping independence. The studentized procedure inverts the undivided moment with Bonferroni calibration; width is the total length when the confidence set is disconnected. At an exactly degenerate moment it reports $[0,1]$.
\end{minipage}
\end{table}

Table \ref{tab:simulation-hlao-inference} shows exceptionally reliable familywise coverage: it equals one for Hoeffding projection in every reported design and ranges from $0.994$ to one for Gaussian projection. The accompanying intervals are wider, especially under weak reach. The studentized undivided-moment procedure is much shorter but relies on regular-case approximation; its familywise coverage ranges from $0.150$ to $0.960$ across the positive-reach designs. H10, which combines concentrated preferences, sparse pairwise support, and low reach, is the most demanding configuration: coverage rises from $0.150$ at $N_S=50$ to $0.866$ at $N_S=500$. At the exactly degenerate H08 boundary, every procedure correctly returns $[0,1]$, making visible the distinction between finite-sample validity and informativeness at zero reach. These results strongly support projection intervals as the primary procedure and studentized intervals as a shorter regular-case diagnostic.

The subject-clustered experiment is calibrated to the repeated-choice structure and numbers of independent subjects in the empirical application. To isolate within-subject dependence, it uses a balanced full-menu design: each subject draws one preference ranking that remains fixed across the full 15-menu domain and makes three choices per menu, while stopping draws vary across choices. We use $G\in\{21,37\}$ subjects, so $N_S=3G\in\{63,111\}$, under either diffuse preferences (C01) or concentrated preferences (C02). The cluster-Hoeffding and cluster-multiplier procedures propagate simultaneous cluster-robust primitive-probability bands; the cluster-studentized procedure applies the shorter undivided-moment inversion with cluster calibration.

\begin{table}[t]
\centering
\caption{Subject-clustered H-LAO pairwise confidence sets}
\label{tab:simulation-hlao-cluster}
\resizebox{\textwidth}{!}{%
\begin{tabular}{llrrrrrrrr}
\toprule
& & & & \multicolumn{2}{c}{Cluster Hoeffding} & \multicolumn{2}{c}{Cluster multiplier} & \multicolumn{2}{c}{Cluster studentized} \\
\cmidrule(lr){5-6} \cmidrule(lr){7-8} \cmidrule(lr){9-10}
Design & Preference heterogeneity & $G$ & $N_S$ & Familywise & Width & Familywise & Width & Familywise & Width \\
\midrule
C01 & Diffuse & 21 & 63 & 1.000 & 0.970 & 1.000 & 0.898 & 0.902 & 0.619 \\
C01 & Diffuse & 37 & 111 & 1.000 & 0.910 & 1.000 & 0.751 & 0.921 & 0.469 \\
\addlinespace[2pt]
C02 & Concentrated & 21 & 63 & 1.000 & 1.000 & 1.000 & 0.947 & 0.703 & 0.312 \\
C02 & Concentrated & 37 & 111 & 1.000 & 0.998 & 1.000 & 0.552 & 0.797 & 0.270 \\
\bottomrule
\end{tabular}
}
\begin{minipage}{0.96\textwidth}
\footnotesize Notes: Each subject draws one preference ranking that is held fixed across the full 15-menu domain and makes three choices per menu; stopping draws vary across choices. Hence $N_S=3G$ and observations are dependent within subject. The values $G=21$ and $G=37$ mirror the complete-design and dominance-consistent empirical samples. Familywise coverage requires simultaneous coverage of all six pairwise preference shares in a replication; width is averaged over the six targets and replications.
\end{minipage}
\end{table}

Table \ref{tab:simulation-hlao-cluster} shows familywise coverage equal to one for cluster-Hoeffding projection and between $0.9995$ and one for cluster-multiplier projection. Cluster-studentized coverage is $0.902$ and $0.921$ under diffuse preferences, but only $0.703$ and $0.797$ under concentrated preferences. Its intervals are shorter, but the loss of simultaneous coverage is material at precisely the cluster counts used in the application. This evidence reinforces the use of cluster projection intervals as the primary empirical results.

\begin{table}[t]
\centering
\caption{Sensitivity to preference--stopping independence and Sequential Path Independence}
\label{tab:simulation-hlao-sensitivity}
\resizebox{\textwidth}{!}{%
\begin{tabular}{lrrrrrrrr}
\toprule
& & \multicolumn{3}{c}{Population identified-set width} & \multicolumn{2}{c}{Dependence robust} & \multicolumn{2}{c}{No SPI} \\
\cmidrule(lr){3-5} \cmidrule(lr){6-7} \cmidrule(lr){8-9}
Design & $N_S$ & Independent & Dependent & No SPI & Coverage & Width & Coverage & Width \\
\midrule
H01 & 100 & 0.000 & 0.190 & 0.590 & 1.000 & 0.675 & 1.000 & 0.749 \\
H01 & 500 & 0.000 & 0.190 & 0.590 & 1.000 & 0.408 & 1.000 & 0.661 \\
\addlinespace[2pt]
H05 & 100 & 0.000 & 0.750 & 0.873 & 1.000 & 0.983 & 1.000 & 0.983 \\
H05 & 500 & 0.000 & 0.750 & 0.873 & 1.000 & 0.909 & 1.000 & 0.921 \\
\addlinespace[2pt]
H07 & 100 & -- & 0.914 & 0.914 & 1.000 & 0.992 & 1.000 & 0.992 \\
H07 & 500 & -- & 0.914 & 0.914 & 1.000 & 0.951 & 1.000 & 0.951 \\
\addlinespace[2pt]
H11 & 100 & -- & -- & 0.456 & -- & -- & 1.000 & 0.608 \\
H11 & 500 & -- & -- & 0.456 & -- & -- & 1.000 & 0.527 \\
\bottomrule
\end{tabular}
}
\begin{minipage}{0.96\textwidth}
\footnotesize Notes: Population columns report the width of the sharp identified interval for the designated preference event. Sampling columns use hybrid Gaussian/exact-binomial probability bands. H07 violates preference--stopping independence; H11 violates Sequential Path Independence while preserving prefix consideration, attention overload, and a stable preference marginal. Dashes mark a misspecified model or an inapplicable coverage calculation.
\end{minipage}
\end{table}

Table \ref{tab:simulation-hlao-sensitivity} shows that the sensitivity designs reproduce the theoretical nesting of the identified sets. In H01, for example, the population width rises from zero under independence to $0.190$ under dependence and $0.590$ without Sequential Path Independence. The corresponding widths in weak-reach H05 are $0$, $0.750$, and $0.873$. The robust projection regions cover in every reported design, and their average widths decrease toward the corresponding population identified-set widths as sample size increases.

\begin{table}[t]
\centering
\caption{H-LAO specification diagnostics}
\label{tab:simulation-hlao-diagnostics}
\begin{tabular}{lrrrrr}
\toprule
Design & $N_S$ & Violation & Hoeffding outer & Gaussian outer & Direct delta \\
\midrule
Null & 100 & 0.000 & 0.000 & 0.000 & 0.018 \\
Null & 200 & 0.000 & 0.000 & 0.000 & 0.013 \\
Null & 500 & 0.000 & 0.000 & 0.000 & 0.011 \\
\addlinespace[2pt]
AO local & 100 & 0.075 & 0.000 & 0.000 & 0.114 \\
AO local & 200 & 0.053 & 0.000 & 0.005 & 0.102 \\
AO local & 500 & 0.034 & 0.000 & 0.005 & 0.092 \\
\addlinespace[2pt]
AO fixed & 100 & 0.080 & 0.000 & 0.000 & 0.144 \\
AO fixed & 200 & 0.080 & 0.000 & 0.020 & 0.593 \\
AO fixed & 500 & 0.080 & 0.000 & 0.874 & 0.986 \\
\addlinespace[2pt]
BM local & 100 & 0.300 & 0.000 & 0.000 & 0.160 \\
BM local & 200 & 0.212 & 0.000 & 0.000 & 0.126 \\
BM local & 500 & 0.134 & 0.000 & 0.000 & 0.142 \\
\addlinespace[2pt]
BM fixed & 100 & 0.300 & 0.000 & 0.000 & 0.171 \\
BM fixed & 200 & 0.300 & 0.000 & 0.000 & 0.502 \\
BM fixed & 500 & 0.300 & 0.000 & 0.034 & 0.977 \\
\bottomrule
\end{tabular}
\begin{minipage}{0.94\textwidth}
\footnotesize Notes: Entries in the final three columns are rejection probabilities at the five-percent level. The null row uses the omnibus diagnostic; AO and BM alternatives use their corresponding targeted restrictions. Local violations shrink at the parametric rate, while fixed violations do not depend on sample size.
\end{minipage}
\end{table}

Table \ref{tab:simulation-hlao-diagnostics} shows that the direct delta-Gaussian diagnostic combines conservative null rejection probabilities of $0.011$--$0.018$ with power of $0.986$ and $0.977$ against the fixed attention-overload and Block--Marschak violations at $N_S=500$. Both outer tests make no false rejections under correct specification in these designs. Among them, the Gaussian outer test reaches power of $0.874$ against the fixed attention-overload violation, while power against the Block--Marschak alternative is limited; the Hoeffding outer test prioritizes finite-sample validity and remains highly conservative at these sample sizes. The results distinguish the roles of the procedures: outer tests provide finite-sample or boundary-robust safeguards, while the direct regular-case diagnostic is highly informative when positive reach and differentiability are credible.

\begin{figure}[t]
\centering
\includegraphics[width=0.85\textwidth]{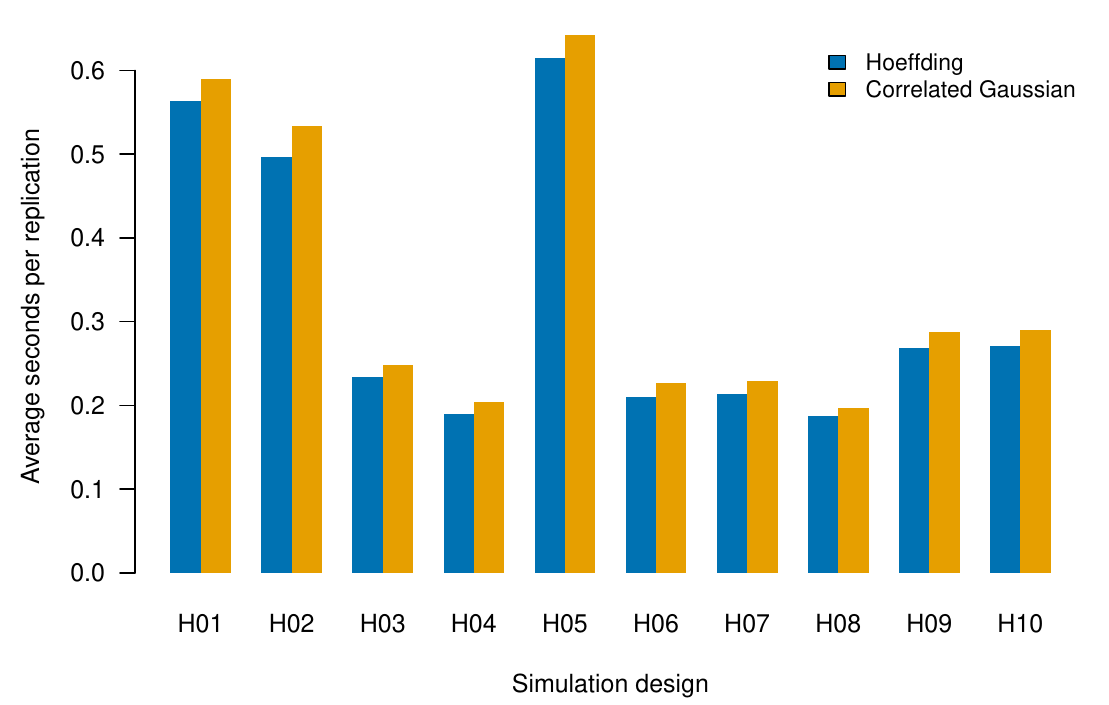}
\caption{Computation time for H-LAO projection inference at $N_S=500$.}
\label{fig:simulation-hlao-runtime}
\end{figure}

For H01--H10 at $N_S=500$, Figure \ref{fig:simulation-hlao-runtime} reports average end-to-end time per replication below $0.65$ seconds for both projection procedures. Runtime is highest in the full-menu designs and lower with prefix-rich or sparse pairwise support. Correlated-Gaussian projection is slightly slower because it estimates and simulates from the joint covariance structure, but the difference is small throughout.

\clearpage

\clearpage
\singlespacing
\addcontentsline{toc}{section}{References}
\bibliographystyle{econometrica}
\bibliography{CCMM_2026_AttOverload--bib}